\documentclass[aps,prx,twocolumn,longbibliography,superscriptaddress,showpacs]{revtex4-2}
\usepackage{times}
\usepackage{graphicx}
\usepackage{latexsym}
\usepackage{amssymb}
\usepackage{amsmath}
\usepackage{amsfonts}
\usepackage{mathtools}
\usepackage{dsfont}
\usepackage{bbm}
\usepackage{bm}
\usepackage{multirow}
\usepackage{array}
\usepackage{color}
\usepackage{tikz}
\usetikzlibrary{shapes}
\usepackage[normalem]{ulem}
\usepackage{comment}
\usepackage{footmisc}
\usepackage{xcolor}
\newcommand{\ket}[1]{\vert#1\rangle}
\newcommand{\bra}[1]{\langle#1\vert}
\usepackage[colorlinks=true, citecolor={blue!80!black}, urlcolor={blue!50!black}, linkcolor = {blue!80!black}]{hyperref}
\allowdisplaybreaks[1]
\usepackage[percent]{overpic}
 \usepackage[utf8]{inputenc}

\newcommand{\Ucal}{
\begin{tikzpicture}[baseline=-0.5ex]
  \draw[line width=0.8pt, fill=none] (0,0) circle (0.19cm);
  \node[color=black] at (0,0) {\small $\mathcal{U}$};
\end{tikzpicture}}

\DeclareMathAlphabet{\mathbbold}{U}{bbold}{m}{n}

\def\i{\textbf{i}}

\definecolor{forestgreen}{rgb}{0.13, 0.55, 0.13}
\definecolor{gr}{rgb}{0,0.82,0.18}
\definecolor{DarkerRed}{rgb}{0.9,0.05,0.25}

\definecolor{darkred_malte}{HTML}{BB5566}
\definecolor{darkblue_malte}{HTML}{004488}
\definecolor{darkyellow_malte}{HTML}{DDAA33}

\definecolor{color1_malte}{HTML}{EE6677}
\definecolor{color2_malte}{HTML}{DDAA33}
\definecolor{color3_malte}{HTML}{4477AA}
\definecolor{color4_malte}{HTML}{82218B}

\DeclareMathOperator\arctanh{arctanh}

\begin{document}

\title{Higher Nishimori Criticality and Exact Results at the Learning Transition of Deformed Toric Codes}

\date{\today}

\author{Rushikesh A. Patil}
\affiliation{Department of Physics, University of California, Santa Barbara, CA 93106, USA}

 \author{Malte P\"utz}
 \affiliation{Institute for Theoretical Physics, University of Cologne, Z\"ulpicher Straße 77, 50937 Cologne, Germany}

 \author{Simon Trebst}
 \affiliation{Institute for Theoretical Physics, University of Cologne, Z\"ulpicher Straße 77, 50937 Cologne, Germany}

 \author{Guo-Yi Zhu}
\affiliation{The Hong Kong University of Science and Technology (Guangzhou), Nansha, Guangzhou, 511400, Guangdong, China}

\author{Andreas W. W. Ludwig}
\affiliation{Department of Physics, University of California, Santa Barbara, CA 93106, USA}


\begin{abstract}
Motivated by the observation of a learning-induced tricritical point, at which three phases with strong, weak, and broken $\mathbb{Z}_2$ symmetry meet,
we revisit the phase diagram of a deformed toric code wavefunction subjected to weak measurements. 
This setting is exactly dual to the classical Bayesian inference problem of the
$2D$ classical Ising model under bond-energy measurements. 
Here we demonstrate that this tricritical point lies on a distinct \textit{higher Nishimori line},
which has an {\it emergent} gauge-invariant formulation, just like the {ordinary} Nishimori 
line but with a higher replica symmetry as a replica stat-mech model in the replica number $R\rightarrow2$ limit, 
where disorder is averaged according to the Born / Bayes rule.
As such, the learning tricritical point is in fact a $\textit{higher Nishimori critical point}$.
Using this identification, we obtain a number of \textit{exact results} at this \textit{higher} Nishimori critical point;
e.g., we show that the power-law exponent of the Edwards-Anderson (EA) correlation function is
exactly equal to that of the spin correlation function at the unmeasured Ising critical point
-- an observation that is readily supported in our numerical simulations.
In addition, we obtain a number of exact bounds on the power-law exponents of higher measurement-averaged moments of the spin-spin correlation function.
We also show that the scaling dimension for the second moment of the \textit{dual} spin correlation function at the {\it higher} Nishimori critical point
vanishes, and that the scaling dimensions of all moments higher than second
are \textit{non-positive}.
An analogous {\it higher} Nishimori critical point exists also
for the Bayesian inference problem of the general $D$-dimensional
classical Ising model when $D>1$, which again allows us to obtain exact
results for the EA correlator, and to
use its existence to determine the 
phase diagram of the classical Ising critical point (or, equivalently, the corresponding conformal
quantum critical 
`Rokhsar-Kivelson' wavefunction) subjected to
bond-energy measurements {in general dimension $D$}.
Finally, coming back to the two-dimensional case, we establish -- using
the tools employed in the proof of 
a
recent $c$-effective theorem [arXiv:2507.07959] -- that the 
Casimir effective central charge {$c_{\rm eff}$}, a characteristic of the universality class, 
\textit{decreases} under the
renormalization group (RG)
flow from the {\it higher} Nishimori critical point 
to the unmeasured $2D$ Ising critical point, and is thus greater than $1/2$.
This is corroborated by extensive numerical simulations finding a Casimir effective central charge $c_{\rm eff} = 0.522(1)$,
and a sharp decrease towards $c_{\rm eff} = 1/2$ as one moves towards the Ising critical point.
The analytical result also explains, with a certain physically motivated assumption, 
the numerically observed \textit{increase} of the Casimir effective central charge under the RG flow from the ordinary Nishimori critical point 
to the clean Ising critical point in the $2D$ random-bond Ising model.

\end{abstract}
\maketitle

\section{Introduction}
\label{SecIntro}

Mappings to statistical mechanics systems with quenched randomness have served as a powerful tool to understand the effects of decoherence or measurements on quantum memories.
At the center of these efforts resides 
the problem of the toric code~\cite{KITAEV20032} with incoherent errors, 
for which the decoding problem can be
mapped to the Nishimori line in the $2D$ random-bond Ising model (RBIM)~\cite{DennisKitaevLandahlPreskill} such that the optimal threshold for quantum error correction (QEC) in the presence of such incoherent errors 
maps exactly to the Nishimori critical point in the $2D$ RBIM~\cite{Nishimori_1980,Nishimori1981,NishimoriBook}.
More recently, the problem of state preparation using quantum measurements has also been mapped  to the Nishimori line in the RBIM, establishing a threshold for successful state preparation~\cite{ZhuTantivasadakarnVishwanthTrebstVerresen,LeeJiFisherBi,Chen2025}. 
In both settings, the \textit{decodability} transition at the threshold  
can be understood as a \textit{mixed-state phase transition}, which has been studied by a number of recent works in the context of 
both decoherence and measurements on the toric code ground state~\cite{FanBaoAltmanVishwanath,BaoFanVishwanathAltman,LeeJianXu,SalaGopalakrishnanOshikawaYou,EllisonCheng,WangWuWang,YHChenGrover,YHChenGroverSeparability,LessaStrongToWeak,SohalPrem,SuYangJian,LeeExactCalculation,EcksteinPRX,DecoherenceWangEtAl,eckstein2025learningtransitionstopologicalsurface}.
Interestingly, the Nishimori line and the Nishimori critical point also permit a {\it classical} Bayesian inference perspective~\cite{Iba_1999}
that predates the quantum measurement/decoherence perspectives -- as it turns out, the problem of an {\it infinite} temperature $2D$ classical Ising model 
under Bayesian bond-energy measurements is exactly dual to the problem of performing Born-rule Pauli-$Z$ measurements on the toric code ground state~\cite{EcksteinPRX,PutzGarrattNishimoriTrebstZhu, ZhuTantivasadakarnVishwanthTrebstVerresen,LeeJiFisherBi, eckstein2025learningtransitionstopologicalsurface}.
Recently, some of us expanded this perspective and studied~\cite{PutzGarrattNishimoriTrebstZhu} a \textit{deformed} toric code wavefunction~\cite{CastelnovoChamon,PapanikolaouRamanFradkin,ArdonneFedleyFradkin,IsakovFendleyLudwigTrebstTroyer, Zhu19deform, Kim23deform, Verresen25deform}, away from its stabilizer limit, subjected to Born-rule measurements with the Pauli-$Z$ operator, which is exactly dual to the classical Bayesian inference problem for the $2D$ classical Ising model but now at \textit{finite} temperatures. 
In this analysis, a novel tricritical point was discovered in this expanded learning/Bayesian inference phase diagram~\cite{PutzGarrattNishimoriTrebstZhu}, 
in parallel to independent work~\cite{NahumJacobsen}
arguing for the existence of an additional critical point in the context of the 2D classical critical Ising model
under Bayesian bond-energy measurements.
In the dual quantum (Rokhsar-Kivelson wavefunction) formulation~\cite{PutzGarrattNishimoriTrebstZhu}, 
this novel \textit{learning} tricritical point sits at the junction of strong, weak and broken $\mathbb{Z}_2$ symmetric phases.
Its universal critical properties  are dominated by the {\it intrinsic randomness of quantum measurements}.

\begin{figure*}[t]
\centering
\includegraphics[width=\textwidth]{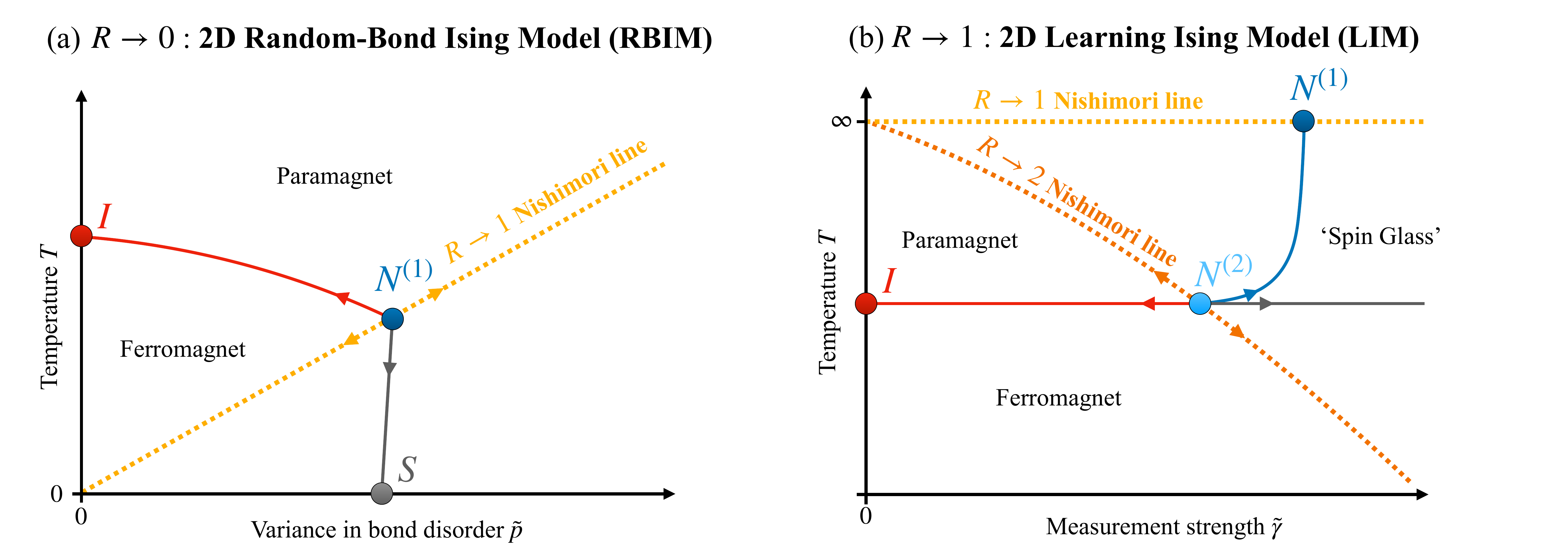}
\caption{
{\bf Comparison of the phase diagram for the $\mathbf{2D}$ RBIM (the replica $R\rightarrow0$ theory) and the 
(classical) $\mathbf{2D}$ Learning Ising Model (LIM) 
(the replica $R\rightarrow1$ theory)}, with Gaussian bond-disorder 
and Gaussian measurements, respectively. 
The latter phase diagram is equivalent to the phase diagram of the deformed toric code wavefunction under Pauli-$Z$ measurements, in which the paramagnetic, `spin glass', and ferromagnetic phases with strong, weak, and broken 
$\mathbb{Z}_2$ symmetry of the dual quantum (Rokhsar-Kivelson) formulation,
correspond to the topological quantum memory, topological classical memory and no-memory phases, respectively 
(see Ref.~\cite{PutzGarrattNishimoriTrebstZhu}). 
The dotted light orange line in the left phase diagram and the dotted orange line in the right phase diagram denote, respectively, 
the ordinary $T=\tilde{p}$ (with $\tilde{p}$ being the variance in bond-disorder of unit mean strength)
and the {\it higher} Nishimori line $T=1/\tilde{\gamma}^{2}$ (where $\tilde{\gamma}^2=\Delta$ 
characterizes the strength of Gaussian measurements, see Eq.~\eqref{eq:PmsigmaGaussian} below).
Note that the ordinary Nishimori line (in light orange) also appears as the $\beta=0$ line in the learning phase diagram, 
and the ordinary Nishimori critical point $N^{(1)}$ also appears in both the left (a) and the right (b) phase diagrams.
The tricritical point $N^{(2)}$ in the (right) learning phase diagram lies on the {\it higher} Nishimori line (dotted orange line), 
and is hence a {\it higher} Nishimori critical point.
Since the universal critical properties are expected to be independent of whether we perform Gaussian or binary measurements [see the discussion 
at the end of  Sec.~\ref{SecEmergentGaugeAndEnlargedPermutationSymmetry}], 
the long-distance properties of the  tricritical point for binary measurements are also governed by the same {\it higher} Nishimori critical point.
\label{FigComparison}}
\end{figure*}

In this work, we revisit this special tricritical point and reveal some of its novel universal critical properties \textit{exactly}.
We show that this tricritical point in the 
{deformed toric code or $2D$ classical}
Ising learning phase diagram is a \textit{higher Nishimori critical point}, which
{resides}
in a different universality class than the {\it ordinary} Nishimori critical point.
{The familiar ordinary Nishimori line~\cite{Nishimori_1980,Nishimori1981} and the 
corresponding ordinary {Nishimori}
critical point, which
appears 
ubiquitously in 
a variety of 
problems
{with criticality in monitored quantum circuits~\cite{ZhuTantivasadakarnVishwanthTrebstVerresen,Chen2025,PutzGarrattNishimoriTrebstZhu,eckstein2025learningtransitionstopologicalsurface,hauser2024informationdynamicsdecoheredquantum,hauser2025informationdynamicssymmetrybreaking}, mixed-state phase transitions in toric code states under decoherence~\cite{LeeJianXu,FanBaoAltmanVishwanath,SalaGopalakrishnanOshikawaYou,EllisonCheng,WangWuWang,YHChenGrover,YHChenGroverSeparability,LessaStrongToWeak,SohalPrem,SuYangJian,LeeExactCalculation,DecoherenceWangEtAl},}
related quantum error correction 
{problems}~\cite{DennisKitaevLandahlPreskill,wan2025revisitingnishimorimulticriticalitylens,MullerEtAll2024,DiehlEtAl2025} and also Bayesian inference problems in
classical statistical mechanics models~\cite{Iba_1999,PutzGarrattNishimoriTrebstZhu,NahumJacobsen,KimKeyserlingkLamacraft},
was
{originally}
understood as a line in the random-bond Ising model (RBIM) that has a gauge
invariant
formulation with 
enlarged replica
permutation symmetry~\cite{LeDoussalHarrisI,LeDoussalHarrisII,GeorgesHanselLeDoussalMaillard,GRL2001}.
More specifically, the phase diagram of the RBIM, illustrated on the left-hand side of Fig.~\ref{FigComparison}(a),
is governed by a $R\rightarrow0$ replica theory and the 
ordinary
Nishimori line in this phase diagram has a gauge-invariant formulation 
as an
enlarged
replica-symmetric theory in the $R\rightarrow1$ limit.
The entire {deformed toric code or $2D$ classical} Ising learning phase diagram discussed in Refs.~\cite{PutzGarrattNishimoriTrebstZhu,NahumJacobsen}, which is a problem with quenched randomness introduced by \textit{measurements} and not a traditional `impurity-type' {uncorrelated} quenched disorder problem like the RBIM, is governed by a replica theory in the 
limit of $R\to 1$ replicas~\cite{JianYouVasseurLudwig2019,BaoChoiAltman2019}.
As illustrated on the right-hand side of Fig.~\ref{FigComparison}(b),
the ordinary Nishimori line appears at inverse temperature $\beta=0$ (undeformed toric code) in the 
learning phase diagram. 
As we will discuss below,
there is a distinct second  
Nishimori line with a
gauge-invariant and enlarged replica permutation-symmetric formulation
that governs the long-distance behavior of a distinct part of
{the deformed toric code or the $2D$ classical}
Ising learning phase diagram.
{In particular, the replica permutation symmetry of the gauge-invariant formulation of this new Nishimori line is {\it larger} than that of the ordinary Nishimori line,}
and we refer to it as the \textit{higher Nishimori line}. 
To contrast with the RBIM (the replica limit $R\rightarrow0$) and the ordinary Nishimori line (the gauge-invariant replica limit $R\rightarrow1$), the {\it higher} Nishimori line has a gauge-invariant
formulation with 
enlarged
replica permutation symmetry described by the  $R\rightarrow2$ replica limit
{in the 
learning phase diagram which is otherwise described by the $R\rightarrow 1$ replica limit.}
We show that the critical point on this {\it higher} Nishimori line governs the universal properties of the tricritical point 
in the 
{learning phase diagram {of Ref.~\cite{PutzGarrattNishimoriTrebstZhu} for}
the deformed toric code or the $2D$ classical Ising model,} 
using a logic analogous to that
found in Ref.~\cite{LeDoussalHarrisI,GeorgesHanselLeDoussalMaillard,LeDoussalHarrisII}.}
That is, we show that the novel tricritical point in the 
{deformed toric code or $2D$ classical}
Ising learning phase diagram is a {\it higher} Nishimori critical point.\\

\subsection*{Summary of results}

Let us summarize some of our key results before going through the analytical arguments in great detail in the following.

\subsubsection*{Power-law decay of Edwards-Anderson correlator
\\ and dual spin correlations}

\begin{figure}[t]
    \centering
    \includegraphics[width=0.9\linewidth]{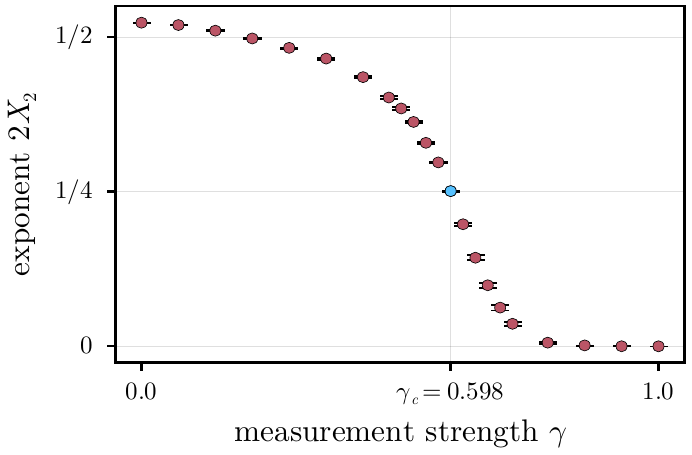}
    \caption{ {\bf Power-law exponent of the EA correlator.}
    Shown is the power-law exponent $2X_2$ 
    of the EA correlator as a function of the measurement strength $\gamma$ for the case of binary measurements. 
    Right at the \textit{higher} Nishimori critical point $N^{(2)}$ (at $\gamma_c = 0.598(2)$ \cite{PutzGarrattNishimoriTrebstZhu}), 
    the power-law exponent takes a value of $2X_2(\gamma_c) = 0.2508(8)$ close to the analytical value $1/4$. 
    The data shown is obtained by fitting the EA correlator in the bulk of a $512\times512$ system with open boundary conditions,
    as illustrated explicitly in Fig.~\ref{fig:EA-IsingTricritical} below. 
    Note that towards the unmeasured $2D$ Ising critical point, (marginally irrelevant) finite-size effects lead to a small deviation
    from the known exact value of 1/2.
    }
    \label{fig:EA-Correlations}
\end{figure}

We use this gauge-invariant and enlarged
replica-symmetric formulation in the $R\rightarrow2$ replica limit
to obtain {\it exact} results for 
the universal properties of the
\textit{learning}
tricritical point, i.e.\ the \textit{higher} Nishimori critical point.
{This}
includes the 
{\it exact}  critical exponent for the power-law decay {of} the Edwards-Anderson (EA) correlator,
an \textit{exact}
equality of
the long-distance properties
of the measurement-averaged $(2k-1)^{\text{th}}$ and
$(2k)^{\text{th}}$ (with $k$ being an arbitrary positive integer) moments of 
any multipoint spin
correlation function,
and
a number of
rigorous bounds on the scaling dimensions 
of
the measurement-averaged higher moments of the spin-spin correlation function, 
measurement-averaged moments of {its}
absolute value, and the typical 
spin-spin correlation function.
{See Table~\eqref{LabelTableSummaryOfScalingDimension} for a summary of our results for various scaling dimensions at the {\it higher} Nishimori critical
 point.}
As we demonstrate,
the emergent gauge {invariant} and enlarged permutation
symmetric formulation allows
us {to} conclude
in particular that the
{power-law}
exponent for the EA correlator is
{exactly} equal to that of the {measurement-averaged} 
expectation value (first moment) of
the 
{spin-spin}
{correlator 
at the {\it higher} Nishimori point.
{In measurement-averaged first moments, averaging over the measurement outcomes gets rid of any conditioning on the measurement outcomes (a consequence of the `POVM~\footnote{standing for `Positive Operator Valued Measure'} condition' satisfied by the Kraus operators in the quantum formulation), and hence}
the power-law of the {measurement-averaged} 
first moment of the spin-spin}
correlator at the {\it higher} Nishimori critical
point is the same {as} 
{that}
at the
{unmeasured}
$2D$ Ising critical point.
Thus, the 
power-law exponent of the
EA 
correlator
at the {\it higher} Nishimori
critical
point 
is the same {as} that of the
{two-point} spin 
{correlator}
at the unmeasured {$2D$} Ising 
critical point,
{i.e.\ }
\begin{equation}
    \nonumber\overline{\langle \sigma_{i}\sigma_{j}\rangle^2_{\vec{m}}}\sim\frac{1}{|i-j|^{1/4}} \,.
\end{equation}
This analytical value of 
the exponent is in excellent agreement with the numerically obtained value presented in Fig.~\ref{fig:EA-Correlations}.
Note that the unmeasured $2D$
classical Ising critical point yields
$\langle \sigma_{i}\sigma_{j}\rangle^2\sim \frac{1}{|i-j|^{1/2}} $ 
instead, also in agreement with
Fig.~\ref{fig:EA-Correlations}.

We also show, following the logic in~\cite{MerzChalker}, that the measurement-averaged {\it second} moment of the \textit{dual} spin correlation function at the {\it higher} Nishimori critical point has \textit{vanishing} scaling dimension, while the scaling dimension of the measurement-averaged {\it first} moment of the dual spin correlation function, analogous to other correlation functions, is equal to its value at the unmeasured Ising critical point~\cite{KadanoffCeva} ($=\frac{1}{8}$).
This, assuming analyticity in the moment order, implies multifractality
in the spectrum of scaling dimensions for the
dual spin correlation function at the {\it higher} Nishimori point, and also implies that the higher moments of the dual spin correlation after the second moment have \textit{non-positive} scaling dimensions.
This result at the {\it higher} Nishimori critical point should be contrasted with that at the ordinary Nishimori critical point in the $2D$ RBIM, where a multifractal spectrum of scaling dimensions has been established for the dual spin correlation function and where all the higher bond-randomness averaged moments 
beyond the \textit{first} moment have non-positive scaling dimensions~\cite{MerzChalker}.

\subsubsection*{Casimir effective central charge, RG flows, and \texorpdfstring{$c$}{Lg}-effective theorem}

In addition, by making use of the tools used in the proof of the $c$-effective theorem in
Ref.~\cite{PatilLudwig20251}, we also demonstrate
 the {\it decrease} of the {Casimir} effective 
central charge under the renormalization group (RG) flow from the {\it higher} Nishimori critical point to the unmeasured  {$2D$} Ising critical point.
That is, we show that the Casimir effective central charge of the {\it higher} Nishimori point, which characterizes the universal finite-size scaling of the Shannon entropy of the measurement record~\cite{NahumJacobsen,ZabaloGullansWilsonVasseurLudwigGopalakrishnanHusePixley,KumarKemalChakrabortyLudwigGopalakrishnanPixleyVasseur,DecoherenceWangEtAl} at this point, is greater than $1/2$, the central charge of the Ising critical point.
Interestingly, this result together with a certain physically motivated assumption, also explains the numerically observed~\cite{HoneckerPiccoPujol2001,PiccoPujolHonecher2006} \textit{increase} of the Casimir effective central charge under the RG flow from the ordinary Nishimori critical point to the \textit{clean} Ising critical point in the $2D$ RBIM (see App.~\ref{AppCommentOnTheCeffAtOrdinaryNishimori}, particularly Fig.~\ref{fig:Reversal-of-Central-Charge-Change}, for details of this argument).

\begin{figure}[t]
    \centering
    \includegraphics[width=0.9\linewidth]{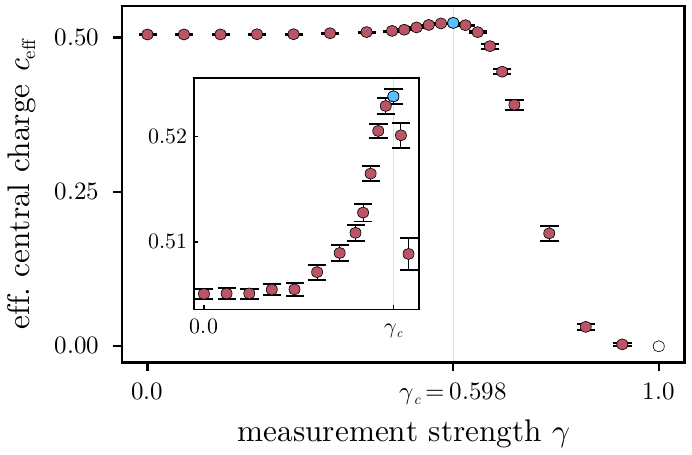}
    \caption{
    {\bf Casimir effective central charge and RG flows.} 
    Shown is the Casimir effective
    central charge along the $\beta = \beta_c$ line of our numerical phase diagram.   
    Right at the \textit{higher} Nishimori critical point $N^{(2)}$ in this numerical phase diagram
    (at $\gamma_c = 0.598(2)$ \cite{PutzGarrattNishimoriTrebstZhu}), 
    the Casimir effective central charge takes a value of $c_{\rm eff}(\gamma_c) = 0.524(1)$.
    Moving away from the critical measurement strength, the Casimir
    effective  central charge quickly drops, as shown in the inset, indicating
    renormalization group (RG) flows away from the higher Nishimori point.
    This data is obtained by fitting the slope of the free energy density as a function of $\pi / (6L^2)$ in the bulk of a cylinder with circumference $L$ 
    and length $1000L$ for system sizes of $L=8$ to $L=24$, as illustrated for a number of representative points in Fig.~\ref{fig:FreeEnergy} below.
    {[Note that the Casimir effective central charge estimated here is slightly different from the one obtained from fits to system sizes $L=8, \ldots, 32$
    shown in Fig.~\ref{fig:FreeEnergy} below, which we otherwise refer to.]}
    }
    \label{fig:CasimirCentralCharge_betac}
\end{figure}

\subsubsection*{General \texorpdfstring{$D>1$}{Lg} spatial dimensions}

{Let us further note that using {the} analogous argument, we can establish the existence of the
{\it higher} Nishimori critical point, which admits a formulation as an emergent gauge-invariant and enlarged replica-symmetric theory in the $R\rightarrow2$ replica limit, in the phase diagram of the 
classical Ising model {in
arbitrary $D>1$ dimensions}
under Bayesian bond-energy
measurements.}
{This 
{{\it higher} Nishimori critical point is the 
generalization
to arbitrary dimensions $D>1$
of the novel 
tricritical
point found in the learning phase diagram of the $2D$ classical Ising model in Ref.~\cite{PutzGarrattNishimoriTrebstZhu},
where the}
classical Bayesian inference problem {in any 
$D$ 
dimensions}
is equivalent to
the
problem
of performing Born-rule measurements with the bond-energy operator on the quantum 
Rokhsar-Kivelson (RK) wavefunction~\cite{CLHenley_2004,IsakovFendleyLudwigTrebstTroyer}
corresponding to the $D$-dimensional classical Ising model 
(see, e.g.,~\cite{PutzGarrattNishimoriTrebstZhu} and Appendix C of \cite{PatilLudwig20251}).}
{While the ordinary Nishimori critical point appears at a finite measurement strength and zero inverse-temperature in
the classical Ising learning
phase diagram, the latter 
{{\it higher} Nishimori critical}
point 
appears
at a suitable finite measurement strength and at the temperature corresponding to the critical point of the unmeasured $D$-dimensional classical Ising model.}
{In particular, 
for Gaussian measurements, given the location of the critical point in the (unmeasured) $D$-dimensional classical Ising model, we can also
determine 
{the}
exact location {of the {\it higher} Nishimori critical point in the learning phase diagram}.}
{Moreover,}
using exactly the same argument
as that used in 
two dimensions,
we can show that the power-law exponent for the EA correlator at the 
{{\it higher} Nishimori critical point}
in the learning phase diagram of {the} $D>2$ dimensional {classical} Ising model (or of the corresponding RK wavefunction) is also equal to 
the power-law exponent
of the spin-spin correlator
at the 
(unmeasured) {$D$-dimensional classical} Ising critical point.
Of course,
the scaling dimension of the spin operator at the (unmeasured) $3D$ Ising critical point is not known exactly, however, we can use the numerical conformal bootstrap~\cite{Chang2025,PhysRevD.86.025022} estimate for the latter, and that then also serves 
at the same time
as an accurate estimate for the power-law exponent of the EA correlator at the 
{{\it higher} Nishimori critical point}
in the 3D Ising or the corresponding RK wavefunction learning phase diagram.
For $D\geq 4$, the scaling dimension of the spin-operator at the (unmeasured) {classical} Ising critical point is given by its mean field value, and from the above
argument that also gives us the power law exponent of the EA correlator at the 
{{\it higher} Nishimori critical point}
in the $D\geq4$ dimensional
learning phase diagram.~\footnote{Note that the logarithmic corrections
to scaling
{of}
the spin-spin correlation function at the (unmeasured) {classical} Ising critical point in $D=4$ dimensions, will also appear in the EA correlation function
{at}
the 
{{\it higher} Nishimori critical point}
{in the $D=4$ dimensional}
learning phase diagram.}

Finally, we use 
the existence of the {\it higher} Nishimori critical point
to understand the structure of the phase diagram for 
the classical Ising critical point
under  Bayesian bond-energy measurements  in {\it arbitrary} dimension $D$.
In particular, two
{logical} possibilities for the Bayesian inference/learning
phase diagram
of the 
$D$-dimensional
{classical}
Ising critical point under 
bond-energy measurements were discussed in Ref.~\cite{NahumJacobsen} (see Fig.~$4$ in Ref.~\cite{NahumJacobsen}),
and 
we can rule out one of the possibilities by 
{using the existence of the {\it higher} Nishimori critical point at the}
temperature corresponding to the unmeasured critical point and at a 
finite strength of measurements in all dimensions $D>1$.
In particular, building on the discussion in Ref.~\cite{NahumJacobsen}, and using the fact that the
{\it higher} Nishimori critical point occurs at
decreasing measurement strengths with increasing dimension $D$, we arrive at the phase diagram
shown in our Fig.~\ref{fig:IsingLearningPDGeneral}.

{The remaining parts of the paper are structured as follows: 
In Sec.~\ref{SecModel}, following Ref.~\cite{PutzGarrattNishimoriTrebstZhu}, we discuss {the setup of three equivalent measurement problems:
of the deformed toric code wavefunction, of the quantum RK wavefunction corresponding to the $2D$ classical Ising model, and the Bayesian inference problem for the $2D$ classical Ising model. Then we discuss the replica theory that governs the long-distance properties of these 
measurement problems.}
In Sec.~\ref{SecEmergentGaugeAndEnlargedPermutationSymmetry}, we discuss the 
gauge-invariant and higher replica-symmetric formulation of a 
line 
in this replica theory 
and we argue that the critical point on this line{, the {\it higher} Nishimori critical point,} governs the long-distance physics of the novel 
\textit{learning} tricritical point.
In Sec.~\ref{SecExactResults}, we discuss 
a number of exact 
results for the universal critical properties at this {\it higher} Nishimori  critical point.
In Sec.~\ref{SecNumericalResults}, we present our numerical results for the {\it higher} Nishimori 
critical
point.
In Sec.~\ref{SecLearningTransitionsInHigherDimensional}, we discuss the phase diagram of the 
arbitrary 
$D>1$ dimensional {classical} Ising critical point under bond-energy measurements.}

\setcounter{tocdepth}{0}
\tableofcontents

\section{The Setup}
\label{SecModel}

{Following Ref.~\cite{PutzGarrattNishimoriTrebstZhu}, 
in this section
we will 
discuss the 
deformed toric code wavefunction and the
setup of
performing
weak measurements on it. 
As discussed below, this {problem} is 
{dual}
to the \textit{classical}
Bayesian inference
problem of performing
\textit{Bayesian} 
bond-energy measurements on the $2D$ classical Ising model.}
{Let us first consider a square lattice on a cylinder with qubits on the links $\langle ij\rangle$ of the square lattice.
The usual toric code (undeformed)
Hamiltonian
can be defined on this cylinder with suitable boundary conditions at the open boundaries, and given any toric code ground state $\ket{TC}$, we can define the deformed toric code wavefunction as~\cite{CastelnovoChamon,PapanikolaouRamanFradkin,IsakovFendleyLudwigTrebstTroyer,Zhu19deform,putz2025flownishimoriuniversalityweakly,ZhuTantivasadakarnVishwanthTrebstVerresen},
\begin{equation}
{\mathbf{\ket{TC}}_{\beta}} 
\propto\exp\big\{\frac{\beta}{2}\sum_{\langle ij\rangle}\hat{\sigma}^{z}_{ij}\big\}\ket{TC},\label{EqDeformedToricCodeWF}
\end{equation}
{where $\hat{\sigma}^{z}_{ij}$ is the Pauli-$Z$ operator on the link $\langle ij\rangle$.}
We want to study the effects of weak
Born-rule 
measurements with the $\hat{\sigma}^z_{ij}$ operator on the qubits of the above deformed toric code wavefunction. The Kraus operator for these measurements is given by
\begin{equation}\label{EqToricCodeKrausBinary}
    \hat{\mathcal{K}}_{\vec{m}}
    =\frac{e^{\frac{\tilde{\gamma}}{2}\sum_{\langle ij\rangle}m_{ij}\hat{\sigma}^z_{ij}}}{(2\cosh\tilde{\gamma})^{N_{b}/2}},\quad
    {m_{ij}=\pm 1,}
\end{equation}
where $\vec{m}=\{m_{ij}\}$ denotes the measurement outcomes,
{$\tilde{\gamma}\in [0,\infty]$ quantifies the measurement strength,} 
and $N_b$ is the number of
{bonds/links}
on the square lattice.
It is easily verified that the {above} Kraus operators $ \hat{\mathcal{K}}_{\vec{m}}$ satisfy the `POVM'~\cite{Note1} condition: $\sum_{\vec{m}}\hat{\mathcal{K}}_{\vec{m}}^{\dagger} \hat{\mathcal{K}}_{\vec{m}}=\hat{1}$,
where $\hat{1}$ is the identity operator. 

To study the above problem, we note that the wavefunction in Eq.~\eqref{EqDeformedToricCodeWF}
is exactly dual~\footnote{{The (undeformed) toric code wavefunction can be written as an equal weight superposition of certain eigenstates of the $\{\hat{\sigma}^{z}_{ij}\}$ operators~\cite{KITAEV20032}, where each 
eigenstate in the superposition
corresponds to
a loop configuration on the dual square lattice
{where}
the loops cut-through the links $\langle ij\rangle$ with $\hat{\sigma}^z_{ij}=-1$ on the original lattice. 
The factor of $e^{\frac{\beta}{2}\sum_{\langle ij\rangle}{\hat{\sigma}^{z}_{ij}} }$ acting on the (undeformed) toric code ground state in Eq~\eqref{EqDeformedToricCodeWF} assigns (up to an overall constant) a factor of $e^{-\beta \times(\text{length of the loop})}$ to each loop in every loop configuration state. 
Each loop configuration state can be associated to a domain wall configuration of 
classical 
spins $\{\sigma_i\}$ defined on the sites (instead of the links) of the original square lattice, where {the} configuration of spins $\{\sigma_i\}$ is uniquely defined up to a global 
{$\mathbb{Z}_2$ flip}, and the weight $e^{-{\beta} (\text{length of the loop})}$ is precisely the Boltzmann weight of the domain wall 
{in}
the $2D$ {classical} Ising 
{model}
in Eq.~\eqref{Eq2DIsingHamiltonian} at (inverse) temperature $\beta/2$.
Therefore, the deformed toric code wavefunction in Eq.~\eqref{EqDeformedToricCodeWF} and the RK wavefunction in Eq.~\eqref{EqRKWF} for the $2D$ classical Ising model are exactly dual to each other. 
Finally, following the above mapping, it is also clear that the operator $\hat{Z}_{ij}$ in the deformed toric code wavefunction in Eq.~\eqref{EqDeformedToricCodeWF} corresponds to the bond-energy operator $\hat{\sigma}^z_{i}\hat{\sigma}^z_{j}$ for the RK wavefunction in Eq.~\eqref{EqRKWF}.}}
{(see e.g.~\cite{putz2025flownishimoriuniversalityweakly,ZhuTantivasadakarnVishwanthTrebstVerresen})}
to the `Rokhsar-Kivelson' (RK) wavefunction~\cite{CLHenley_2004}
corresponding to
the 
$2D$ classical Ising model with {different} spins
{$\hat{\sigma}^z_i$}
{defined on the sites  $i$, as opposed to the links, of the square lattice}~\footnote{{The operator $\hat{\sigma}^z_{k}$ and
{its} eigenstates $\ket{\{\sigma_i\}}$ are naturally defined as $\hat{\sigma}^z_k\ket{\{\sigma_i\}}=\sigma_k\ket{\{\sigma_i\}}$.}}
\begin{align}
   \ket{RK}
   &=\frac{1}{\sqrt{Z}}\sum_{\{\sigma_{i}\}} 
   {e^{-
   {\beta\over 2}H[\{\sigma_i\}]}}
   \ket{\{\sigma_i\}}\label{EqRKWF}\\
   Z&=\sum_{\{\sigma_{i}\}}e^{-{\beta H[\{\sigma_i\}]}}
\end{align}
where $H[\{\sigma_i\}]$ ($\sigma_i=\pm 1$) is the Hamiltonian of the $2D$ classical Ising model given by
\begin{equation}
     H[\{\sigma_i\}]=-\sum_{\langle ij\rangle}\sigma_i\sigma_j\label{Eq2DIsingHamiltonian}
\end{equation}
with suitable boundary conditions at the open boundaries.}
{[In this work, we will be interested in the bulk critical phenomena and hence not worry about boundary conditions~\footnote{{Although, we note that the boundary conditions decide the ground state of the toric code on the cylinder (surface code), and hence are important in evaluations of properties like coherent information (see the discussion in Ref.~\cite{PutzGarrattNishimoriTrebstZhu}). 
In this work, we will be only interested in calculating bulk correlation functions that distinguish different phases in the phase diagram in Fig.~\ref{fig:2DGaussianMeasurements}}}.]}
{In turn, 
the 
problem of performing weak 
Born-rule 
measurements with the
{$\hat{\sigma}^z_{ij}$}
operator on the deformed toric code wavefunction in 
Eq.~\eqref{EqDeformedToricCodeWF}
is 
exactly dual~\cite{Note3} to 
the problem of weak 
Born-rule measurements with the bond-energy operator $\hat{\sigma}^z_{i}\hat{\sigma}^z_{j}$~\cite{Note4} on the RK wavefunction in Eq.~\eqref{EqRKWF}. Analogously to Eq.~\eqref{EqToricCodeKrausBinary}, the
Kraus operator for weak measurements with the bond-energy operator $\hat{\sigma}^z_{i}\hat{\sigma}^z_{j}$ on all links $\langle ij\rangle$ of the RK wavefunction in Eq.~\eqref{EqRKWF} is given by
\begin{equation}\label{EqKrausBinary}
    \hat{K}_{\vec{m}}
    =\frac{e^{\frac{\tilde{\gamma}}{2}\sum_{\langle ij\rangle}m_{ij}\hat{\sigma}^z_i\hat{\sigma}^z_j}}{(2\cosh\tilde{\gamma})^{N_{b}/2}},\quad
    {m_{ij}=\pm 1,}
\end{equation}
which also clearly satisfy the `POVM'~\cite{Note1} condition,
\begin{equation}
     \label{EqPOVMBinary}\sum_{\vec{m}}\hat{K}_{\vec{m}}^{\dagger} \hat{K}_{\vec{m}}=\hat{1}.
\end{equation}
We can thus focus on studying the latter equivalent measurement problem for the RK wavefunction, where
the post-measurement state of the RK wavefunction
corresponding to measurement outcomes $\vec{m}$ is given by,
\begin{equation}
\ket{\Psi_{\vec m}}
=
{{\hat {K}}_{\vec m} \ket{RK}
\over
\sqrt{\bra{RK}
{\hat { K}}^\dagger_{\vec m}
 {\hat { K}}_{\vec m}
\ket{RK}}},\label{EqPostMeasurementEnsemble}
\end{equation}
{and the Born rule probability of obtaining these measurement outcomes is
\begin{align}
    \tilde{P}(\vec{m})&=\bra{RK}
{\hat { K}}^\dagger_{\vec m}
 {\hat { K}}_{\vec m}
\ket{RK}\nonumber\\
&=\frac{{(2\cosh(\tilde{\gamma}))}^{-N_b}}{Z}\sum_{\{\sigma_i\}}\;e^{-\beta H[\{\sigma_i\}]+\tilde{\gamma}\sum_{\langle ij\rangle}m_{ij}\sigma_i\sigma_j}.\label{EqQuantumProbability}
\end{align}}
Finally, using Eq.~\eqref{EqRKWF} and~\eqref{EqPostMeasurementEnsemble}, the expectation value of a general operator 
$\hat{\mathcal{O}}_1=\hat{\sigma}^z_{i_1}\hat{\sigma}^z_{i_2}\hat{\sigma}^z_{i_3} \cdots\hat{\sigma}^z_{i_k}$, which
is diagonal in $\ket{\{\sigma_i\}}$ basis, 
in the above post-measurement state $\ket{\Psi_{\vec m}}$ is given by
\begin{align}
   \langle \hat{\mathcal{O}}_1\rangle_{\vec{m}} &=\bra{\Psi_{\vec m}}\hat{\mathcal{O}}_1\ket{\Psi_{\vec m}}\nonumber\\
   &=\frac{\sum_{\{\sigma_i\}}\mathcal{O}_1\;e^{-\beta H[\{\sigma_i\}]+\tilde{\gamma}\sum_{\langle ij\rangle}m_{ij}\sigma_i\sigma_j}}{Z[\vec{m}]}\label{EqQuantumExpValue}
\end{align}
where $\mathcal{O}_1={\sigma}_{i_1} {\sigma}_{i_2}{\sigma}_{i_3} \cdots{\sigma}_{i_k}$ is the eigenvalue of the operator $\hat{\mathcal{O}}_1=\hat{\sigma}^z_{i_1} \hat{\sigma}^z_{i_2}\hat{\sigma}^z_{i_3} \cdots\hat{\sigma}^z_{i_k}$ in the basis state $\ket{\{\sigma_i\}}$,
and we have defined $Z[\vec{m}]$ as
\begin{equation}
    Z[\vec{m}]=\sum_{\{\sigma_i\}}e^{-\beta H[\{\sigma_i\}]+\tilde{\gamma}\sum_{\langle ij\rangle}m_{ij}\sigma_i\sigma_j}\label{EqZ[m]}.
\end{equation}
{{Since the measurement operator $\hat{\sigma}^z_{i}\hat{\sigma}^z_{j}$ is diagonal in the $\ket{\{\sigma_i\}}$ basis,
the above discussed
problem
of performing Born-rule measurements on the RK wavefunction
{for the $2D$ classical Ising model}
is in turn 
equivalent~\cite{PutzGarrattNishimoriTrebstZhu,PatilLudwig20251} to a
classical 
Bayesian inference problem~\cite{Iba_1999,Sourlas_1994,NishimoriBook} {for}
the classical model, and we will now discuss 
the classical inference problem following Ref.~\cite{PutzGarrattNishimoriTrebstZhu} and also Ref.~\cite{NahumJacobsen}.
Note that,
the probability of obtaining a spin configuration $\{\sigma_{i}\}$ in the $2D$ classical Ising model is given by the Boltzmann weight,
\begin{equation}
    P(\{\sigma_i\})=\frac{e^{-H[\{\sigma_i\}]}}{Z},
\end{equation}
where $H[\{\sigma_i\}]$ is the Hamiltonian for the $2D$ classical Ising model 
given in Eq.~\eqref{Eq2DIsingHamiltonian}.
Now consider performing 
weak
`classical'  measurements of the bond-energies on the {classical} Ising model, such that given a spin configuration $\{\sigma_{i}\}$, the probability of receiving binary measurement outcomes $\{m_{ij}=\pm 1\}$  
is given by
\begin{equation}
    P(\{m_{ij}\}|\{\sigma_{i}\})=\prod_{\langle ij\rangle}\left(\frac{1+m_{ij}
    \gamma
    \sigma_i\sigma_j}{2}\right),
    \label{eq:Pmsigma}
\end{equation}
where we have defined 
\begin{equation}
    \gamma:=\tanh\tilde{\gamma}.\label{EqDefofGamma}
\end{equation}
Then, we can consider the `updated' ensemble of spin configurations $\{\sigma_{i}\}$ conditioned on the received measurement outcomes $\{m_{ij}\}$, and from Bayes' theorem the conditional probabilities are given by 
\begin{equation}
    P(\{\sigma_i\}|\{m_{ij}\})=\frac{e^{-\beta H[\{\sigma_i\}]+\tilde{\gamma}\sum_{\langle ij\rangle}m_{ij}\sigma_i\sigma_j}}{Z[\{m_{ij}\}]},\label{EqProbabilitySpinGivenMeasBetaVersion}
\end{equation}
{where}
$Z[\{m_{ij}\}]$ {[Eq.~\eqref{EqZ[m]}]} is proportional to the probability of obtaining measurement outcomes $\vec{m}=\{m_{ij}\}$:
\begin{align}
    P(\vec{m})&=\frac{Z[\{m_{ij}\}]}{\sum\limits_{\{m_{ij}=\pm 1\}}Z[\{m_{ij}\}]}\nonumber\\
&=\frac{{(2\cosh(\tilde{\gamma}))}^{-N_b}}{Z}\sum_{\{\sigma_i\}}\;e^{-\beta H[\{\sigma_i\}]+\tilde{\gamma}\sum_{\langle ij\rangle}m_{ij}\sigma_i\sigma_j}.\label{EqClassicalProbability}
\end{align}
Given any measurement trajectory $\vec{m}=\{m_{ij}\}$, we can calculate the expectation value of 
a general classical
observable
$\mathcal{O}_1={\sigma}_{i_1} {\sigma}_{i_2}{\sigma}_{i_3} \cdots{\sigma}_{i_k}$ 
{conditioned on the measurement outcomes $\vec{m}$ as follows}
\begin{align}
    \langle \mathcal{O}_1\rangle_{\vec{m}}&=\sum_{\{\sigma_i\}}P(\{\sigma_i\}|\vec{m})\mathcal{O}_1\nonumber\\&=\frac{\sum_{\{\sigma_i\}}\mathcal{O}_1\;e^{-\beta H[\{\sigma_i\}]+\tilde{\gamma}\sum_{\langle ij\rangle}m_{ij}\sigma_i\sigma_j}}{Z[\vec{m}]}.\label{EqClassicalExpValue}
\end{align}
}}

{Clearly, in 
any
given measurement trajectory $\vec{m}$, the expectation value of 
a classical
observable made out of product of spins $\mathcal{O}_1={\sigma}_{i_1} {\sigma}_{i_2}{\sigma}_{i_3} \cdots{\sigma}_{i_k}$ in Eq.~\eqref{EqClassicalExpValue} in the classical inference problem and the expectation value of the corresponding operator $\hat{\mathcal{O}}_1=\hat{\sigma}^z_{i_1} \hat{\sigma}^z_{i_2}\hat{\sigma}^z_{i_3} \cdots\hat{\sigma}^z_{i_k}$ {in Eq.~\eqref{EqQuantumExpValue} for} 
the quantum measurement problem on the RK wavefunction
match, and so do the expressions for the classical probability ${P}[\vec{m}]$ in Eq.~\eqref{EqClassicalProbability} and the Born-rule probability $\tilde{P}[\vec{m}]$ in Eq.~\eqref{EqQuantumProbability}. 
We will then
just study the classical inference problem and consider
calculating the following measurement-averaged 
product of expectation values for general observables $\mathcal{O}_1$, $\mathcal{O}_2$, $\cdots$ and $\mathcal{O}_N$ in the classical inference problem,
\begin{align}
\label{EqDefOfMeasurementAveragedMoments}
&\overline{
\langle { {\cal O}}_1\rangle_{\vec m}
\langle { {\cal O}}_2\rangle_{\vec m}
...
\langle {{\cal O}}_N\rangle_{\vec m}
}=
\sum_{\vec{m}} P(\vec{m})\langle { {\cal O}}_1\rangle_{\vec m}
\langle { {\cal O}}_2\rangle_{\vec m}
...
\langle {{\cal O}}_N\rangle_{\vec m}.
\end{align}}
{Using the replica trick, as discussed in App.~\ref{AppDerivationOfReplicaTheory}, 
the above measurement-averaged product of expectation values can be written as}
\begin{align}
    &\overline{
\langle { {\cal O}}_1\rangle_{\vec m}
\langle { {\cal O}}_2\rangle_{\vec m}
...
\langle {{\cal O}}_N\rangle_{\vec m}
}\propto\lim_{R\rightarrow 1}
\sum_{\{\sigma_i^{(a)}\}}\mathcal{O}_1^{(1)}\mathcal{O}_2^{(2)}\cdots \mathcal{O}_N^{(N)}\times\nonumber\\&\;\;\;\times
 \exp{\Big\{\beta\sum_{\langle
ij \rangle} \sum_{a=1}^{R} \sigma_i^{(a)}\sigma_j^{(a)}+\frac{\tilde{\gamma}^2}{2}\sum_{\langle ij\rangle}\sum_{\substack{a,b=1\\ a\neq b}}^{R}\sigma_i^{(a)}\sigma_j^{(a)}\sigma_i^{(b)}\sigma_j^{(b)}}\nonumber\\
&\qquad\qquad\qquad\qquad\qquad\qquad\qquad\qquad\qquad\qquad+\mathcal{O}(\tilde{\gamma}^4)\Big\}.\label{EqLatticeReplicaTheoryBetaV}
\end{align}
The above replica theory allows {us} to calculate all the measurement-averaged products of expectation values 
and study
the effects of 
`learning' 
local energy densities in the $2D$ {classical} Ising model.
{The latter problem, as discussed above, is equivalent to the problem of Born-rule bond-energy measurements on the corresponding RK wavefunction [Eq.~\eqref{EqRKWF}], which is in turn dual to 
the problem of
Born-rule measurements 
with {the} Pauli-$Z$ operator on the deformed toric code wavefunction [Eq.~\eqref{EqDeformedToricCodeWF}].}
We will now drop the terms of order $\mathcal{O}(\tilde{\gamma}^4)$ from the weight in the exponent of the replica theory in Eq.~\eqref{EqLatticeReplicaTheoryBetaV} and study the following replica theory in the $R\rightarrow1$ replica limit
\begin{equation}
    -\mathcal{H}=\beta \sum_{\langle
ij \rangle}\sum_{a=1}^{R} \sigma_i^{(a)}\sigma_j^{(a)}+\frac{\Delta}{2}\sum_{\langle ij\rangle}\sum_{\substack{a,b=1\\ a\neq b}}^{R}\sigma_i^{(a)}\sigma_j^{(a)}\sigma_i^{(b)}\sigma_j^{(b)},\label{EqLatticeReplicaTheory}
\end{equation}
where we have defined $\Delta:={\tilde{\gamma}^2}$.
We expect the replica theory in Eq.~\eqref{EqLatticeReplicaTheory} to capture all the 
key long-distance physics of measurement-averaged products of expectation values in Eq.~\eqref{EqLatticeReplicaTheoryBetaV}, i.e.\ we expect the `higher-spin' terms 
with couplings of order
$\mathcal{O}(\tilde{\gamma}^4)$ appearing in the exponent on the RHS of Eq.~\eqref{EqLatticeReplicaTheoryBetaV} to not change the long-distance physics in the phase diagram. 
{This is because in the case of 
Gaussian measurements with continuous measurement outcomes
of the bond-energies in the $2D$ classical Ising model, 
i.e.\ replacing Eq.~\eqref{eq:Pmsigma} by 
\begin{equation}
P(\{\mathfrak{m}_{ij}\}|\{\sigma_i\}) \propto  e^{-
\frac{\Delta}{2} 
\sum_{\langle ij\rangle}(\mathfrak{m}_{ij}-\sigma_i\sigma_j)^2} \ ,\label{eq:PmsigmaGaussian}
\end{equation}
for $\mathfrak{m}_{ij} \in (-\infty,\infty)$,
in contrast to binary measurements $m_{ij}=\pm 1$, the replica theory in Eq.~\eqref{EqLatticeReplicaTheory} 
serves as
an exact replica theory description of the 
{measurement}
problem in the $R\rightarrow1$ replica limit.
We 
discuss the case of Gaussian measurements
in App.~\ref{AppGaussianMeasurements}.
We do \textit{not} expect the universal properties of phases and critical points, including 
those of the
tricritical point, in the 
deformed toric code or the
$2D$ {classical} Ising 
`learning'
phase diagram {[see Fig.~\ref{fig:2DGaussianMeasurements}]} to depend on 
{whether}
we consider Gaussian or binary measurements. Therefore, the replica theory in Eq.~\eqref{EqLatticeReplicaTheory},
{with all ${\cal O}({\tilde \gamma}^4)$ terms dropped,}
should also capture the long-distance physics in the case of binary measurements.}

\begin{figure}
   \centering
   \includegraphics[width=0.9\linewidth]{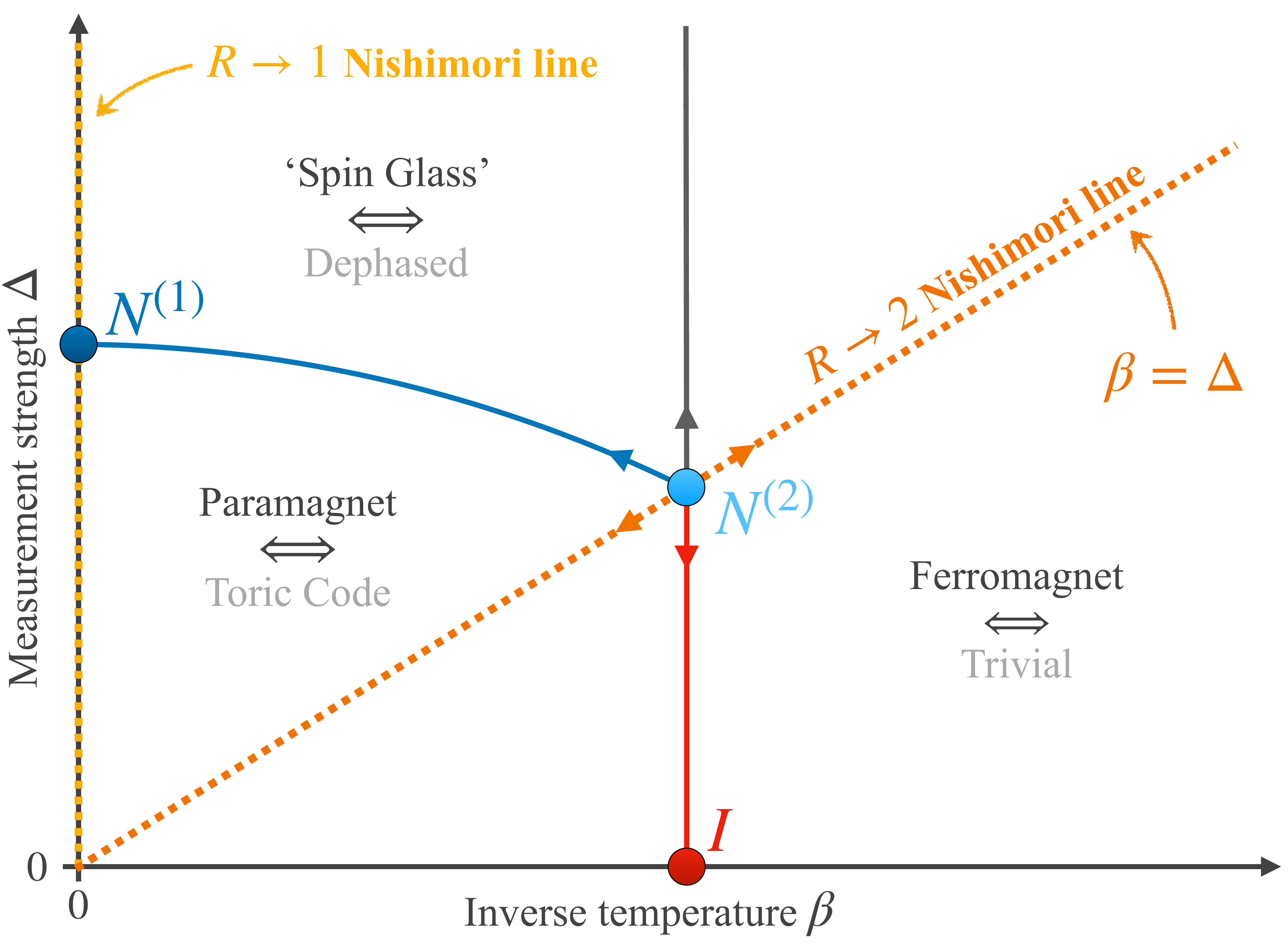}
   \caption{
   {{\bf Phase diagram and renormalization group (RG) flows for the replica theory} in Eq.~\eqref{EqLatticeReplicaTheory} in the $R\rightarrow1$ replica limit. 
   As discussed in Sec.~\ref{SecModel}, in the $R\rightarrow1$ replica limit, we expect this replica theory to govern the long distance physics of the 
   {deformed toric code 
 under Pauli-$Z$ measurements or the}
   $2D$ {classical} Ising model under {Bayesian} bond energy measurements,
   and this sketch of the phase diagram is based on the
   phase diagram {obtained} in Ref.~\cite{PutzGarrattNishimoriTrebstZhu} for the case of binary measurements. 
   [For Gaussian measurements, 
   the replica theory forms an exact description of the measurement problem(s) valid at all length scales.]
   The novel tricritical point in the 
   \textit{learning}
   phase diagram of the 
   deformed toric code or the
   $2D$ {classical} Ising model is governed by the {\it higher} Nishimori critical point  $N^{(2)}$ 
    lying on the {\it higher} Nishimori line $\beta=\Delta$ 
   [dotted orange line]
   of the replica theory {[Eq.~\eqref{EqLatticeReplicaTheory}]} in the $R\rightarrow1$ replica limit.}}
   \label{fig:2DGaussianMeasurements}
\end{figure}

{We will now briefly discuss the sketch in Fig.~\eqref{fig:2DGaussianMeasurements} of the phase diagram of the replica theory in Eq.~\eqref{EqLatticeReplicaTheory} in the $R\rightarrow1$ replica limit that governs the problem of performing discussed measurements on the deformed toric code wavefunction and the $2D$ classical Ising model, i.e.\ we discuss their `learning' phase diagram. 
This sketch is based on the phase diagram obtained in Ref.~\cite{PutzGarrattNishimoriTrebstZhu} for binary measurements. 
At small deformation for the toric code corresponding to small inverse-temperature $\beta$ in the $2D$ {classical} Ising model, and at low measurement strength quantified by $\Delta$,
we have a topological `quantum memory' phase that corresponds to the paramagnetic (PM) phase characterized by the \textit{absence} of long-range order in both the measurement-averaged first and
second moment of the spin-spin correlation function $\langle\sigma_i\sigma_j\rangle_{\vec{m}}$ in the dual $2D$ classical Ising 
Bayesian inference problem. 
At small deformation $\beta$ for the wavefunction and large measurement strength $\Delta$, we have a  topological `classical memory' phase, and that corresponds to the 
`spin glass' (`SG') phase,
characterized by 
the short-ranged first moment $\overline{\langle\sigma_i\sigma_j\rangle_{\vec{m}}}$ and long-range order in the second moment i.e.\ the Edwards-Anderson (EA) correlator $\overline{\langle\sigma_i\sigma_j\rangle_{\vec{m}}^2}$, in the Bayesian inference problem. 
For large deformations corresponding to large inverse-temperature, the
`memory' 
is lost for the toric code wavefunction and that corresponds to the ferromagnetic (FM) phase of the Bayesian inference problem characterized by the presence of long-range order in both {the} first moment and {the}
EA correlator. 
We refer the reader to Ref.~\cite{PutzGarrattNishimoriTrebstZhu} for a more detailed discussion of this `learning' phase diagram, including an information theoretic perspective on the phases using the coherent information~\cite{gullans2020scalable,FanBaoAltmanVishwanath}, and also a strong-to-weak spontaneous symmetry breaking (SWSSB) perspective~\cite{SalaGopalakrishnanOshikawaYou,LessaStrongToWeak,zhang2025strongtoweakspontaneousbreaking1form}.
We will be primarily interested in the tricritical point that lies at the `heart' of the phase diagram where the
deformed toric code's ($2D$ Ising model's)
topological `quantum memory' (PM), dephased `classical memory' (`SG') and the trivial `no memory' (FM) phases meet.
In particular, 
the line of transitions between the 
 topological `quantum memory'
(PM) and the trivial `no memory' phase (FM) characterized by the $2D$ Ising universality class [solid 
red line in Fig.~\ref{fig:2DGaussianMeasurements}] and the line of transitions between the topological `quantum memory' phase (PM)
and the dephased `classical memory' phase (`SG')
characterized by the $2D$ ordinary Nishimori critical universality class~\cite{Nishimori_1980,Nishimori1981} [solid dark-blue line in Fig.~\ref{fig:2DGaussianMeasurements}] meet at this tricritical point~\footnote{{The line of transitions between the dephased `classical memory' (`spin glass') phase and the trivial `no memory' (ferromagnet) phase [gray line in Fig.~\ref{fig:2DGaussianMeasurements}] also meets the other two line of transitions at the tricritical point. This line of transitions flows under RG to the point of 
{perfect}
(projective) measurements corresponding to $\Delta=\infty$ at the $2D$ classical Ising critical point~\cite{PutzGarrattNishimoriTrebstZhu,NahumJacobsen}.}}.
In the next section, we will demonstrate that this novel \textit{learning} tricritical point is a \textit{higher} Nishimori critical point with a higher replica symmetry than the ordinary Nishimori critical point.}

\section{A  {{\it higher}} Nishimori Critical Point: Emergent Gauge Invariance and Enlarged Permutation Symmetry\label{SecEmergentGaugeAndEnlargedPermutationSymmetry}}

{In this section, we will demonstrate that the tricritical point 
$N^{(2)}$ in Fig.~\ref{fig:2DGaussianMeasurements} for 
the $R\rightarrow1$ replica theory 
in Eq.~\eqref{EqLatticeReplicaTheory}
is a {\it higher} Nishimori critical point, which admits a gauge-invariant and higher replica-symmetric formulation as a replica theory in the $R\rightarrow2$ replica limit.
To demonstrate this, we will use the logic used in Ref.~\cite{GeorgesHanselLeDoussalMaillard,LeDoussalHarrisI,LeDoussalHarrisII} (also Ref.~\cite{GRL2001}) for the ordinary Nishimori critical point.
To this end, we will study the replica theory in the $R\rightarrow1$ replica limit in Eq.~\eqref{EqLatticeReplicaTheory} on the line $\beta=\Delta$.}

Note that when $\beta=\Delta$, the replica theory in Eq.~\eqref{EqLatticeReplicaTheory} can be written as 
\begin{equation}
-\mathcal{H}=\frac{\beta}{2} \sum_{\langle
ij \rangle}\Big(\sum_{a=1}^{R} \sigma_i^{(a)}\sigma_j^{(a)}+1\Big)^2\
+ \, {\rm const.},
\ \ (\Delta=\beta) \,.
\label{EqHamiltonianOnNishimoriStarLineUngauged}
\end{equation}
Following Ref.~\cite{GeorgesHanselLeDoussalMaillard,LeDoussalHarrisI,LeDoussalHarrisII}
and also
the discussion in Ref.~\cite{GRL2001},
we can write the partition function $Z_{R}$ for the above $R$-replica theory 
{in Eq.~\eqref{EqHamiltonianOnNishimoriStarLineUngauged}, which
is thus that of 
Eq.~\eqref{EqLatticeReplicaTheory}}
on the line $\beta=\Delta$, as the partition function for a theory with $R+1$ replicas with enlarged permutation symmetry as follows: Note that the partition function $Z_{R}$ for the replica theory in Eq. \eqref{EqHamiltonianOnNishimoriStarLineUngauged} can be written as
\begin{equation}
    Z_{R}=\sum_{\{\sigma^{(a)}_i\}_{a=1}^R} \exp{\big\{ \frac{\beta}{2} \sum_{\langle
ij \rangle}\Big(\sum_{a=1}^{R} \sigma_i^{(a)}\sigma_j^{(a)}+1\Big)^2\big\}}\label{EqPartionFunctionWithoutExtraSpinOnNishimoriStarLine}
\end{equation}
Now consider the variable transformation
\begin{equation}
    \tilde{\sigma}_{i}^{(a)}=s_{i} {\sigma}_{i}^{(a)} \Leftrightarrow{\sigma}_{i}^{(a)} =s_{i}\tilde{\sigma}_{i}^{(a)}\label{EqVariableChange}
\end{equation}
where we choose a 
fixed but arbitrary value
$s_{i}=\pm 1$ at each site $i$ on the square lattice. Then the partition function $Z_{R}$ can be written as 
\begin{align}
    Z_{R}&=\sum_{\{\sigma^{(a)}_i\}_{a=1}^R} \exp{\big\{ \frac{\beta}{2} \sum_{\langle
ij \rangle}\Big(\sum_{a=1}^{R} s_{i} \tilde{\sigma}_{i}^{(a)}s_{j}\tilde{\sigma}_j^{(a)}+1\Big)^2\big\}}\nonumber\\
&=\sum_{\{\sigma^{(a)}_i\}_{a=1}^R} \exp{\big\{ \frac{\beta}{2} \sum_{\langle
ij \rangle}\Big(\sum_{a=1}^{R} \tilde{\sigma}_{i}^{(a)}\tilde{\sigma}_j^{(a)}+s_is_j\Big)^2\big\}}\nonumber\\
&=\sum_{\{\tilde{\sigma}^{(a)}_i\}_{a=1}^R} \exp{\big\{ \frac{\beta}{2} \sum_{\langle
ij \rangle}\Big(\sum_{a=1}^{R} \tilde{\sigma}_{i}^{(a)}\tilde{\sigma}_j^{(a)}+s_is_j\Big)^2\big\}}\nonumber\\
&=\sum_{\{{\sigma}^{(a)}_i\}_{a=1}^R} \exp{\big\{ \frac{\beta}{2} \sum_{\langle
ij \rangle}\Big(\sum_{a=1}^{R} {\sigma}_{i}^{(a)}{\sigma}_j^{(a)}+s_is_j\Big)^2\big\}} \,,
\label{EqTransformedPartFunExtraSpin}
\end{align}
where in the second-to-last step above, we have changed the sum over spins $\sigma_{i}^{(a)}$ from $a=1$ to $R$ to the sum over transformed spins
$\tilde{\sigma}_{i}^{(a)}$ from $a=1$ to $R$, and in the last step we have dropped the
`tilde' notation
over the dummy variables. 
{Clearly, 
Eq.~\eqref{EqTransformedPartFunExtraSpin} is independent of the values for $\{s_{i}\}$.}
Let us then define $s_{i}$ as a spin in an extra `$(R+1)^{\text{th}}$' replica copy
\begin{equation}
    \sigma_i^{(R+1)}:=s_{i}\label{EqExtraSpinDef}
\end{equation}
and sum the RHS of Eq.~\eqref{EqTransformedPartFunExtraSpin} over all choices of $\sigma_i^{(R+1)}:=s_{i}$ for all $i$, such that
\begin{align}
	&Z_{R}=\frac{1}{2^{N_{s}}}\sum_{\{{\sigma}^{(R+1)}_i\}}\sum_{\{{\sigma}^{(a)}_i\}_{a=1}^{R}} \exp{\Big\{ \frac{\beta}{2} \sum_{\langle
	ij \rangle}\Big(\sum_{a=1}^{R} {\sigma}_{i}^{(a)}{\sigma}_j^{(a)}+}\nonumber\\
	&\qquad\qquad\qquad\qquad\qquad\qquad\qquad+\sigma_i^{(R+1)}\sigma_j^{(R+1)}\Big)^2\Big\} \nonumber \\
	&\Rightarrow Z_{R}=\frac{1}{2^{N_{s}}}\sum_{\{{\sigma}^{(a)}_i\}_{a=1}^{R+1}} \exp{\Big\{ \frac{\beta}{2} \sum_{\langle
	ij \rangle}\Big(\sum_{a=1}^{R+1} {\sigma}_{i}^{(a)}{\sigma}_j^{(a)}\Big)^2\Big\}}
	\label{EqGaugeSymmetricPartionFunction}
\end{align}
{Note that in the above equation, Eq.~\eqref{EqGaugeSymmetricPartionFunction},
at each site $i$
{there is a number of $(R+1)$}
replicas of Ising spins, $\sigma^{(a)}_i$, where $a=1, 2, ..., R, (R+1)$.
This is in contrast to 
Eq.~\eqref{EqLatticeReplicaTheory}, where at each site there are only $R$ replicas,  $\sigma^{(a)}_i$, where $a=1, 2, ..., R$.}
The above representation, Eq.~\eqref{EqGaugeSymmetricPartionFunction},
of the partition function allows us to see that the replica theory in Eq.~\eqref{EqLatticeReplicaTheory},
{which}
we want to study {on the line $\beta=\Delta$ [dotted
orange line in Fig.~\ref{fig:2DGaussianMeasurements}] and} in the $R\rightarrow1$ replica limit,
{has}
a formulation with larger replica permutation symmetry
$S_{R+1}$ instead 
of
{just}
$S_{R}$. 
Moreover, the above
{{formulation}
in Eq.~\eqref{EqGaugeSymmetricPartionFunction}}
{of the replica theory in  Eq.~\eqref{EqLatticeReplicaTheory} on the line $\beta=\Delta$ also}
has 
gauge invariance,
where the weight in the 
{exponential} of Eq.~\eqref{EqGaugeSymmetricPartionFunction} is invariant under the following local (gauge) transformation at any site $i$:
\begin{equation}
    \sigma_{i}^{(a)}\rightarrow-\sigma_{i}^{(a)} \;\;\forall a=1,2, \cdots, R, R+1 \,.
\end{equation}
We note that indeed, if we were interested in studying the replica theory in Eq.~\eqref{EqLatticeReplicaTheory} in the $R\rightarrow0$ replica limit, the line $\beta=\Delta$ is nothing but the {ordinary}  Nishimori line~\cite{Nishimori_1980,Nishimori1981} and the above described logic was used to study the properties of
the 
{ordinary}
Nishimori line in the $R\rightarrow0$ replica limit in Ref.~\cite{LeDoussalHarrisI,LeDoussalHarrisII,GeorgesHanselLeDoussalMaillard,GRL2001}. 
That is, it was demonstrated in Ref.~\cite{LeDoussalHarrisI,LeDoussalHarrisII,GeorgesHanselLeDoussalMaillard,GRL2001} that the $R\rightarrow0$ replica limit random-bond Ising model  (RBIM),
{with partition function {in} Eq.~\eqref{EqLatticeReplicaTheory}},
on the
{ordinary}
Nishimori line $\beta=\Delta$ 
can be written as a gauge-invariant $R\rightarrow1$ replica theory {in Eq.~\eqref{EqGaugeSymmetricPartionFunction}.}
Since
in the present paper we are
already interested in studying the replica theory in Eq.~\eqref{EqLatticeReplicaTheory} in the $R\rightarrow1$ replica limit, the above representation in Eq.~\eqref{EqGaugeSymmetricPartionFunction} 
allows us to study
the $R\to 1$ replica limit of Eq.~\eqref{EqLatticeReplicaTheory} 
{on the $\beta=\Delta$ line}
as 
{the
equivalent gauge-invariant and replica-symmetric theory 
[Eq.~\eqref{EqGaugeSymmetricPartionFunction}] in the $R \to 2$ replica limit.}
[The gauge-invariant {$R$-}replica theory in the 
{exponential} of  Eq.~\eqref{EqGaugeSymmetricPartionFunction} is the
same as that for the
`spin glass'.]
{Since the 
ordinary
Nishimori line already occurs as the $\beta=0$ line in the phase diagram of the replica theory in Eq.~\eqref{EqLatticeReplicaTheory} in the replica limit $R\rightarrow1$ {[see Fig.~\ref{fig:2DGaussianMeasurements}]},}
we will refer to the line $\beta=\Delta$
{[dotted orange line in Fig.~\ref{fig:2DGaussianMeasurements}]}
of the replica limit $R\rightarrow1$ replica theory in Eq.~\eqref{EqLatticeReplicaTheory} as the 
{{\it higher} Nishimori line}
\begin{equation}\label{EqNishimoriStarLine}
    \beta=\Delta \;\;\;:\;\;
    \text{{{\it higher} Nishimori} line}.
\end{equation}
 
{We will now demonstrate that there is a novel critical point of the replica theory in Eq.~\eqref{EqLatticeReplicaTheory} in the replica $R\rightarrow1$ limit on the {\it higher} Nishimori line in Eq.~\eqref{EqNishimoriStarLine}, which we will refer to as {\it higher} Nishimori critical point.}
{{As discussed in Sec.~\ref{SecModel}, since}
the replica theory in the $R\rightarrow1$ replica limit governs the long-distance physics of the 
`learning'
phase diagram {for the deformed toric code or the $2D$ Ising model},}
{the}
location of {this}
{{\it higher} Nishimori critical point}
can be identified by noting that the measurement-averaged first moment of the expectation value for any observable
{$\mathcal{O}_1$}
is equal to its 
expectation value in the {unmeasured} {classical} Ising model.
{This 
{follows}
from Eq.~\eqref{EqDefOfMeasurementAveragedMoments}:
{for}
$N=1$ the RHS of Eq.~\eqref{EqDefOfMeasurementAveragedMoments}, upon using Eq.~\eqref{EqClassicalProbability} and Eq.~\eqref{EqClassicalExpValue}, reduces to} 
\begin{align}
    \overline{\langle \mathcal{O}_1\rangle_{\vec{m}}}&=\frac{1}{\sum_{\vec{m}}Z[\vec{m}]}\sum_{\vec{m}} \big( \sum_{\{\sigma_i\}}\mathcal{O}_1\;e^{-\beta H[\{\sigma_i\}]+\tilde{\gamma}\sum_{\langle ij\rangle}m_{ij}\sigma_i\sigma_j}\big)\nonumber
\end{align}
{Then using Eq.~\eqref{EqZ[m]} for 
{$Z[\vec{m}]$,}
and exchanging the order of summations over $\{\sigma_i\}$ and {$\vec{m}=\{m_{ij}\}$}, we obtain}
\begin{align}
     \overline{\langle \mathcal{O}_1\rangle_{\vec{m}}}&=\frac{ \sum_{\{\sigma_i\}}\big(\mathcal{O}_1\;e^{-\beta H[\{\sigma_i\}]}\sum_{\vec{m}}e^{\tilde{\gamma}\sum_{\langle ij\rangle}m_{ij}\sigma_i\sigma_j}\big)}{\sum_{\{\sigma_i\}}\big(e^{-\beta H[\{\sigma_i\}]}\sum_{\vec{m}}e^{\tilde{\gamma}\sum_{\langle ij\rangle}m_{ij}\sigma_i\sigma_j}\big)}\,.
     \label{EqIntermediateStepFirstMomentAverage}
\end{align}
Now, 
\begin{align}
	\sum_{m_{ij}=\pm 1}e^{\tilde{\gamma}m_{ij}\sigma_i\sigma_j}&=2 \cosh(\tilde{\gamma}\sigma_i\sigma_j)
	=2\cosh(\tilde{\gamma}) \,,
	\label{EqAverageOverMeasurementOutcomesFirstMoment}
\end{align}
where 
the last equality follows because $\cosh$ is an even function and $\sigma_i\sigma_j=\pm 1$. 
Then from Eq.~\eqref{EqIntermediateStepFirstMomentAverage} and Eq.~\eqref{EqAverageOverMeasurementOutcomesFirstMoment},
\begin{align}
    \overline{\langle \mathcal{O}_1\rangle_{\vec{m}}}
    &=\frac{ \sum_{\{\sigma_i\}}\mathcal{O}_1\;e^{-\beta H[\{\sigma_i\}]}}{\sum_{\{\sigma_i\}}e^{-\beta H[\{\sigma_i\}]}}=\langle \mathcal{O}_1\rangle \,,	
    \label{EqFirstMomentIsEqualToTheUnmeasured}
\end{align}
where the expectation value $\langle \mathcal{O}_1\rangle$
is evaluated in the unmeasured Ising model.
{[In the quantum RK formulation, Eq.~\eqref{EqFirstMomentIsEqualToTheUnmeasured} can be understood as a simple consequence of the POVM condition (Eq.~\eqref{EqPOVMBinary})~\footnote{
Note that
{\begin{align*}
&\overline{\langle \hat{\mathcal{O}}_1\rangle_{\vec{m}}}=\sum_{\vec{m}}\tilde{P}(\vec{m})\bra{\Psi_{\vec{m}}}\hat{\mathcal{O}}_1\ket{\Psi_{\vec{m}}}
\\&=\sum_{\vec{m}}\bra{RK}\hat{K}^{\dagger}_{\vec{m}}\hat{K}_{\vec{m}}\ket{RK}\frac{\bra{RK}\hat{K}^{\dagger}_{\vec{m}}\hat{\mathcal{O}}_1\hat{K}_{\vec{m}}\ket{RK}}{\bra{RK}\hat{K}^{\dagger}_{\vec{m}}\hat{K}_{\vec{m}}\ket{RK}}\\
&=\sum_{\vec{m}}{\bra{RK}\hat{K}^{\dagger}_{\vec{m}}\hat{\mathcal{O}}_1\hat{K}_{\vec{m}}\ket{RK}}\\&={\bra{RK}\hat{\mathcal{O}}_1\sum_{\vec{m}}\hat{K}^{\dagger}_{\vec{m}}\hat{K}_{\vec{m}}\ket{RK}}={\bra{RK}\hat{\mathcal{O}}_1\ket{RK}}=\langle \hat{\mathcal{O}}_1\rangle,
\end{align*}
where we have commuted $\hat{\mathcal{O}}_1$ through the Kraus operator $\hat{K}_{\vec{m}}$, since both are diagonal in $\ket{\{\sigma_i\}}$ basis, and {then} used the POVM condition~[Eq.~\eqref{EqFirstMomentIsEqualToTheUnmeasured}].}}.]}
Coming back to the {{\it higher} Nishimori line} $\beta=\Delta$,
the location of the
{{\it higher} Nishimori critical point on it,}
can be identified 
{by}
requiring that the {measurement-averaged} first moment
of observables such as the spin-spin correlation function are power-law decaying. 
Now since the first moments of expectation values are equal to the unmeasured expectation values (corresponding to $\Delta=0$), we immediately see that the 
{{\it higher} Nishimori critical point} occurs at the point where $\beta=\beta_{c}$, the (inverse) temperature corresponding to the critical point of the unmeasured $2D$ 
{classical}
Ising model. Therefore, 
the 
{{\it higher} Nishimori critical point}
of the 
$R\rightarrow1$
replica theory in Eq.~\eqref{EqLatticeReplicaTheory} 
lies on the {{\it higher} Nishimori line} at the following point
\begin{equation}\label{EqLocationFixedPoint}
    \beta=\Delta=\beta_{c}=\frac{1}{2}\ln\left(1+\sqrt{2}\right) \,,
\end{equation}
where in the last equality we have used the famous result of Kramers and Wannier~\cite{KramersWannier1941} that gives the location of the $2D$ Ising critical point. 

{The gauge-invariant and higher replica-symmetric formulation Eq.~\eqref{EqGaugeSymmetricPartionFunction} of the {\it higher} Nishimori line, 
and hence of the above 
{\it higher} Nishimori critical point 
{[Eq.~\eqref{EqLocationFixedPoint}] for}
the replica theory in Eq.~\eqref{EqLatticeReplicaTheory} in the $R\rightarrow1$ replica limit is invariant under 
RG flow.
Therefore, the universality class for the {\it higher} Nishimori critical point is clearly distinct from the ordinary Nishimori critical point
that occurs at
$(\beta=0,\Delta\approx1)$~\footnote{{The location $(\beta=0,\Delta\approx1)$ of the ordinary Nishimori critical point can be obtained by exact mapping of the replica $R\rightarrow1$ theory in Eq.~\eqref{EqLatticeReplicaTheory} on the $\beta=0$ line to the Nishimori line in the replica $R\rightarrow0$ RBIM with \textit{Gaussian} bond disorder~\cite{McMillan,PiccoPujolHonecher2006}. 
This should be contrasted with the location of the ordinary Nishimori critical point in the phase diagram of Ref.~\cite{PutzGarrattNishimoriTrebstZhu} for binary measurements, where it
occurs at $(\beta=0, \tilde{\gamma}\approx0.782)$. The latter difference in the locations of the critical point is of course non-universal.}}~\cite{McMillan,PiccoPujolHonecher2006}
and the (unmeasured) $2D$ Ising critical point that occurs at $(\beta=\beta_c=\frac{1}{2}\ln(1+\sqrt{2}), \Delta=0)$ for the $R\rightarrow1$ replica theory in Eq.~\eqref{EqLatticeReplicaTheory}. 
This can also be verified by noting
{that}
the universal critical properties at the  {\it higher} Nishimori critical point, some of which we will derive exactly in the next section [Sec.~\ref{SecExactResults}], 
are 
clearly distinct from the 
{numerically known (see e.g.~\cite{PiccoPujolHonecher2006}) universal critical properties of the ordinary Nishimori critical point and also the exactly known critical properties of the}
(unmeasured) $2D$ Ising critical point.
As discussed above, apart from the ordinary Nishimori critical point and the
`unmeasured' 
$2D$ Ising critical point, Ref.~\cite{PutzGarrattNishimoriTrebstZhu} discovered a novel unstable tricritical point that occurs at
{the
temperature corresponding to that of the unmeasured  $2D$ Ising critical point} and {a} finite 
strength of measurements.
Thus, the universal critical properties of this novel \textit{learning} tricritical point must be governed by the 
{above-discussed}
{\it higher} Nishimori critical point 
[Eq.~\eqref{EqLocationFixedPoint}] {of the replica theory in Eq.~\eqref{EqLatticeReplicaTheory} in the $R\rightarrow1$ replica limit}.}

{{We}
note that in the case of binary measurements, the 
{{\it higher} Nishimori line $\beta=\Delta$}
does not itself lie in the 
deformed toric code or the 2D Ising
learning phase diagram, because 
{at}
the lattice level{,} 
the terms of order $\mathcal{O}(\tilde{\gamma}^4)$ in Eq.~\eqref{EqLatticeReplicaTheoryBetaV} 
spoil the 
{higher replica-symmetric formulation [Eq.~\eqref{EqGaugeSymmetricPartionFunction}] of}
the replica theory in
Eq.~\eqref{EqLatticeReplicaTheory} on the  
{$\beta=\Delta$}
{{\it higher} Nishimori line.}
{Consequently, we have not been able to determine the exact location of the tricritical point in the learning phase diagram with binary measurements (this was determined numerically in Ref.~\cite{PutzGarrattNishimoriTrebstZhu}), since it is a `non-universal' detail in the RG sense.}
However, since the replica theory {in} Eq.~\eqref{EqLatticeReplicaTheory} in the $R\rightarrow1$ replica limit 
is expected to capture
the long-distance physics of the 
learning phase diagram,
the long-distance physics at the
tricritical point in the case of binary measurements should also be
the same as that of the 
{\textit{higher Nishimori critical point},}
the critical point {[Eq.~\eqref{EqLocationFixedPoint}]} on the {\it higher} Nishimori line $\beta=\Delta$ of the 
$R\rightarrow1$
replica theory in Eq.~\eqref{EqLatticeReplicaTheory}.
{The replica theory {in} Eq.~\eqref{EqLatticeReplicaTheory} is an exact description of the measurement problem in the case of Gaussian measurements.
{For Gaussian measurements {of strength $\Delta$} [Eq.~\eqref{eq:PmsigmaGaussian}], as}
discussed in App.~\ref{AppGaussianMeasurements},
the entire 
{{\it higher} Nishimori line} {$\beta=\Delta$}
lies in the
learning phase diagram, and that also allows us to determine the exact location of the tricritical point in the Gaussian measurement case, which is exactly the same as the location of the
{{\it higher} Nishimori critical point} {given}
in Eq.~\eqref{EqLocationFixedPoint}.
Since we do not expect the universality class of the tricritical point in the 
learning phase diagram to depend on whether we use binary or Gaussian measurements, the tricritical point in the case of binary measurements should also lie in the universality class of the tricritical point in the case of Gaussian measurements. 
Thus, we expect the long-distance physics at the tricritical point in the case of binary measurements will also be governed by the 
{{\it higher} Nishimori critical point}
Eq.~\eqref{EqLocationFixedPoint} of the {$R\rightarrow1$} replica theory in Eq.~\eqref{EqLatticeReplicaTheory}, which is equivalent to saying that the order $\mathcal{O}(\tilde{\gamma}^4)$ terms in Eq.~\eqref{EqLatticeReplicaTheoryBetaV}  are irrelevant at the 
{{\it higher} Nishimori critical point}
in the RG sense.}}
The agreement between the exact result discussed in Sec.~\ref{SubSecExactResultsA} for the power-law exponent of the EA correlator [Eq.~\ref{EqSecondMomentSpinSpinCorrFunAtTricPoint}] and the numerically obtained value for the same with binary measurements [Sec.~\ref{SecNumericalResults}, Fig.~\ref{fig:EA-IsingTricritical}] is a testament for the RG irrelevance of the latter terms, which quantify the microscopic deviation between binary and Gaussian measurement protocols.

\section{Exact Results for Critical Exponents}
\label{SecExactResults}

{In this section, we discuss various exact results for the novel 
\textit{learning}
tricritical point, which, as we discussed in the previous section, is a {\it higher} Nishimori critical point. 
In Sec.~\ref{SubSecExactResultsA}, we discuss exact 
equivalence of
the long-distance properties of the measurement-averaged $(2q)^{\text{th}}$ and  $(2q-1)^{\text{th}}$ moments of any multipoint spin correlation at the tricritical point. 
In particular, we use the latter
{equivalence}
to obtain the exact power-law exponent for the measurement-averaged second moment, i.e.\ the Edwards-Anderson correlation function at the novel tricritical point. 
In Sec.~\ref{SubSecExactResultsB}, we discuss various rigorous bounds on the power-law exponents for the measurement-averaged higher ($>2$) moments of the spin-spin correlation function, moments of its absolute value, and the typical spin-spin correlation function.
{In}
Sec.~\ref{SubSecExactResultsC}, we use the tools used in the proof of the recent $c$-effective theorem~\cite{PatilLudwig20251} for $2D$ Bayesian inference problems to establish the decrease of the Casimir effective central charge under the RG flow from the 
novel tricritical point, i.e.\ the {\it higher} Nishimori {critical} point,} to the unmeasured $2D$ Ising critical point.
{Finally, in Sec.~\ref{SubSecOnDualSpinCorrFun}, we discuss various exact results for the measurement-averaged moments of the dual spin correlation function at the 
novel 
{\it higher} Nishimori} critical point.

\subsection{Measurement-averaged First and Second Moments of Spin Correlation Functions
\label{SubSecExactResultsA}}

{From Eq.~\eqref{EqFirstMomentIsEqualToTheUnmeasured} {[or using the POVM condition~\cite{Note7} in quantum RK formulation]}, 
the}
measurement-averaged first moment of the expectation value for any {observable} 
$\mathcal{O}_1=\sigma_{i_1}\cdots\sigma_{i_k}$ 
at the 
\textit{learning}
{tricritical point, 
which at long-distances is governed by a
 {\it higher} Nishimori critical point,}
is equal to 
its expectation value at the unmeasured $2D$ Ising critical point.
{In}
particular, at long-distances
\begin{equation}\
    \overline{\langle \sigma_{i_1}\cdots\sigma_{i_k}\rangle_{\vec{m}}}\sim\langle \sigma_{i_1}\cdots\sigma_{i_k}\rangle_{\text{Ising critical}} \,,\label{EqFirstMomentOfGeneralObservableAtTricPoint}
\end{equation}
where 
`$\sim$' denotes 
equivalence at long-distances.
For example, the long-distance behavior of the measurement-averaged expectation value of the spin-spin correlation function at the {\it higher} Nishimori
critical
point 
is given by
\begin{equation}
    \overline{\langle \sigma_{i}\sigma_{j}\rangle_{\vec{m}}}\sim \frac{1}{|i-j|^{2X_{1}}}=\frac{1}{|i-j|^{1/4}}\Rightarrow X_1=X_{\sigma}=\frac{1}{8} \,,
    \label{EqFirstMomentSpinSpinCorrFunAtTricPoint}
\end{equation}
where $X_{\sigma}=1/8$ is the scaling dimension of the spin operator at the {$2D$} {classical} Ising critical point.

The gauge 
{invariant formulation of the {\it higher} Nishimori critical point, which governs the long-distance behavior of the tricritical point, also}
allows us to 
obtain the long-distance behavior of 
the 
measurement-averaged
second moment
of 
correlation functions at the tricritical point.
As is well-known, the {spin} correlation functions on the 
ordinary
Nishimori line {in the RBIM~\cite{Nishimori1981,LeDoussalHarrisI,PiccoPujolHonecher2006}} {[}$R\rightarrow0$ replica limit
{of}
Eq.~\eqref{EqLatticeReplicaTheory} {on the line $\beta=\Delta$]}, 
satisfy the following equality~\cite{Nishimori_1980,Nishimori1981} 
\begin{equation}\label{EqNishimoriEquality}
    \left[\left\langle \sigma_{\boldsymbol{i}_1} \ldots \sigma_{\boldsymbol{i}_k}\right\rangle^{2 q-1}\right]=\left[\left\langle \sigma_{\boldsymbol{i}_1} \ldots \sigma_{\boldsymbol{i}_k}\right\rangle^{2 q}\right]
\end{equation}
where the square bracket $[\cdots]$ denotes average over {quenched} bond randomness in the random-bond Ising model (RBIM). The analogous equality is also satisfied by the correlation functions on the
{{\it higher} Nishimori line $\beta=\Delta$}
in the $R\rightarrow1$ replica limit of the replica theory in Eq.~\eqref{EqLatticeReplicaTheory}, which is equivalent to the $R\rightarrow2$ replica limit of the gauge 
{invariant}
and enlarged 
permutation symmetric theory in Eq.~\eqref{EqGaugeSymmetricPartionFunction}.
{To this end, we note the following identity for the $R$-replica theory in Eq.~\eqref{EqLatticeReplicaTheory} on the $\beta=\Delta$ line}
\begin{align}
    &\langle (\sigma_{i_1}^{(a_1)}\cdots\sigma_{i_k}^{(a_1)})\cdots(\sigma_{i_1}^{(a_{2q-1})}\cdots\sigma_{i_k}^{(a_{2q-1})})\rangle_{R}=\nonumber\\ &\langle (\sigma_{i_1}^{(a_1)}\cdots\sigma_{i_k}^{(a_1)})\cdots(\sigma_{i_1}^{(a_{2q-1})}\cdots\sigma_{i_k}^{(a_{2q-1})})(\sigma_{i_1}^{(a_{2q})}\cdots\sigma_{i_k}^{(a_{2q})})\rangle_{R}\label{EqNishimoriEqualitiesReplicatedVersion} \,,
\end{align}
{where the replica indices
$a_1, a_2, ..., a_{2q-1}, a_{2q}$ are all pairwise unequal, and}
where the correlation functions $\langle \cdots\rangle_{R}$ are evaluated in the $R$-replica theory of Eq.~\eqref{EqLatticeReplicaTheory} 
{on the {$\beta=\Delta$} line.}
This equality can be obtained by following the argument in Ref.~\cite{GeorgesHanselLeDoussalMaillard,LeDoussalHarrisII,GRL2001} and the details are given in 
{our}
App.~\ref{AppDerivationOfIdentityForMoments}.
The replica limit $R\rightarrow0$ version of the equality in Eq.~\eqref{EqNishimoriEqualitiesReplicatedVersion} is precisely the famous equality in Eq.~\eqref{EqNishimoriEquality} {due to Nishimori~\cite{Nishimori1981}}. However, the equality in Eq.~\eqref{EqNishimoriEqualitiesReplicatedVersion} exists for the replica theory {in} Eq.~\eqref{EqLatticeReplicaTheory} for any number of replicas $R$  when $\beta=\Delta$, and we can apply it in the $R\rightarrow1$ replica limit
which is of relevance to the 
learning phase diagram {of the deformed toric code wavefunction or the $2D$ classical Ising model}. 
{Since the {\it higher} Nishimori critical point [Eq.~\eqref{EqLocationFixedPoint}] on the {\it higher} Nishimori line 
 describes the long-distance physics of the tricritical point in the 
learning phase diagram, using Eq.~\eqref{EqNishimoriEqualitiesReplicatedVersion} for the replica theory [Eq.~\eqref{EqLatticeReplicaTheory}] on the  $\beta=\Delta$ line in the $R\rightarrow1$ replica limit, we obtain the following equality at the learning tricritical point
\begin{equation}
 	\overline{\left\langle \sigma_{\boldsymbol{i}_1} \ldots \sigma_{\boldsymbol{i}_k}\right\rangle_{{\vec{m}}}^{2 q-1}}\sim\overline{\left\langle \sigma_{\boldsymbol{i}_1} \ldots \sigma_{\boldsymbol{i}_k}\right\rangle_{{\vec{m}}}^{2 q}} \, ,
 \label{EqMeasurementAveragedOddAndEvenMomentsAreEquivalent}
\end{equation}
where $q$ is an arbitrary positive integer.
That is, the long-distance behavior of the measurement-averaged $(2q-1)^{\text{th}}$ and $(2q)^{\text{th}}$ moment of any multipoint spin correlation is equivalent at the learning tricritical point.
In particular, the long-distance behavior of the measurement-averaged second moment of the expectation value for any observable 
{$O_1=\sigma_{i_1}\cdots\sigma_{i_k}$} at the tricritical point must be equal to {that}
of the measurement-averaged first moment of the expectation value for the same observable.}
This relation between the {measurement-averaged} 
first and second
moments at the tricritical point in the 
learning phase diagram {for the deformed toric code or the $2D$ classical Ising model} is very powerful. 
This is because we already know, from Eq.~\eqref{EqFirstMomentIsEqualToTheUnmeasured}, that the measurement-averaged first moment of 
the expectation value for
any observable 
at the tricritical point
is equal to its
expectation value 
at the
(unmeasured) $2D$ Ising
critical point.
Therefore, we obtain 
the result 
that 
the long-distance behavior of the
measurement-averaged
second moment of the expectation value for any observable 
{$\mathcal{O}_1=\sigma_{i_1}\cdots\sigma_{i_k}$}
at the 
\textit{learning}
tricritical point,
is 
equal to its expectation value at the (unmeasured) {$2D$} Ising critical point,
\begin{equation}
   \overline{ \langle\sigma_{i_1}\cdots\sigma_{i_k}\rangle_{\vec{m}}^2}\sim \langle \sigma_{i_1}\cdots\sigma_{i_k}\rangle_{\text{Ising critical}}\label{EqSecondMomentOfGeneralObservableAtTricPoint}
\end{equation}
where 
again 
`$\sim$' denotes 
equivalence at long-distances.
{In contrast to binary measurements, 
{for}
{the}
tricritical point 
in the phase diagram 
{with}
Gaussian measurements, the above equation indicating identical long-distance behavior becomes an exact equality at all length scales, and, as demonstrated in App.~\ref{AppGaussianMeasurements}, 
the latter
equality in the case of Gaussian measurements can be
{shown}
without introduction of replicas.}
{Eq.~\eqref{EqSecondMomentOfGeneralObservableAtTricPoint}} implies that the 
power-law exponent
of the Edwards-Anderson (EA) correlation function at the tricritical point in the 
learning phase diagram, 
in either the case of binary measurements or Gaussian measurements, is given by the scaling dimension of the spin ($X_{\sigma}=1/8$) 
{at}
the (unmeasured) {$2D$} Ising critical point, i.e.
\begin{align}
    \overline{\langle \sigma_{i}\sigma_{j}\rangle^2_{\vec{m}}}\sim \frac{1}{|i-j|^{2X_{2}}}=\frac{1}{|i-j|^{1/4}}\Rightarrow X_2=X_{\sigma}=\frac{1}{8}\,.\label{EqSecondMomentSpinSpinCorrFunAtTricPoint}
\end{align}
The above equation is one of the key results of 
the present
paper that shows that the gauge 
{invariant}
and enlarged
{replica-symmetric formulation of the {\it higher} Nishimori critical point [Eq.~\eqref{EqLocationFixedPoint}], which
governs the long-distance physics of
the tricritical point, allows us to}
calculate the exact 
{power-law}
exponent for the EA correlation function at 
this `frustrated' {and} `finite-measurement-strength'
tricritical point in the 
learning phase diagram.

{Our {\it analytical} results in Eqs.~\eqref{EqFirstMomentSpinSpinCorrFunAtTricPoint}
{and}
\eqref{EqSecondMomentSpinSpinCorrFunAtTricPoint}{, for the measurement-averaged moments at the {\it higher} Nishimori critical point that describes the tricritical point in the learning phase diagram,} are to be contrasted with  the {\it numerical} results for the
bond-randomness averaged
moments at the 
ordinary
Nishimori 
{critical}
point
{in the {$2D$} RBIM}
reported in 
Ref.~\cite{PiccoPujolHonecher2006}.}
{In particular, the numerically obtained values in Ref.~\cite{PiccoPujolHonecher2006} (also compare other numerical studies \cite{HoneckerPiccoPujol2001,deQueirozStinchcombe2003}) for scaling dimensions $X_1'$ and $X_2'$  of the 
bond-randomness averaged
first and second moment of the spin-spin correlation function, respectively, at the
{\it 
ordinary}
Nishimori critical point are given by
\begin{equation}\label{EqOrdinaryNishimoriScalingDimensions}
    X'_1=X'_2 \approx 0.0924 \,.
\end{equation}
{These}
scaling dimensions for the bond-randomness averaged first and 
second moments at the 
ordinary
Nishimori critical point 
{in the {$2D$} RBIM} are also equal to each other due to the analogous gauge and replica-symmetric formulation 
{of}
the 
{ordinary}
Nishimori critical point 
that lets us write it as a $R\rightarrow1$ {gauge-invariant} replica theory {starting} from the RBIM which is
a $R\rightarrow0$ replica theory~\cite{Nishimori_1980,Nishimori1981,GeorgesHanselLeDoussalMaillard,LeDoussalHarrisI,LeDoussalHarrisII}.
{The latter $R\rightarrow1$ gauge-invariant replica theory is precisely the line $\beta=0$ in the learning phase diagram in Fig.~\ref{fig:2DGaussianMeasurements}.} 
{Note that the scaling dimension in Eq.~\eqref{EqOrdinaryNishimoriScalingDimensions}  of both the \textit{bond-randomness} averaged first and second moment of the spin-spin correlation function at the ordinary  Nishimori critical point in the RBIM correspond to the \textit{measurement-averaged} second moment of the spin-spin correlation function in the Bayesian inference (replica limit $R\rightarrow1$) formulation of the ordinary Nishimori critical point as the critical point on the line $\beta=0$ in {the} learning phase diagram of Fig.~\ref{fig:2DGaussianMeasurements}~\cite{NahumJacobsen,PutzGarrattNishimoriTrebstZhu,ZhuTantivasadakarnVishwanthTrebstVerresen}
(see App.~\ref{AppTranslatingBetweenRBIMandBayesianInference}).}
{[The \textit{measurement-averaged} first moment of the spin-spin correlation function vanishes identically everywhere on the line $\beta=0$ in the learning phase diagram of Fig.~\ref{fig:2DGaussianMeasurements}, because it is equal to the spin-spin correlation function in the unmeasured model (Eq.~\eqref{EqFirstMomentIsEqualToTheUnmeasured}) and
the unmeasured model is at infinite temperature.]}
The 
numerical value {in Eq.~\eqref{EqOrdinaryNishimoriScalingDimensions}} for 
{the scaling dimension of the measurement-averaged second moment of the spin-spin correlation function at the ordinary Nishimori critical point on the $\beta=0$ line in the learning phase diagram {[Fig.~\ref{fig:2DGaussianMeasurements}]},}
is clearly distinct from the exact scaling dimension 
$=1/8=0.125$ in
Eq.~\eqref{EqSecondMomentSpinSpinCorrFunAtTricPoint} for the measurement-averaged 
second moment
{of}
the spin-spin correlation function at the {\it tricritical} point in the
learning phase diagram~{[Fig.~\ref{fig:2DGaussianMeasurements}]}, where the tricritical point, as discussed above, 
{is a {\it higher} Nishimori critical point.}
This clearly implies that the tricritical point and the ordinary Nishimori critical point are in two different critical universality classes.
{In fact, 
in the deformed toric code or the $2D$ {classical} Ising learning phase diagram of Ref.~\cite{PutzGarrattNishimoriTrebstZhu} {(see the sketch in Fig.~\ref{fig:2DGaussianMeasurements})}, 
it has been observed numerically that under a particular RG relevant perturbation
the tricritical point, which lies at a finite value of temperature and measurement strength, flows 
under RG
to the ordinary Nishimori critical point, which lies at infinite temperature ($\beta=0$) but at a finite measurement strength
in the 
learning phase diagram.}

Of course,
one of the key-points here is that we knew the measurement-averaged first moments at the
\textit{learning}
tricritical point because of {the}
equality in Eq.~\eqref{EqFirstMomentIsEqualToTheUnmeasured}, {which} is seen as a consequence of the `POVM' condition~\cite{Note7} in the quantum RK formulation of the problem. 
{Then}
the emergent gauge {invariant} and enlarged {replica} permutation symmetric
{formulation}
{at the 
\textit{learning} tricritical point, i.e.\ the \textit{higher} Nishimori critical point,
 allowed us to evaluate the long-distance behavior of the EA correlation function. 
On the other hand,
for the 
ordinary
Nishimori 
{critical} point {in the {$2D$} RBIM},
we do 
{not} know the {bond-}randomness averaged first 
{moment}
of the spin-spin correlation function
analytically, and consequently 
one has to rely on numerical results for the scaling dimension of the bond-randomness averaged
first moment to calculate the scaling dimension of the 
bond-randomness averaged
second moment,
i.e.\ the EA correlation function.

\subsection{Bounds on the Exponents for the Measurement-averaged Higher \texorpdfstring{$(n>2)$}{Lg} Moments
\label{SubSecExactResultsB}}

Now,
knowing the scaling dimension for the first [Eq.~\eqref{EqFirstMomentSpinSpinCorrFunAtTricPoint}] and the second 
{[Eq.~\eqref{EqSecondMomentSpinSpinCorrFunAtTricPoint}]}
moment
allows us to obtain bounds on the 
{power-law}
exponents for 
higher measurement-averaged
 moments of the spin-spin correlation function {at the 
 \textit{learning}
 tricritical point{, i.e.\ the {\it higher} Nishimori critical point}}. In particular, note that from Jensen's inequality~\cite{fellerintroduction} 
\begin{align}
\overline{f(\langle\sigma_i\sigma_j\rangle_{\vec{m}})}\geq f\Big(\overline{\langle\sigma_i\sigma_j\rangle_{\vec{m}}}\Big)\label{EqJensensInequality}
\end{align}
where $f(x)$ is a convex function. 
For example, if we take $f(x)=x^{2k}$, where $k\geq1$ is a positive integer, we obtain the following inequality
\begin{equation}
\overline{(\langle\sigma_i\sigma_j\rangle_{\vec{m}})^{2k}}\geq \Big(\overline{\langle\sigma_i\sigma_j\rangle_{\vec{m}}}\Big)^{2k}.\label{EqJensenInequalityEvenMoments}
\end{equation}
{At} the novel tricritical point
in the learning phase diagram, 
which occurs 
{at the intersection 
of 
lines of continuous phase transitions
in the $2D$ Ising 
[solid red line in Fig.~\ref{fig:2DGaussianMeasurements}] and the $2D$ ordinary Nishimori critical universality 
[solid dark-blue line in Fig.~\ref{fig:2DGaussianMeasurements}]
{classes},}
we expect 
{that}
all the moments of the spin-spin correlation function 
{exhibit}
power-law {decay} at long-distances, i.e.
\begin{equation}
    \overline{\langle\sigma_i\sigma_j\rangle_{\vec{m}}^{q}} \sim \frac{1}{|i-j|^{2X_{q}}} \,.
    \label{EqDefScalingDimofQthMomentofSpinSpinCorrFun}
\end{equation}
Then from Eq.~\eqref{EqJensenInequalityEvenMoments}, the power-law exponent $X_{2k}$ for the measurement-averaged $(2k)^{\text{th}}$ moment of the spin-spin correlation function must be less than $2kX_{1}$, where $X_{1}=1/8$ from Eq.~\eqref{EqFirstMomentSpinSpinCorrFunAtTricPoint}. Therefore, we obtain that
\begin{equation}
    X_{2k}\leq \frac{2k}{8},\;\;\; k\in \mathbb{Z}^{+}.\label{EqBoundOnExponentForEvenMomentsBeta}
\end{equation}
Clearly, the exact power-law exponent $X_{2}=1/8$ [Eq.~\eqref{EqSecondMomentSpinSpinCorrFunAtTricPoint}] for the Edwards-Anderson correlation function (second moment, $2k=2$) satisfies the above bound.
{{In fact, we can obtain a tighter bound on the scaling dimensions of the measurement-averaged even moments of the spin-spin correlation function. 
To this end,
we apply Jensen's inequality in Eq.~\eqref{EqJensensInequality} to the square of the expectation value of the spin-spin correlation function $\langle\sigma_i\sigma_j\rangle_{\vec{m}}^2$, as follows
\begin{equation}
    \overline{f(\langle\sigma_i\sigma_j\rangle_{\vec{m}}^2)}\geq f\Big(\overline{\langle\sigma_i\sigma_j\rangle_{\vec{m}}^2}\Big)\label{EqJensensInequalitySquare},
\end{equation}
where $f(x)$ is again a convex function. Note that $\langle\sigma_i\sigma_j\rangle_{\vec{m}}^2\geq0$, and for $x\geq 0$ $f(x)=x^{r}$ is a convex function for any $r\geq1$. Therefore, we obtain the following inequality
\begin{equation}
\overline{|\langle\sigma_i\sigma_j\rangle_{\vec{m}}|^{2r}}\geq \Big(\overline{\langle\sigma_i\sigma_j\rangle_{\vec{m}}^2}\Big)^{r},\;\;\;\;\; r\geq 1.\label{EqJensenInequalityEvenMoment}
\end{equation}}
 Let us consider the 
 {above}
 inequality in Eq.~\eqref{EqJensenInequalityEvenMoment} with $r=k$ where $k\geq 1$ is a positive integer,
\begin{equation}
\overline{(\langle\sigma_i\sigma_j\rangle_{\vec{m}})^{2k}}\geq \Big(\overline{\langle\sigma_i\sigma_j\rangle_{\vec{m}}^2}\Big)^{k},\;\;\; k\in \mathbb{Z}^{+}.
\end{equation}
Therefore, analogous to Eq.~\eqref{EqBoundOnExponentForEvenMomentsBeta} {and}
using Eq.~\eqref{EqSecondMomentSpinSpinCorrFunAtTricPoint}, we
{obtain}
the following bound on {the} power-law exponent $X_{2k}$ for the
measurement-averaged $(2k)^{\text{th}}$ moment of the spin-spin correlation function,
\begin{equation}
    X_{2k}\leq \frac{k}{8},\;\;\; k\in \mathbb{Z}^{+},\label{EqBoundOnExponentForEvenMoments}
\end{equation}
which is a tighter bound than the one in Eq.~\eqref{EqBoundOnExponentForEvenMomentsBeta}.
{Moreover, from Eq.~\eqref{EqMeasurementAveragedOddAndEvenMomentsAreEquivalent},}
we know that at the tricritical point
the scaling dimension of the measurement-averaged $(2k)^{\text{th}}$ moment of the spin-spin correlation function must be equal to the scaling dimension of the measurement-averaged $(2k-1)^{\text{th}}$ moment of the spin-spin correlation function. Therefore, from Eq.~\eqref{EqBoundOnExponentForEvenMoments}, we also obtain a bound on the scaling dimension of the measurement-averaged $(2k-1)^{\text{th}}$ moment of the spin-spin correlation function,
\begin{equation}
    X_{2k-1}=X_{2k}\leq \frac{k}{8},\;\;\; k\in \mathbb{Z}^{+}.\label{EqBoundOnExponentForOddMoments}
\end{equation}
Of course, 
{from Eq.~\eqref{EqFirstMomentSpinSpinCorrFunAtTricPoint} and Eq.~\eqref{EqSecondMomentSpinSpinCorrFunAtTricPoint},}
the inequality in Eq.~\eqref{EqBoundOnExponentForEvenMoments} {and Eq.~\eqref{EqBoundOnExponentForOddMoments}}  is saturated for $k=1$.}}

{Using}
Eq.~\eqref{EqJensenInequalityEvenMoment}, we
can also obtain 
a bound on the 
{power-law} exponent $\tilde{X}_{2k-1}$
for the measurement-averaged 
$(2k-1)^{\text{th}}$ odd moment
of the 
\textit{absolute} value
of the spin-spin correlation function, where $\tilde{X}_{q}$, analogous to Eq.~\eqref{EqDefScalingDimofQthMomentofSpinSpinCorrFun}, is defined as
\begin{equation}
    \overline{|\langle\sigma_i\sigma_j\rangle_{\vec{m}}|^{q}} \sim \frac{1}{|i-j|^{2\tilde{X}_{q}}} .\label{EqDefScalingDimofQthMomentofAbsValueofSpinSpinCorrFun}
\end{equation}
In particular,
taking $r=\frac{2k-1}{2}$ {in Eq.~\eqref{EqJensenInequalityEvenMoment},} where $k\geq 2$ 
{is}
a positive integer, we obtain that
\begin{equation}
    \overline{|\langle\sigma_i\sigma_j\rangle_{\vec{m}}|^{2k-1}}\geq \Big(\overline{\langle\sigma_i\sigma_j\rangle_{\vec{m}}^2}\Big)^{\frac{2k-1}{2}},\;\;\; k\geq2 \,.
\end{equation}
Therefore, using Eq.~\eqref{EqSecondMomentSpinSpinCorrFunAtTricPoint}, we find that the power-law exponent $\tilde{X}_{2k-1}$ for
the measurement-averaged $(2k-1)^{\text{th}}$ moment of the
absolute value
of {the} spin-spin correlation function 
must be less than $(\frac{2k-1}{2})X_{2}=\frac{2k-1}{2}\frac{1}{8}$, i.e.\ 
\begin{equation}
    \tilde{X}_{2k-1}\leq \frac{2k-1}{16},\;\;\; k\geq 2 \,.
    \label{EqBoundOnExponentForOddMomentsAbsolute}
\end{equation}
{The case {of} $k=1$, 
i.e., the scaling dimension $\tilde{X}_1$ of the measurement-averaged (first moment of) absolute value of the spin-spin correlation function is 
{slightly}
more interesting. 
Firstly, note
that
\begin{equation}
     \overline{\langle \sigma_i\sigma_j\rangle_{\vec{m}}} \leq\overline{|\langle \sigma_i\sigma_j\rangle_{\vec{m}}|} \,.
     \label{EqInequalityIForFirstAbsoluteValue}
\end{equation}
Now for $0<r<1$, $f(x)=x^r$ is a concave function for $x\geq 0$, and consequently the
inequality sign 
in Eq.~\eqref{EqJensenInequalityEvenMoment} gets flipped, e.g. for $r=1/2$ we obtain
\begin{equation}
	\overline{|\langle\sigma_i\sigma_j\rangle_{\vec{m}}|}\leq \Big(\overline{\langle\sigma_i\sigma_j\rangle_{\vec{m}}^2}\Big)^{1/2}\,.	
	\label{EqInequalityIIForFirstAbsoluteValue}
\end{equation}
Then using Eqs.~\eqref{EqInequalityIForFirstAbsoluteValue} and~\eqref{EqInequalityIIForFirstAbsoluteValue}, we obtain the following upper and lower bound for the scaling dimension $\tilde{X}_1$ of the measurement-averaged absolute value of the spin-spin correlation function
\begin{equation}
	\label{EqBoundOnModCorrelator}
  	1/16 \leq \tilde{X}_1\leq 1/8\,,
\end{equation}
where we have used the  exact scaling dimension for the measurement-averaged first and second moment of the spin-spin correlation function from Eq.~\eqref{EqFirstMomentSpinSpinCorrFunAtTricPoint} and~\eqref{EqSecondMomentSpinSpinCorrFunAtTricPoint}, respectively.
{Finally, we can also obtain a lower bound on $\tilde{X}_{\text{typ}}$, the typical scaling dimension for $|\langle\sigma_i\sigma_j\rangle_{\vec{m}}|$ correlation function, defined as {follows}
\begin{equation}
    \overline{\log|\langle\sigma_i\sigma_j\rangle_{\vec{m}}|}\sim -2\tilde{X}_{\text{typ}}\,{\log}|i-j|,\;\;\;|i-j| \gg 1 \,.
\end{equation}
As noted above, 
$f(x)=x^r$ is a concave function for $0<r<1$, and
\begin{equation}
   \overline{|\langle \sigma_i\sigma_j\rangle_{\vec{m}}|^r}\leq \Big (\overline{|\langle \sigma_i\sigma_j\rangle_{\vec{m}}|}\Big )^r,\;\;\,r<1 \,,
\end{equation}
and using Eq.~\eqref{EqBoundOnModCorrelator}, the 
power-law exponent
$\tilde{X}_{r}$ 
{of}
$\overline{|\langle \sigma_i\sigma_j\rangle_{\vec{m}}|^r}$ for $r<1$ must be greater than $\frac{r}{16}$.
The typical scaling dimension is given by
\begin{equation}
    \tilde{X}_{\text{typ}}=\frac{{d}\tilde{X}_r}{{d}r}\Big|_{r=0}=\lim_{r\rightarrow0}\frac{\tilde{X}_r}{r} \,,
\end{equation}
and hence,
\begin{equation}
    \tilde{X}_{\text{typ}}\geq \frac{1}{16}\,.
\end{equation}}
The summary of above results is presented in Table~\ref{LabelTableSummaryOfScalingDimension}.}
\begin{table}
\begin{center}
\caption{{\bf Summary of exact results for power-law exponents}
		of various moments of the spin-spin correlation function and its absolute value at the {\it higher}
        Nishimori critical point.}
\label{LabelTableSummaryOfScalingDimension}
\begin{tabular}{ m{5.3cm} |  m{3.2cm} } 
\hline
{\bf correlation function} & {\bf scaling dimension} \\
\hline \multicolumn{2}{c}{}\\[-2.5mm] \hline
moments $\overline{\langle \sigma_i\sigma_j\rangle_{{\vec{m}}}^q}$ &\\ \hline
\hspace{4mm}
first moment $X_1$ & $X_1=1/8$ \\ 
\hline
\hspace{4mm}
 second moment $X_2$ &$X_2= 1/8$\\
\hline
\hspace{4mm}
 $(2k)^{\text{th}}$ and $(2k-1)^{\text{th}}$ moments ($k\geq 1$) & $X_{2k}=X_{2k-1}$\\
\hline
\hspace{4mm} bound on power-law exponents
& $X_{2k}=X_{2k-1}\leq k/8$\\
 \hline
moments of absolute values $\overline{|\langle \sigma_i\sigma_j\rangle_{{\vec{m}}}|^q}$ & \\
 \hline
 \hspace{4mm} first moment 
 $\tilde{X}_1$
 & $1/16\leq 
\tilde{X}_1
\leq1/8 $\\
 \hline
 \hspace{4mm} other odd moments 
 $\tilde{X}_{2k-1}$
 ($k\geq 2$) & 
 $\tilde{X}_{2k-1}\leq(2k-1)/16$
 \\
 \hline
 \hspace{4mm} typical scaling dimension 
  $\tilde{X}_{\text{typ}}$
 & 
 $\tilde{X}_{\text{typ}} \geq 1/16 $
 \\
 \hline
\end{tabular}
\end{center}
\end{table}

\subsection{Monotonicity of the Casimir  Effective Central Charge
\texorpdfstring{$\mathbf{c_{\text{eff}}> c_{\text{Ising}}=1/2}$}{Lg}
\label{SecCeffThmBound}\label{SubSecExactResultsC}}
{We discussed in Sec.~\ref{SecEmergentGaugeAndEnlargedPermutationSymmetry} that the {\it higher} Nishimori critical point Eq.~\eqref{EqLocationFixedPoint} of the replica theory in Eq.~\eqref{EqLatticeReplicaTheory} in the $R\rightarrow1$ replica limit describes the long-distance properties of the tricritical point.}
Due to the replica limit, the tricritical point  is described by a non-unitary CFT
and we can associate a Casimir effective central charge~\cite{LUDWIGCardy,CardyLog2013}
with this non-unitary CFT.
{As demonstrated in Ref.~\cite{NahumJacobsen}} (also see Ref.~\cite{ZabaloGullansWilsonVasseurLudwigGopalakrishnanHusePixley,KumarKemalChakrabortyLudwigGopalakrishnanPixleyVasseur,DecoherenceWangEtAl}), this Casimir effective central charge characterizes the finite-size scaling of the Shannon entropy of the measurement record for critical points in $2D$ Bayesian inference problems{, which can equivalently be thought of as the problems of performing Born-rule measurements on Rokhsar-Kivelson wavefunctions~\cite{PutzGarrattNishimoriTrebstZhu,PatilLudwig20251}.}
Recently, some of us derived a $c$-effective theorem~\cite{PatilLudwig20251} 
that non-perturbatively demonstrates the monotonic decrease of the Casimir effective central charge under a relevant RG flow induced by performing weak Bayesian measurements on a $2D$ classical critical point.
As we discuss in App.~\ref{AppDerivationOfCeffBound},
{we can use the 
{tools}
used in the proof of the $c$-effective theorem
to 
{demonstrate}
that the 
{Casimir effective central charge decreases under RG flow from the tricritical point to the (unmeasured) $2D$ Ising critical point [see Fig.~\ref{fig:2DGaussianMeasurements}]. 
That is, we can show that the} {Casimir}
effective central charge $c_{\text{eff}}$ of the tricritical point (the UV fixed point) {\it must be greater} than the central charge of the 
{$2D$ Ising critical point}
(the IR fixed point), i.e.
\begin{equation}
    c_{\text{eff}}>\frac{1}{2} \,.
    \label{EqCeffDecrease}
\end{equation}}

{To this end, we first note that the {$c$-effective theorem in}}
Ref.~\cite{PatilLudwig20251} 
{is directly applicable to setups where a relevant RG flow is caused by performing weak measurements on a unitary (translationally-invariant)
$2D$ classical critical point, which in the infrared takes us to some non-unitary
fixed point at non-vanishing measurement strength. 
However, the present setup of interest is different, because \textit{reducing} the strength of measurements at the novel tricritical point is relevant in an RG sense, and that in the infrared takes us to a relatively simple
fixed point -- 
the
(unmeasured) $2D$ Ising critical point.}
{So we cannot directly apply the
$c$-effective theorem in Ref.~\cite{PatilLudwig20251} to this problem.}
{Nevertheless, the tools used in the proof of the $c$-effective theorem~\cite{PatilLudwig20251} turn out to be useful also in demonstrating the decrease of the Casimir effective central charge 
under the RG flow
from the novel tricritical point to the (unmeasured) $2D$ Ising critical point.}
Under fairly standard assumptions listed at the start of App.~\ref{AppDerivationOfCeffBound}, the derivation of 
{this result [Eq.~\eqref{EqCeffDecrease}]}
is given in App.~\ref{AppDerivationOfCeffBound}.

\begin{figure}[t!]
    \centering
    \includegraphics[width=\linewidth]{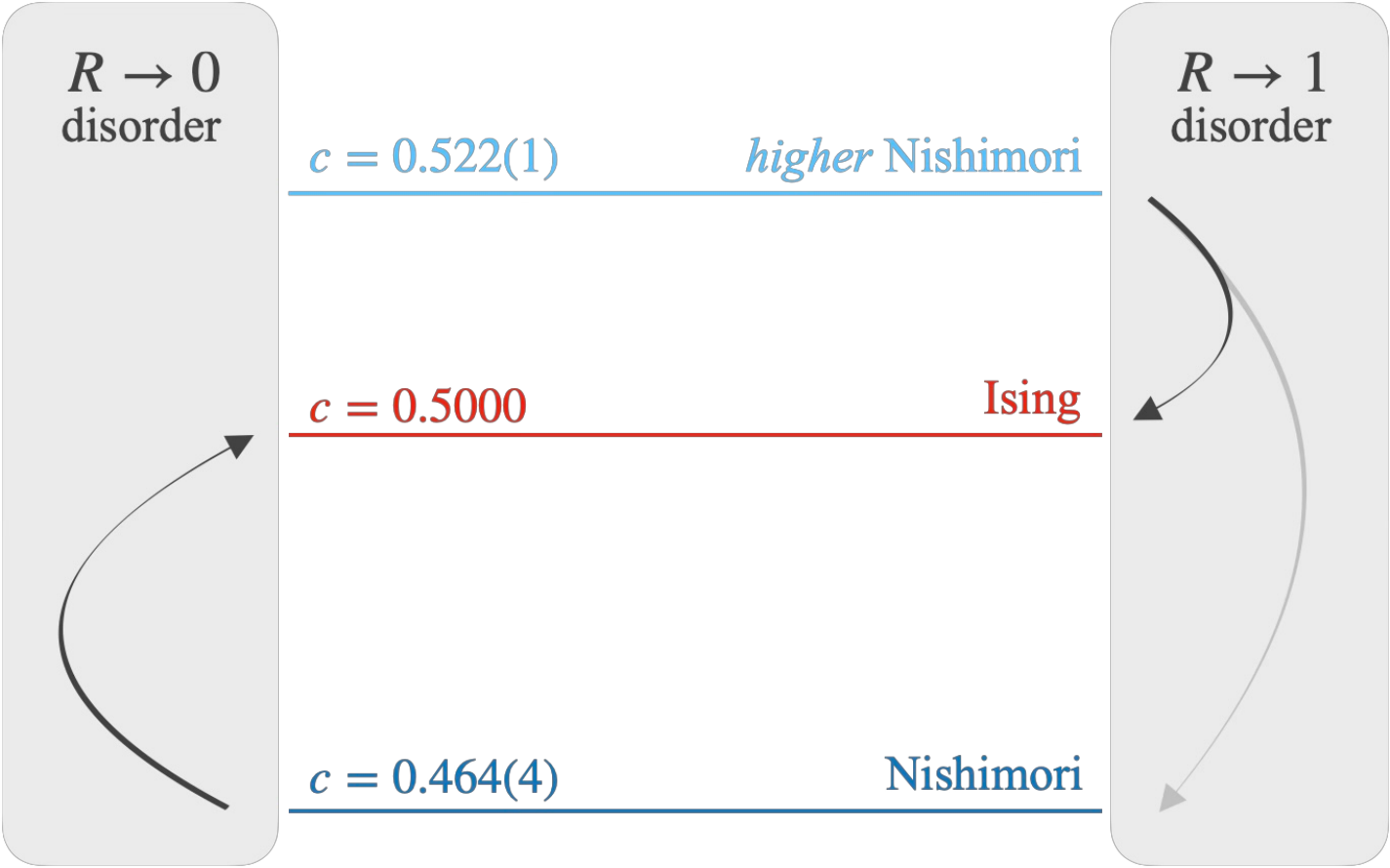}
    \caption{{\bf Hierarchy of Casimir central charges.} 
        Shown are the Casimir effective central charges of the three principal critical points of the learning phase diagram as a ladder.
        Renormalization group (RG) flows between the critical points are shown on the left and right, 
        where the left-hand side shows RG flows in the presence of $R \to 0$, i.e.\ conventional {\it quenched} `impurity-type' disorder, 
        while the right-hand side shows RG flows in the presence of $R \to 1$, i.e.\ {\it learning/measurement-induced} disorder.
        The RG flow from the ordinary Nishimori point to Ising criticality in the random-bond Ising model (with $R \to 0$ disorder) 
        was discussed, e.g., in the early Refs.~\onlinecite{McMillan}, and \onlinecite{Georges1985b},
        for a slightly more recent discussion see, e.g., the summary presented in the Introduction of Ref.~\onlinecite{GRL2001}.
        The RG flows from the {\it higher} Nishimori point introduced in this work were numerically identified in the context of the learning phase diagram
        of Ref.~\onlinecite{PutzGarrattNishimoriTrebstZhu}, 
        with the RG flow from 
        our
        {\it higher} Nishimori point to the conventional Ising criticality concurrently discussed 
        in Ref.~\onlinecite{NahumJacobsen}.
    }
    \label{fig:LadderPlot}
\end{figure}

{{We 
also note
that}
in contrast to the RG flow in the learning phase diagram [Fig.~\ref{fig:2DGaussianMeasurements}]
from the tricritical point, i.e.\ the {\it higher} Nishimori critical point, to the 
unmeasured Ising critical point, for the RG flow in the $2D$ RBIM from the ordinary Nishimori critical point~\cite{Nishimori_1980} to the `clean' Ising critical point
the corresponding (replica $R\rightarrow0$) Casimir effective central charge is known to increase~\cite{HoneckerPiccoPujol2001,PiccoPujolHonecher2006}.
{Using the decrease of the replica $R\rightarrow1$ Casimir effective central charge established in Eq.~\eqref{EqCeffDecrease} and a certain physically motivated assumption, an argument can be made to understand this increase of the replica $R\rightarrow0$ Casimir  effective central charge~\cite{PatilLudwig20251}. 
We refer the reader to App.~\ref{AppCommentOnTheCeffAtOrdinaryNishimori} (see, e.g., Fig.~\ref{fig:Reversal-of-Central-Charge-Change}) for the discussion on
the RG flow from the ordinary Nishimori critical point to the `clean' $2D$ Ising critical point 
in the $2D$ RBIM
and the argument for the corresponding increase of the (replica $R\rightarrow0$)  Casimir effective central charge.
The different RG flows between the three principal critical points of our manuscript in the presence of either
Born-rule measurement randomness (replica $R\rightarrow1$) or quenched impurity-type disorder ($R\rightarrow0$)
are summarized in the `ladder plot' of Fig.~\ref{fig:LadderPlot}}.
}}

\subsubsection*{An {extension} of the \texorpdfstring{$c$}{Lg}-effective theorem for \texorpdfstring{$R\to1$}{Lg} disorder}

Finally, coming back to the demonstrated decrease of the replica $R\rightarrow1$ Casimir effective central charge 
in Eq.~\eqref{EqCeffDecrease}, we note that the corresponding argument that uses the tools in the proof of the $c$-effective theorem~\cite{PatilLudwig20251} straightforwardly generalizes to other scenarios where an unstable novel fixed point can be obtained by performing Bayesian measurements (say Gaussian analogous to Ref.~\cite{NahumJacobsen}) of suitable strength with a local observable on a $2D$ (unmeasured) classical critical point:
\begin{itemize} 
\item  
	If there is an RG flow from an unstable fixed point at finite measurement strength back to an unmeasured critical point
	(i.e.\ an RG flow triggered by reducing the corresponding measurement strength),
        then we can non-perturbatively demonstrate that the Casimir effective central charge $c_{\text{eff}}$ at the unstable fixed point is 
        always {\it greater} than the central charge of the unmeasured critical point and monotonically decreases along the RG flow.
\end{itemize}
This can be viewed as a generalization/extension of the above mentioned $c$-effective theorem derived in Ref.~\cite{PatilLudwig20251}, 
which non-perturbatively demonstrated the decrease of the Casimir effective central charge under a relevant RG flow caused by performing
 measurements on a $2D$ (unitary) classical critical point (or the corresponding RK wavefunction) and that in the infrared takes us 
 to a non-unitary fixed point. 
 That is, the $c$-effective theorem of Ref.~\cite{PatilLudwig20251} concerns an RG flow in the opposite direction, namely from a unitary (unmeasured) theory to a non-unitary theory at finite measurement strength, whereas here we instead consider a flow from a non-unitary theory to a unitary theory.
An analytical derivation of Eq.~\eqref{EqCeffDecrease} is presented in detail in App.~\ref{AppDerivationOfCeffBound}.
It uses
the tools employed in Ref.~\cite{PatilLudwig20251}, which straightforwardly 
generalize to other scenarios and demonstrate the
decrease of the Casimir effective central charge
under the RG flow triggered by reducing the measurement strength
from a non-unitary critical point, occurring
at a finite strength of the corresponding measurements, to an unmeasured (unitary) classical critical point.

A general discussion of the RG monotonicity of the Casimir effective central charge for measurement problems will be presented in an upcoming paper by two of the present authors~\cite{PatilLudwig20262}.

\subsection{Dual Spin Correlation Functions
\label{SubSecOnDualSpinCorrFun}}

Merz and Chalker~\cite{MerzChalker} demonstrated that the scaling dimensions of the bond-randomness averaged moments of the dual spin correlation 
{function}~\cite{ReadLudwig2001} at the ordinary Nishimori critical point in the $2D$ RBIM are symmetric about the \textit{moment order $q=1/2$}. This consequently implied that the bond-randomness averaged \textit{first} moment of the dual spin correlation function at the ordinary Nishimori critical point in the $2D$ RBIM has zero scaling dimension and that 
the 
higher moments after the first moment have \textit{non-positive} scaling dimensions.
{At the learning tricritical point [Fig.~\ref{fig:2DGaussianMeasurements}], i.e.\ the {\it higher} Nishimori critical point, we can demonstrate that the scaling 
dimensions 
$\eta_q$ of the measurement-averaged $q^{\text{th}}$ moments of the dual spin correlation functions are symmetric about the \textit{first moment order $q=1$}. 
We demonstrate this result in App.~\ref{AppDualSpinCorrFun} by suitably defining the dual spin correlation function in the present problem~\cite{KadanoffCeva,ReadLudwig2001}, and then by 
following the logic in~\cite{MerzChalker}.
This result at the {\it higher} Nishimori critical point on the symmetry of scaling dimensions for the measurement-averaged moments of the dual spin correlation function about the \textit{first} moment order should be contrasted with the above-mentioned result of Merz and Chalker at the ordinary Nishimori critical point~\cite{MerzChalker}.}
In particular, our result implies that the measurement-averaged \textit{second} moment of the dual spin correlation function at the learning tricritical point has zero scaling dimension~\footnote{{This is because the zeroth measurement-averaged moment {trivially} has zero scaling dimension.}} and all the higher moments after the second moment have \textit{non-positive} scaling 
{dimensions}, i.e.
\begin{equation}
    \eta_2=0 \,
    \label{EqSecondMomentOfDualSpinCorr}
\end{equation}
and
\begin{equation}
    \eta_q\leq 0\;\;  \text{for}\;\; q> 2 \,.
    \label{EqHigherMomentOfDualSpinCorr}
\end{equation}
We have explicitly derived 
{Eqs.~\eqref{EqSecondMomentOfDualSpinCorr} and~\eqref{EqHigherMomentOfDualSpinCorr}}
in App.~\ref{AppDualSpinCorrFun} for Gaussian measurements. However, analogous to
the
discussion in previous sections, the universal scaling behavior in {Eqs.~\eqref{EqSecondMomentOfDualSpinCorr} and~\eqref{EqHigherMomentOfDualSpinCorr}} is expected to hold also at the learning tricritical point
in the case of binary measurements.
Finally, analogous to other correlation functions [Eq.~\eqref{EqFirstMomentIsEqualToTheUnmeasured}], the scaling dimension $\eta_1$ of the measurement-averaged \textit{first} moment of the dual spin correlation function is equal to the scaling dimension of the dual spin correlation function at the unmeasured $2D$ Ising critical point, and due to the KW duality~\cite{KadanoffCeva}, the latter scaling dimension is equal to the scaling dimension of the spin operator ($=1/8$), and we obtain
\begin{equation}
    \eta_1=\frac{1}{8} \, .
    \label{EqFirstMomentOfDualSpinCorr}
\end{equation}
{Since $\eta_{0}$ is trivially equal to zero~\cite{Note9}, assuming {analyticity}
in the moment order,  the result for $\eta_1$ [Eq.~\eqref{EqFirstMomentOfDualSpinCorr}] and $\eta_2$ [Eq.~\eqref{EqSecondMomentOfDualSpinCorr}] imply a multifractal spectrum
for the scaling dimensions of
the dual spin correlation function at the learning tricritical point, i.e.\ the latter scaling dimensions are non-linear functions of the moment order.}

\section{Numerical Results}
\label{SecNumericalResults}

Let us now turn to a high-level summary of our numerical simulations, which demonstrate 
that the previously identified tricritical point~\cite{PutzGarrattNishimoriTrebstZhu}
in the learning phase diagram with {\it binary} measurement outcomes is a {\it higher} Nishimori critical point. 
Here we concentrate on two principal observables, the decay of the Edwards-Anderson correlators and the Casimir 
effective central charge. 
Our general numerical approach is similar to the one used in Ref.~\cite{PutzGarrattNishimoriTrebstZhu}, 
which combines tensor network methods with Monte Carlo sampling. 
For the numerically inclined reader, we provide further details on our numerical simulations in Appendix~\ref{App:Numerics}.
    
\subsubsection*{Edwards-Anderson correlators}
    
    We compute the Edwards-Anderson correlator $\nonumber\overline{\langle \sigma_{i}\sigma_{j}\rangle^2_{\vec{m}}}$ 
    at various points along the critical $\beta = \beta_c$ line in the learning phase diagram of the $2D$ 
    classical Ising model under binary measurements. 
    Representative data for a $512 \times 512$ square lattice with open boundary conditions are shown in Fig.~\ref{fig:EA-IsingTricritical}.
    We find the power-law exponent of the EA correlator at the tricritical point to be 
    $2X_2(\gamma_c) = 0.2508(8)$.
    This is remarkably close to the analytically predicted value of $1/4$  of Eq.~\eqref{EqSecondMomentSpinSpinCorrFunAtTricPoint}, strongly supporting the identification of the tricritical point 
    as a {\it higher} Nishimori  critical point.
    
\begin{figure}[h!]
    \centering
    \includegraphics[width=0.9\linewidth]{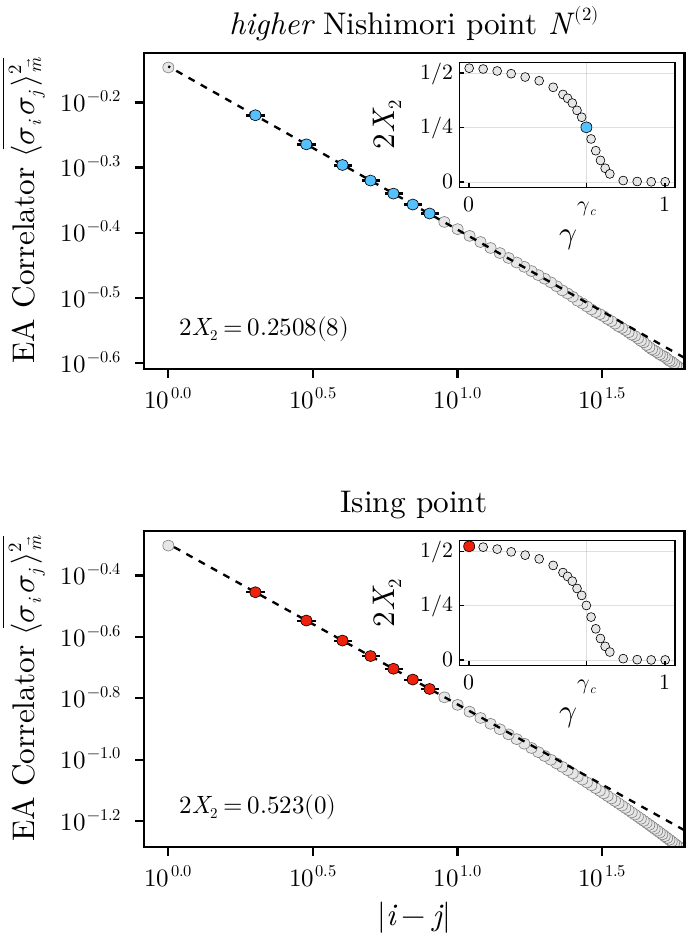}
    \caption{{\bf Edwards-Anderson correlator.} 
    Shown is the Edwards-Anderson correlator at the Ising and \textit{higher} Nishimori critical points in the learning phase diagram 
    of the $2D$ classical Ising model under binary measurements.
    We fit the slope in a log-log plot to extract the power-law exponent $2X_2$ for the EA correlator, for a system size of $512\times 512$.
    Simulations are performed on a $512 \times 512$ square lattice with open boundary conditions. 
    The EA correlator is computed in the bulk of the system, averaging over $\sim 100\,000$ samples. 
    Note that towards the unmeasured $2D$ 
    Ising critical point, (marginally irrelevant) finite-size effects lead to a small deviation
    from the known exact value of 1/2.
    }
    \label{fig:EA-IsingTricritical}
\end{figure}
   
\begin{figure}[t!]
    \centering
    \includegraphics[width=0.9\linewidth]{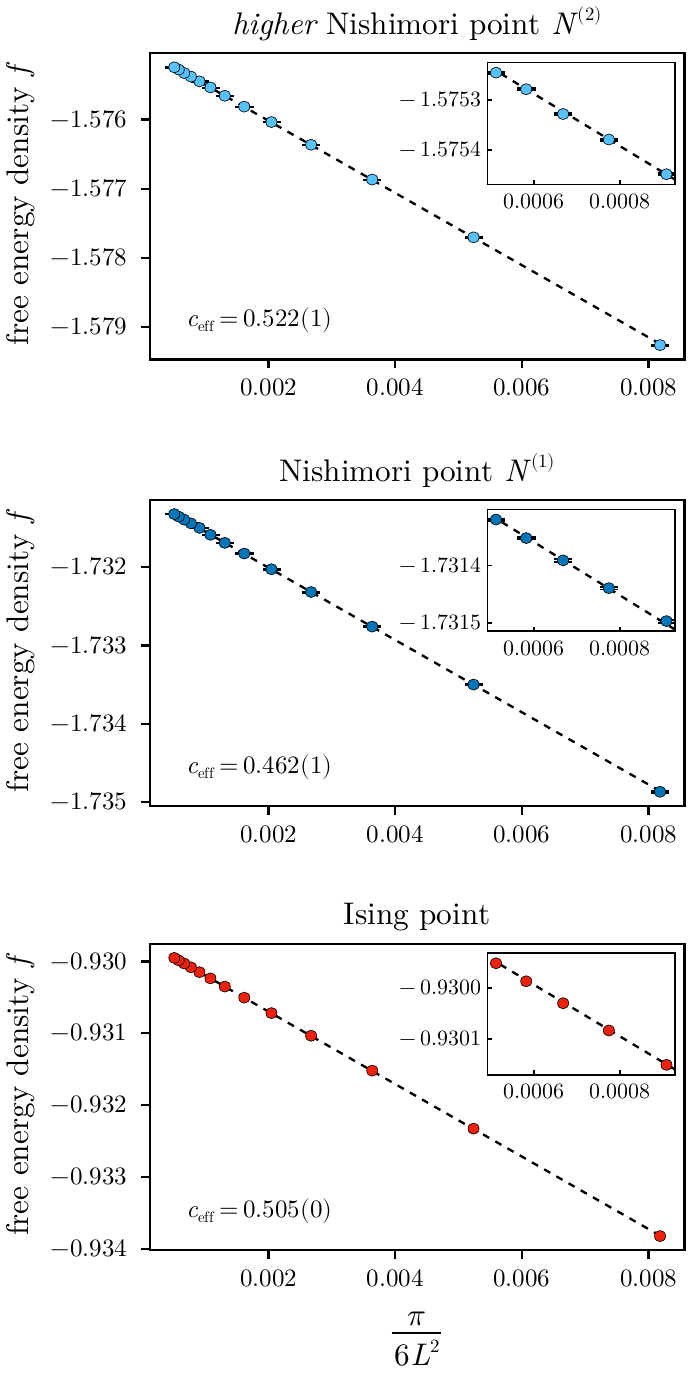}
    \caption{{\bf Casimir central charge estimates.} 
    Shown is the free energy density at the three principal critical points of the learning phase diagram discussed in this manuscript. 
    We fit the slope as a function of $\pi / (6 L^2)$ for system sizes from $L=8$ to $L=32$ 
    to extract the Casimir effective central charge $c_{\text{eff}}$. 
    The free energy density is computed using the transfer matrix method for a cylinder of circumference $L$ and length $1000L$, 
    averaging over $800\,000$ (\textit{higher} Nishimori) and $200\,000$ (Nishimori) samples.
    }
    \label{fig:FreeEnergy}
\end{figure}

\subsubsection*{Casimir effective central charge estimates}

    In order to extract the Casimir effective central charge $c_{\text{eff}}$, we also compute the free energy density $f$ along the critical $\beta = \beta_c$ line in the learning phase diagram with binary measurements. Interpreting the $2+0$ dimensional model as a $1+1$ dimensional imaginary time evolution, we map the model, for any given disorder realization, to a non-interacting Majorana fermion model, derived explicitly in Appendix~\ref{App:Numerics}.
    This mapping allows us to compute the free energy density of a cylinder with circumference $L$ and length $1000L$ 
    by computing the leading Lyapunov exponent
    for the product of random transfer matrices along the long direction of the cylinder (following, e.g., ~\cite{ZabaloGullansWilsonVasseurLudwigGopalakrishnanHusePixley,DecoherenceWangEtAl,MerzChalker}).
    
    By averaging over $200\, 000 - 800\, 000$ samples and fitting system sizes $L = 8$ to $L = 24$ for various points along the critical $\beta = \beta_c$ line, 
    we have produced the evolution of the Casimir
    effective
    central charge from the Ising critical point towards the tricritical / {\it higher} Nishimori 
    critical
    point 
    for increasing measurement strength that we show in Fig.~\ref{fig:CasimirCentralCharge_betac} above.
    Right at 
the {\it higher} Nishimori point we find that the Casimir effective central charge takes a value of 
    \begin{equation}
    	c_{\text{eff}} = 0.522(1) \,,
	\label{eq:CasimirCentralChargeHigherNishimoriPoint}
    \end{equation}
    performing fits up to system size $L=32$, as shown in Fig.~\ref{fig:FreeEnergy} where we also plot fits for the ordinary Nishimori point 
    (on the $\beta=0$ line) and the conventional Ising critical point.

\subsubsection*{Higher Nishimori line}

Finally, we can also identify the location of an 
\textit{emergent}
{\it higher} Nishimori line in the numerical simulations of the learning phase diagram with \textit{binary} measurement outcomes. 
To do so, we calculate the EA correlation and the
measurement-averaged
Ising spin-spin correlation
function (first moment), which is equal to the Ising spin-spin correlation function in the unmeasured model [Eq.~\ref{EqFirstMomentIsEqualToTheUnmeasured}],
for horizontal cuts away from the critical $\beta = \beta_c$ line.  
Then, by matching the long-distance behavior of
the 
averaged
Ising spin-spin correlation and the EA correlation,  we can construct a 
{\it higher Nishimori condition} also away from criticality in the form of 
\begin{equation}
    \overline{\langle \sigma_i\sigma_j\rangle_{\vec{m}}^2} \simeq \overline{\langle \sigma_i\sigma_j\rangle_{\vec{m}}} \ ,
    \label{eq:nishicondition}
\end{equation}    
or
by simply matching their respective correlation lengths, in the paramagnetic phase, at large distance
\begin{align}
    \overline{\langle \sigma_i\sigma_j\rangle_{\vec{m}}^2}\propto e^{-|i-j|/\xi^{(2)}} \ ,\nonumber \\
    \overline{\langle \sigma_i\sigma_j\rangle_{\vec{m}}}\propto e^{-|i-j|/\xi^{(1)}} \ ,\nonumber \\
    \xi^{(2)} = \xi^{(1)} \,. 
\end{align} 
Note that for {\it Gaussian} measurements
the correlation functions in the equations above are {\it exactly} equal even off criticality
along the \textit{higher}
Nishimori line in Fig.~\ref{FigComparison}(b), as stated in App.~\ref{AppGaussianMeasurements}, Eq.~(\ref{EqFinal}). 
Employing the same condition for the {\it binary} measurement phase diagram allows us to pin down the location 
of the {\it emergent} \textit{higher} Nishimori line in the respective \textit{binary} measurement phase diagram, Fig.~\ref{fig:numerical_higher_nishimori_line}, 
as shown in the following.

\begin{figure}[t!]
    \centering
    \includegraphics[width=0.9\linewidth]{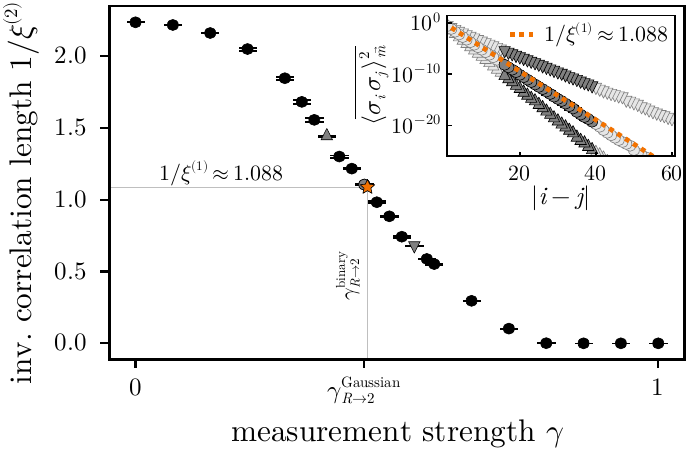}
    \caption{{\bf \textit{Higher} Nishimori condition for $\mathbf{\beta =\beta_c/2}$.}
        Shown is the inverse correlation length  $1 / \xi^{(2)}$ extracted from the EA correlator for a horizontal cut 
        in the paramagnetic phase of the learning phase diagram
        at $\beta=\beta_c/2$ varying the measurement strength $\gamma$.
        The (inverse) EA correlation length is extracted from a fit $\overline{\langle \sigma_i\sigma_j\rangle_{\vec{m}}^2}\propto e^{-|i-j|/\xi^{(2)}}$ 
        and compared to the (inverse) Ising spin-spin correlation length $1 / \xi^{(1)} \approx 1.088$ (horizontal line), 
        which is known analytically~\cite{mccoy2014two}, also from the fermion gap~\cite{SML64isingfermion}, as $1 / \xi^{(1)} = 2\left| \text{KW}(\beta) - \beta \right|$.
        The point where the EA correlation length $\xi^{(2)}$ matches the Ising spin-spin correlation length $\xi^{(1)}$ is indicated by the orange star, 
        which is the numerical estimate for the location of the \textit{higher} Nishimori line for binary measurements.
        The inset shows the EA correlator at the point closest to the \textit{higher} Nishimori condition (circle), as well as one point with larger (up triangle) and smaller (down triangle) inverse correlation length.
        We do this procedure for different cuts (see Fig.~\ref{fig:xi-additional-beta-cuts}) to obtain the numerical estimate for the \textit{higher} Nishimori line in the binary measurement phase diagram, as shown in Fig.~\ref{fig:numerical_higher_nishimori_line}.
     }
    \label{fig:xi-betac-half}
\end{figure}

Figure \ref{fig:xi-betac-half} shows such a cut for $\beta = \beta_c/2$, i.e.\ for a cut going from the paramagnetic phase to the
`spin glass' phase [compare Fig.~\ref{FigComparison}(b)].
Notably, right where this cut hits the predicted location of the {\it higher} Nishimori line for Gaussian measurement outcomes we see that
the EA correlation length is close to 
the value of the Ising spin-spin
correlation length for the unmeasured Ising model, satisfying the 
\textit{higher} Nishimori condition of Eq.~\eqref{eq:nishicondition}.
Additional data along cuts for different inverse temperatures are shown in Fig.~\ref{fig:xi-additional-beta-cuts}  [Appendix \ref{App:Numerics}] 
confirming for each temperature (within the paramagnetic phase) 
that the \textit{higher} Nishimori condition is met in proximity to where the \textit{higher} Nishimori line is predicted for Gaussian measurement outcomes 
[Eq.~\eqref{EqNishimoriStarLine}, with $\Delta=(\arctanh{\gamma})^2$].
Together, this
provides a strong indication
that also the learning phase diagram with
binary
measurement outcomes exhibits a {\it higher Nishimori line} 
with an 
\textit{emergent}
gauge symmetry, though the microscopic model appears to lack that gauge symmetry~\cite{PutzGarrattNishimoriTrebstZhu}.

\begin{figure}[t!]
    \centering
    \includegraphics[width=0.9\linewidth]{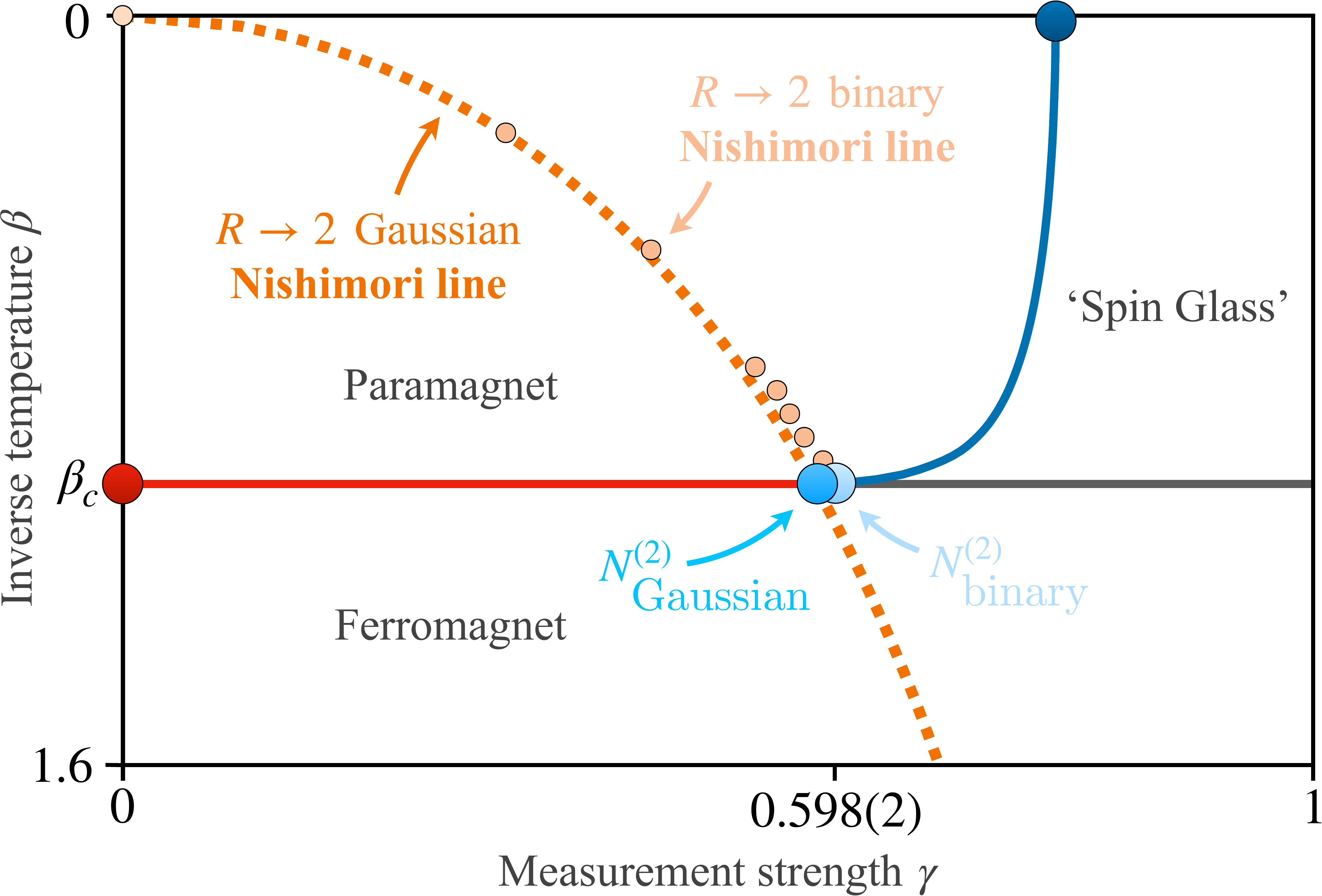}
    \caption{\textbf{Numerical estimate of the \textit{higher} Nishimori line.}
        Shown is the numerical phase diagram with binary measurement outcomes. 
        The orange \textit{higher} Nishimori line is drawn for Gaussian measurement outcomes, 
        while the light orange dots are the numerical estimates for the location of the \textit{higher} Nishimori line in the binary measurement phase diagram.
        The latter is obtained by checking where the
        \textit{higher} Nishimori condition in the paramagnetic phase $\xi^{(2)}(\gamma) = \xi^{(1)}(\gamma)$ 
        is satisfied for cuts at different inverse temperatures $\beta$, as shown, e.g., in Fig.~\ref{fig:xi-betac-half} above for the cut at $\beta = \beta_c/2$  
        or in Fig.~\ref{fig:xi-additional-beta-cuts} in Appendix~\ref{App:Numerics} for additional cuts.
    }
    \label{fig:numerical_higher_nishimori_line}
\end{figure}

\section{Learning Transitions in \texorpdfstring{$\mathbf{D}$}{Lg}-Dimensional (\texorpdfstring{$\mathbf{D > 1}$}{Lg}) Ising Models
\label{SecLearningTransitionsInHigherDimensional}}

Let us now broaden our perspective and discuss the existence of a {\it higher} Nishimori critical point in the 
Bayesian inference/learning phase diagram for the 
$D$-dimensional ($D>1$) classical Ising model under bond-energy measurements, 
or equivalently that of the corresponding RK wavefunction under Born-rule measurements with the bond-energy operator.
We will also discuss
exact results for the Edwards-Anderson correlator at this {\it higher} Nishimori critical point
{that are analogous {to} those in $D=2$ dimensions}.
Finally, we discuss the phase diagram and the RG flows for the 
arbitrary $D>1$
dimensional {classical} Ising critical point (or the corresponding conformal quantum critical RK wavefunction) under Bayesian (Born-rule) bond-energy measurements.

{As long as we have a critical point in the unmeasured {classical} Ising model,
we can follow the
 argument in Sec.~\ref{SecEmergentGaugeAndEnlargedPermutationSymmetry} to establish the existence of a {\it higher} Nishimori
 critical point that admits an emergent formulation as a gauge-invariant and
enlarged replica permutation symmetric theory in the $R\rightarrow2$ replica limit in the learning phase diagram of the classical
Ising model under Bayesian bond energy measurements governed by the $R\rightarrow1$ replica limit. That is, the argument discussed in Sec.~\ref{SecEmergentGaugeAndEnlargedPermutationSymmetry} is independent of the number of spatial dimensions provided the number of dimensions is $D>1$.}
{The latter {\it higher} Nishimori critical point 
{is the 
arbitrary $D>1$ dimensional generalization of the novel tricritical point found in the learning phase diagram of the $2D$ classical Ising model in Ref.~\cite{PutzGarrattNishimoriTrebstZhu}, and it}
will also appear in the learning phase diagram of the equivalent problem of the}
 quantum RK wavefunction [Eq.~\eqref{EqRKWF} and Eq.~\eqref{Eq2DIsingHamiltonian} on {the} $D$-dimensional hypercubic lattice] under Born-rule bond-energy measurements.
{With Gaussian measurements of strength $\Delta$ [Eq.~\eqref{eq:PmsigmaGaussian}], analogous to our discussion in Sec.~\ref{SecEmergentGaugeAndEnlargedPermutationSymmetry} (also see App.~\ref{AppGaussianMeasurements}),
the {\it higher} Nishimori line {$\beta=\Delta$  ($\beta$ is the inverse temperature)} exists in the learning phase diagram of the $D$-dimensional {classical} Ising model,
and analogous to Eq.~\eqref{EqLocationFixedPoint}, 
the 
{{\it higher} Nishimori critical}
point is the critical point
on this line}
{that}
occurs at
\begin{equation}
    \Delta=\beta=\beta^{(D)}_c,\label{EqStrengthOfTheMeasurementsGneralDimensions}
\end{equation}
where $\beta=\beta^{(D)}_c>0$ is the location of the critical point for the unmeasured $D$-dimensional ($D>1$) {classical} Ising model. 

In Ref.~\cite{NahumJacobsen}, the 
Bayesian inference/learning phase diagram of the $D$-dimensional {classical} critical Ising model under Bayesian bond-energy measurements, {which is equivalent to the 
phase diagram~\cite{PutzGarrattNishimoriTrebstZhu,PatilLudwig20251} of the corresponding conformal quantum critical RK wavefunction~\cite{ArdonneFedleyFradkin} under Born-rule bond-energy measurements,} was examined; however,
{their}
study did not yield a conclusive phase diagram, 
and instead two alternative possibilities for the phase diagram were presented.
As we will discuss now,
{realizing that the unstable fixed point $\Ucal$ discussed in Ref.~\cite{NahumJacobsen}}
is a {\it higher} Nishimori critical point with an
emergent gauge-invariant and enlarged replica-symmetric formulation,
we can exclude one of the discussed possibilities of the phase diagram in Ref.~\cite{NahumJacobsen}.

From the study of the $D$-dimensional random bond Ising model {(see, e.g., Ref.~\cite{Cardy_book})}, one can establish, in the $R\rightarrow0$ replica limit, 
an attractive fixed point in spatial dimensions  $D=2+|\epsilon|$, {whose}
universal properties 
can be studied perturbatively in $\epsilon$~\cite{KomargodskiSimmons-Duffin}.
An analogous fixed point can be seen to govern the long-distance properties of the $D=2+|\epsilon|$ dimensional classical Ising critical point under weak measurements of the bond-energies~\cite{NahumJacobsen}, which is governed by
an analogous
replica theory, but in the $R\rightarrow1$ replica limit.
Moreover{, as discussed in Ref.~\cite{NahumJacobsen}}, for 
a classical
Ising critical point in $D=4+|\epsilon|$ dimensions under weak measurements of
bond-energies, there
exists another
{repulsive}
fixed point,
whose universal properties 
can 
be {again} studied perturbatively in $\epsilon$~\cite{NahumJacobsen}.
{Both of these perturbative fixed points, the one in $D=2+|\epsilon|$ dimensions and the one in $D=4+|\epsilon|$ dimensions, are clearly distinct from the {\it higher} Nishimori critical point~\footnote{{One way to see this distinction is to note that at both of the perturbative fixed points the scaling dimension for the measurement-averaged second moment is given, in the respective epsilon expansions, by $2X^{D}_{\sigma}+O(\epsilon)$, where $X^{D}_{\sigma}$ is the scaling dimension of the spin operator at the unmeasured $D$-dimensional classical Ising critical point. While, on the other hand, we know \textit{exactly} that the scaling dimension of the measurement-averaged second moment of the spin-spin correlation function is equal to just $X^{D}_{\sigma}$ at the {\it higher} Nishimori critical point in $D$-dimensions.}}.}
So we know the existence of 
(i) a perturbatively accessible \textit{attractive} fixed point in $D=2+|\epsilon|$ spatial dimensions, 
(ii) a perturbatively accessible \textit{repulsive} fixed point in $D=4+|\epsilon|$ spatial dimensions, 
and also of
(iii) the unstable, {\it higher} Nishimori critical point, which corresponds to 
the $\Ucal$ fixed point in the notation of Ref.~\cite{NahumJacobsen}, and
which, as argued above, exists for all spatial dimensions $D>1$ at the Gaussian
measurement strength {$\Delta$} given by Eq.~\eqref{EqStrengthOfTheMeasurementsGneralDimensions}.
Then, assuming analytical 
continuity in the number of dimensions $D$ and assuming the fewest possible number of fixed points, 
we 
arrive
at the
possible 
phase diagram in Fig.~\ref{fig:IsingLearningPDGeneral} for the 
{classical}
Ising \textit{critical} point in $D>1$ dimensions under bond-energy measurements as a function of dimensions $D$ and measurement strength characterized by $\Delta$. 

\begin{figure}
    \centering
    \includegraphics[width=\linewidth]{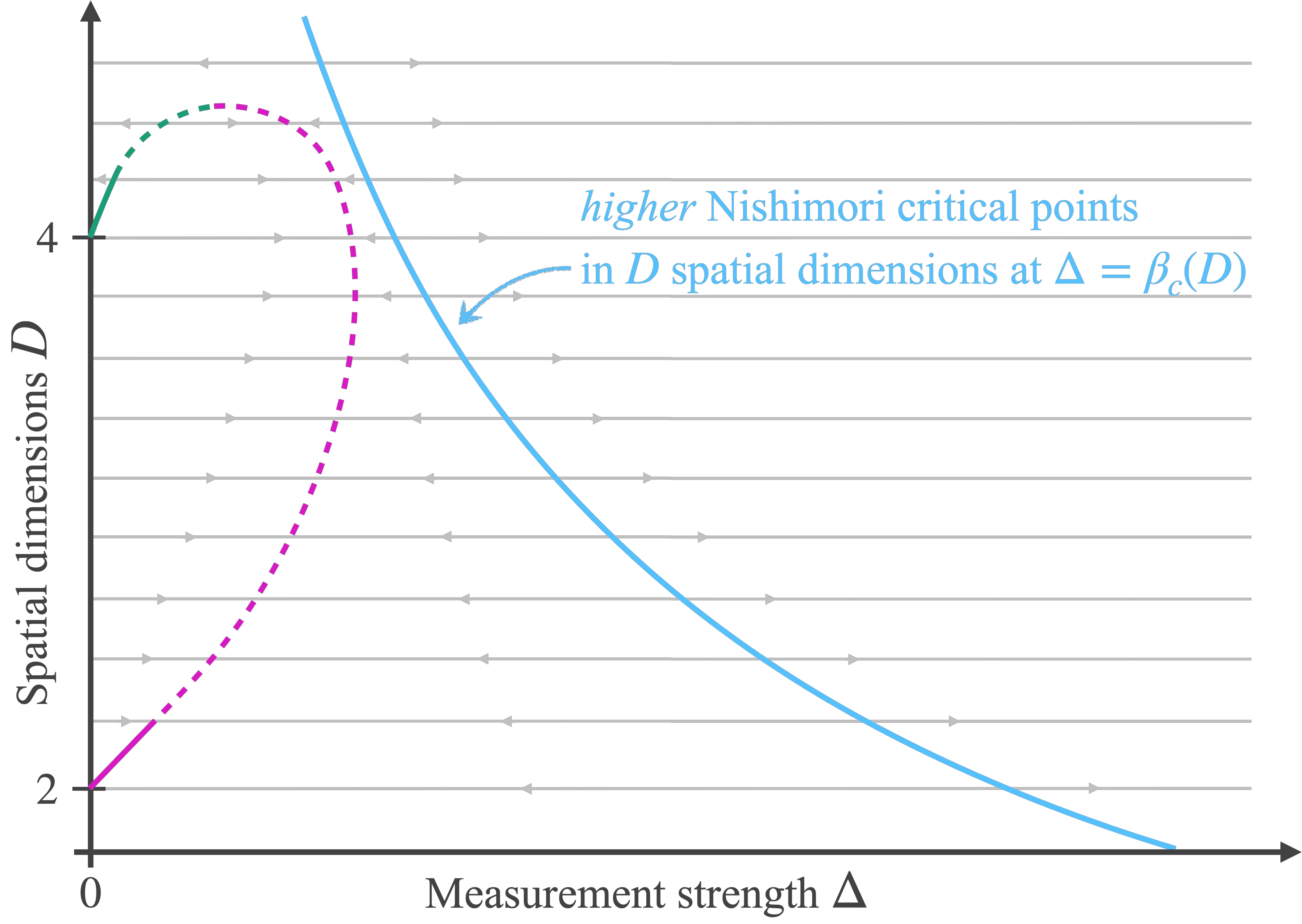}
    \caption{{\bf Learning Transitions in $\mathbf{D}$-Dimensional Ising Models.}
    {Sketch of fixed points and RG flows for the $D$-dimensional classical Ising critical point
    under {Bayesian}
    bond-energy measurements}, 
    {which
    is {also} equivalent to 
    the phase diagram for the corresponding conformal quantum critical RK wavefunction under
    Born-rule bond-energy measurements.} 
    The {magenta}
    curve denotes the attractive fixed point known to exist near 
        {$D=2+|\epsilon|$} spatial
    dimensions and we have shown its extrapolation using a dashed line. 
    The 
    {teal}
    curve that starts at $(D=4,\Delta=0)$ denotes {the} repulsive fixed point known to exist near 
    {$D=4+|\epsilon|$} spatial
    dimensions, and again we have shown its extrapolation using a dashed line. 
    The 
    solid light-blue
    curve denotes the 
    {$D$-dimensional}
    {{\it higher} Nishimori critical point}
    whose location is given by Eq.~\eqref{EqStrengthOfTheMeasurementsGneralDimensions} for Gaussian measurements.}
    \label{fig:IsingLearningPDGeneral}
\end{figure}

The sketch in Fig.~\ref{fig:IsingLearningPDGeneral} is based on
the second possibility discussed in Ref.~\cite{NahumJacobsen} (see their Fig.~$4$). 
Here we have, however, used the fact that the critical inverse-temperature {$\beta_c^{(D)}$} of the $D$-dimensional classical Ising model
should decrease with increasing spatial dimension $D$, and therefore, from Eq.~\eqref{EqStrengthOfTheMeasurementsGneralDimensions}, 
the strength $\Delta$ of Gaussian measurements corresponding to the {\it higher} Nishimori critical points
(solid light-blue line in Fig.~\ref{fig:IsingLearningPDGeneral}) should also decrease with increasing dimension $D$.

We can also obtain that, exactly analogous to Eq.~\eqref{EqFirstMomentOfGeneralObservableAtTricPoint} and Eq.~\eqref{EqSecondMomentOfGeneralObservableAtTricPoint}, the long-distance properties of the measurement-averaged first and 
second moment of any multipoint spin correlation function at the 
{{\it higher} Nishimori critical}
point should be identical to that of the given
correlation function at the 
{unmeasured}
critical point of the $D$-dimensional {classical} Ising model.
{More generally, using the 
arbitrary
$D$-dimensional version of 
Eq.~\eqref{EqMeasurementAveragedOddAndEvenMomentsAreEquivalent},
we again obtain that the long-distance behavior of the measurement-averaged $(2q)^{\text{th}}$ and $(2q-1)^{\text{th}}$ moments of any multipoint spin correlation function at the $D$-dimensional {\it higher} Nishimori critical point are also identical.}
In particular, the power-law exponent of the Edwards-Anderson correlation function at the
{{\it higher} Nishimori critical point}
in the learning phase diagram of the $D$-dimensional {classical} Ising model
(or of the corresponding quantum RK wavefunction) should be equal to the scaling dimension of the spin operator at the unmeasured $D$-dimensional {classical} Ising critical point. 
In spatial dimensions $D\geq 4$, the 
power-law exponent
of the Edwards-Anderson correlation function at the
{{\it higher} Nishimori critical point}
is then equal to the mean-field 
scaling dimension of the spin {operator},
equal to 
$(D-2)/2$. 
{In particular, in dimensions $D=4$, the subleading logarithmic corrections that are present in the spin-spin correlation function at the 
(unmeasured) {classical} Ising critical point are also present in the Edwards-Anderson correlation function at the 
{{\it higher} Nishimori critical point}.}
For the $3D$ {classical} Ising critical point, we do not know the exact scaling dimension of the spin operator {analytically}, but we know a very accurate numerical
{estimate}
for it 
$X_1=0.518148806(24)$
from {numerical} conformal bootstrap calculations~\cite{Chang2025,PhysRevD.86.025022}.
The latter value {then} also provides a very accurate numerical estimate for the 
power-law exponent of the
 Edwards-Anderson correlation function at the
 {{\it higher} Nishimori critical point}
 in the $3D$ {classical}  Ising Bayesian inference/learning phase diagram.

\section{Discussion and Outlook\label{SecOutlook}}

{\it Exact} results for universal critical phenomena in systems with quenched randomness are rather scarce in (quantum) statistical physics. 
One such ground-breaking exact result is the existence of the well-known Nishimori line~\cite{Nishimori_1980,Nishimori1981} that admits a 
gauge-invariant formulation as a replica theory in the $R\rightarrow1$ replica limit 
in the phase diagram of the random-bond Ising model (RBIM), which is otherwise governed by the $R\rightarrow0$ replica limit~\cite{LeDoussalHarrisI,LeDoussalHarrisII,GeorgesHanselLeDoussalMaillard,GRL2001}.
This Nishimori line in the 
$2D$ classical RBIM has been connected, in later ground-breaking work~\cite{DennisKitaevLandahlPreskill}, to the problem of
quantum error-correction (QEC), where the critical point on the Nishimori line captures the criticality of the
threshold for the optimal decoder of the toric code ground state with incoherent errors.
More recently, it has also been appreciated that it also characterizes the mixed-state {phase} transitions of {the} toric code ground state 
under decoherence or weak measurements~\cite{LeeJianXu,FanBaoAltmanVishwanath,SalaGopalakrishnanOshikawaYou,EllisonCheng,WangWuWang,YHChenGrover,YHChenGroverSeparability,LessaStrongToWeak,SohalPrem,SuYangJian,LeeExactCalculation,EcksteinPRX,DecoherenceWangEtAl,eckstein2025learningtransitionstopologicalsurface,DiehlEtAl2025,wan2025revisitingnishimorimulticriticalitylens}.
The appearance of the Nishimori line in quantum measurement problems is {\it not} a coincidence, as the correlated randomness generated by quantum entanglement through Born-rule measurements naturally 
corresponds~\cite{JianYouVasseurLudwig2019,BaoChoiAltman2019} to the $R\rightarrow1$ replica limit instead of the $R\rightarrow0$ replica limit corresponding 
to 
traditional uncorrelated quenched impurity-type disorder.
In this work, we revisited the phase diagram of the deformed toric code wavefunction (away from its stabilizer limit)~\cite{CastelnovoChamon,PapanikolaouRamanFradkin,ArdonneFedleyFradkin,IsakovFendleyLudwigTrebstTroyer, Zhu19deform, Kim23deform, Verresen25deform} subjected to Pauli-$Z$
measurements that some of us had recently studied~\cite{PutzGarrattNishimoriTrebstZhu}, 
which is dual to a Bayesian  
inference problem with the 2D classical Ising model subjected to bond-energy measurements. 
This problem, characterized, in its entirety, by the randomness generated by Born's rule (or Baysian randomness~\cite{Iba_1999,Sourlas_1994,NishimoriBook} in the dual picture), 
is naturally a problem in the $R\rightarrow1$ replica limit, and the \textit{ordinary} Nishimori line also appears in this problem in the undeformed limit of the quantum problem 
and at the infinite temperature in the dual classical Bayesian inference problem.
In this work, we demonstrated the existence of a new, distinct \textit{higher Nishimori line} that admits a gauge-invariant formulation as a replica theory in the $R\rightarrow2$ replica limit, 
and which governs the long-distance physics of a distinct part of the \textit{learning} phase diagram [Fig.~\ref{FigComparison} (b)], which is otherwise governed by the replica limit $R\rightarrow1$.
In particular, we have been able to demonstrate that the novel tricritical point discovered in Ref.~\cite{PutzGarrattNishimoriTrebstZhu} 
(and independently in Ref.~\cite{NahumJacobsen}), 
which sits at the interface of three phases with strong, weak,
and broken $\mathbb{Z}_2$ symmetry in the learning phase diagram [Fig.~\ref{FigComparison} (b)], 
is a \textit{higher} Nishimori critical point.
This \textit{higher} Nishimori criticality is a 
universality class in its own right and, in particular, distinct from the ordinary Nishimori critical point.

The gauge-invariant and enlarged replica-symmetric formulation
in the replica $R\rightarrow2$ limit
allowed us to obtain a 
number
of exact results for universal critical exponents 
at the \textit{learning} tricritical point, i.e.\ the \textit{higher} Nishimori critical point.
In particular, we showed that the power-law exponent for the Edwards-Anderson (EA) correlator (i.e.\ the measurement-averaged second moment of the spin-spin correlation function) 
at this {\it higher} Nishimori critical point is
{exactly} equal to 
that of the spin-spin correlation function
at the critical point in the unmeasured model, which is equal {to $1/8$} 
for the (unmeasured)
$2D$ Ising critical point.
More generally, we proved that the long-distance behavior of the measurement-averaged $(2q)^{\text{th}}$ even moment of any general multipoint spin correlation function is equal to 
that of
the   
$(2q-1)^{\text{th}}$ odd moment at the tricritical point. {Recall that
such an equality of the  $(2q)^{\text{th}}$ even moment and the $(2q-1)^{\text{th}}$ odd moment had been known~\cite{Nishimori1981}
for the bond-disorder averaged moments at the
ordinary Nishimori critical point in the RBIM {(also see App.~\ref{AppTranslatingBetweenRBIMandBayesianInference}).}
Our work thus demonstrates corresponding equalities for the measurement-averaged moments at 
the previously identified {tricritical} point in the learning phase diagram which, as we discovered, is a \textit{higher} Nishimori critical point.  
In particular, since the measurement-averaged first moment of any general multipoint {spin} correlation function at the tricritical point is equal to the
correlation function at the unmeasured Ising critical point [see Eq.~\eqref{EqFirstMomentIsEqualToTheUnmeasured}], the long-distance behavior of the measurement-averaged second moment is also given by {that}
correlation function at the unmeasured Ising critical point.
Using 
this exact result for the equality of the power-law exponents of
the first and 
second moment of the 
(two-point) 
spin-spin correlation function, 
we have also obtained
various rigorous bounds
{on}
the power-law exponents 
{for}
the measurement-averaged higher $(>2)$ moments of the spin-spin correlation function, the measurement-averaged moments of the absolute value
{of}
the correlation function, and also the typical
{spin-spin}
correlation function. See 
{Table}~\ref{LabelTableSummaryOfScalingDimension} for a summary of the {above} results.

We have further demonstrated, following the logic of Merz and Chalker~\cite{MerzChalker}, 
that the scaling dimension of the measurement-averaged second moment of the \textit{dual} spin correlation function at the {\it higher} Nishimori
critical point
vanishes,
and consequently, all the higher moments of the dual spin correlation function 
beyond the second moment have \textit{non-positive} scaling dimensions.}
{Assuming 
{analyticity}
in the moment order, 
our results 
also imply {\it multifractality}
in scaling dimensions
for the moments
of the dual spin correlation function at the {\it higher} Nishimori critical point, and they should be contrasted with the multifractal spectrum of scaling dimensions of the dual spin correlation function at the ordinary Nishimori critical point in the $2D$ RBIM, for which scaling dimensions of all the bond-randomness averaged higher moments beyond the \textit{first} moment are \textit{non-positive}~\cite{MerzChalker}.}
We also note that, in the quantum formulation(s), the multifractality at the \textit{higher} Nishimori critical point is a result of the Born-rule randomness of
quantum measurements
dominating its scaling behavior, 
which breaks translational symmetry in any given realization,  and 
such scaling behavior is, in particular,  qualitatively far richer
(see, e.g., \cite{ZabaloGullansWilsonVasseurLudwigGopalakrishnanHusePixley,putz2025flownishimoriuniversalityweakly,PatilLudwig2024,LiVasseurFisherLudwig,JianShapourianBauerLudwig})
than that familiar from translationally invariant critical points.

In addition, using the tools employed
in the proof of the $c$-effective theorem in Ref.~\cite{PatilLudwig20251}, 
we have also shown that the Casimir effective central charge decreases along the RG flow from the 
{\it higher} Nishimori point to the unmeasured Ising critical point, i.e.\ 
the Casimir effective central charge at the {\it higher} Nishimori critical point, which characterizes the finite-size scaling of the Shannon entropy of the measurement record~\cite{NahumJacobsen,PatilLudwig20251} at this point, is greater than $1/2$.
Our extensive numerical simulations finding $c_{\text{eff}}=0.522(1)$ at the higher Nishimori critical point thus observe the monotonic behavior of the Casimir effective central charge upon RG flow, as expressed by the $c$-effective theorem result.
Interestingly, combined with a certain physically motivated assumption, the above $c$-effective theorem
result also explains the numerically observed~\cite{HoneckerPiccoPujol2001,PiccoPujolHonecher2006} \textit{increase} of the Casimir effective central charge
upon RG flow from the ordinary Nishimori critical point to the clean Ising critical point in the $2D$ RBIM (see App.~\ref{AppCommentOnTheCeffAtOrdinaryNishimori}, particularly Fig.~\ref{fig:Reversal-of-Central-Charge-Change}, 
for details of this argument).

Lastly, we discussed the existence of the corresponding {\it higher} Nishimori critical point in the 
Bayesian inference/learning phase diagram of the $D$-dimensional classical Ising model with $D>1$ subjected to bond-energy measurements which, as we have shown in this work,
has an analogous gauge-invariant and enlarged replica-symmetric formulation as a $R\rightarrow2$ replica theory.
This {\it higher} Nishimori critical point is the 
$D$-dimensional  generalization of the novel tricritical point found in the learning phase diagram of the 2D classical Ising model in Ref.~\cite{PutzGarrattNishimoriTrebstZhu}, and it also appears in the \textit{learning} phase diagram of the RK wavefunction
(e.g.~\cite{CLHenley_2004,IsakovFendleyLudwigTrebstTroyer,putz2025flownishimoriuniversalityweakly}, Appendix C of \cite {PatilLudwig20251})
corresponding to the $D$-dimensional {($D>1$)} classical Ising model
subjected to Born-rule bond-energy measurements, which is an equivalent problem.
{We showed that at this {\it higher} Nishimori critical point in arbitrary dimension $D>1$,}
the long-distance behavior of both the measurement-averaged first and 
second moment of any multipoint spin correlation function 
is identical to that correlation function at the (unmeasured) {$D$-dimensional classical} Ising critical point.
Finally, using the demonstrated existence of the {\it higher} Nishimori critical point {in all spatial dimensions $D>1$,
we can also understand the structure of  the 
Bayesian inference/learning phase diagram
{for}
the 
{general}
$D$-dimensional ($D\geq2$)  {classical} Ising critical point {or  that for the corresponding conformal quantum critical RK wavefunction~\cite{ArdonneFedleyFradkin}} under bond-energy measurements.
{Ref.~\cite{NahumJacobsen} discusses two 
{logical possibilities {(see Fig.~$4$ of Ref.~\cite{NahumJacobsen})} for the latter}
phase diagram, but their study could not rule out one in the favor of the other.
Using the existence of the {\it higher} Nishimori critical point in all dimensions $D >1$, we can exclude one of the possible phase diagrams discussed in Ref.~\cite{NahumJacobsen}, and 
{using the fact that the {\it higher} Nishimori critical point occurs at decreasing measurement strengths with increasing dimension $D$,}
we arrive at the phase diagram shown in our Fig.~\ref{fig:IsingLearningPDGeneral}.\\[-3mm]

Taking a broader perspective, we expect that --
analogous to our discussion of the classical \textit{Ising} model under Bayesian measurements of 
bond-energies in two or arbitrary $D>1$ spatial dimensions --
{{\it higher} Nishimori critical points}
should appear generally 
{in the Bayesian inference/learning
phase diagrams 
for measurements of local energy densities on}
{\textit{other}} 
{classical} statistical {mechanics} models
{at}
critical points 
in 
{two dimensions and also in 
arbitrary $D>1$
dimensions.}
As a point of comparison, it is known that, {\it e.g.},
the phase diagram of the 
{$2D$}
random-bond Potts model 
{governed}
by a $R\rightarrow0$ replica theory {also} 
has an ordinary Nishimori critical point, which 
admits a representation as
a
$R\rightarrow1$ replica theory with
gauge-invariance and enlarged replica {permutation} symmetry~\cite{NishimoriStephan,OzekiNishimori,JacobsenPicco,HoneckerJacobsenPiccoPujol}.
Analogously,
we expect 
that
{certain}
{Bayesian inference/}learning phase diagrams for classical stat-mech models
under 
measurements of 
local energy-densities,
which are governed by $R\rightarrow1$ replica theories, will also 
{likely}
have a 
{{\it higher} Nishimori critical point, whose}
long-distance physics 
is governed by
a gauge-invariant and replica permutation-symmetric $R\rightarrow2$ replica theory.
Such {\it higher} Nishimori critical points, which are
unstable, 
{finite-measurement-strength}
critical 
points,
can be obtained 
by {tuning}
the unmeasured system 
to its critical point and then by cranking up the measurement strength
to a 
suitable finite
value. 
{These}
{{\it higher} Nishimori critical points}
should possess a version of our Eqs.~\eqref{EqFirstMomentOfGeneralObservableAtTricPoint} and~\eqref{EqSecondMomentOfGeneralObservableAtTricPoint}, which identify the long-distance behavior 
{of}
the measurement-averaged first and second moments of correlation functions
for suitable observables
to that of the corresponding 
correlation functions at the unmeasured critical point.
{For}
example, we expect the unstable fixed point 
$\Ucal$
that was discussed in Ref.~\cite{NahumJacobsen} for the two-dimensional
{\textit{classical critical $Q$-state Potts model}}
($2<Q\leq 4$)
under {Bayesian} 
bond-energy measurements
to be such a
{\it higher} Nishimori critical point.
It would be interesting to test this conjecture.\\

{\it Data availability}.-- 
The numerical data shown in the figures is available on Zenodo~\cite{zenodo_higher_nishimori}.\\

\begin{acknowledgments}
We are grateful to the Kavli Institute for Theoretical Physics (KITP), which is supported by the National Science Foundation by grant NSF PHY-2309135, and the KITP Programs ``Noise-robust Phases of Quantum Matter" and ``Learning the Fine Structure
of Quantum Dynamics in Programmable Quantum Matter," where part of this work was initiated.
The Cologne research group is supported, in part, by the Deutsche Forschungsgemeinschaft (DFG, German Research Foundation) under Germany’s Excellence Strategy—Cluster of Excellence Matter
and Light for Quantum Computing (ML4Q) EXC 2004/1 -- 390534769 as well as within the CRC network TR 183 (Project Grant
No. 277101999) as part of subproject B01.
GYZ acknowledge the support of NSFC-Young Scientists Fund (grant no.~12504181) and Start-up Fund of HKUST(GZ) (grant no.~G0101000221) and Guangdong provincial project (grant no.~2024QN11X201). 
Our numerical simulations were performed on the Otus cluster at PC2 in Paderborn and the RAMSES cluster at RRZK Cologne.
\end{acknowledgments}

\appendix

\section{Derivation of Replica Theory for Binary Measurements\label{AppDerivationOfReplicaTheory}}

{In this appendix, we will derive the replica theory expression in Eq.~\eqref{EqLatticeReplicaTheoryBetaV} for measurement-averaged moments of expectation values. As defined in Eq.~\eqref{EqDefOfMeasurementAveragedMoments},}
\begin{align}
&\overline{
\langle { {\cal O}}_1\rangle_{\vec m}
\langle { {\cal O}}_2\rangle_{\vec m}
...
\langle {{\cal O}}_N\rangle_{\vec m}
}=
\sum_{\vec{m}} P(\vec{m})\langle { {\cal O}}_1\rangle_{\vec m}
\langle { {\cal O}}_2\rangle_{\vec m}
...
\langle {{\cal O}}_N\rangle_{\vec m}\nonumber
\\
&=\frac{1}{\sum_{\vec{m}}Z[\vec{m}]}\sum_{\vec{m}} (Z[\vec{m}])^{1-N}\times \nonumber\\
&\quad\quad\quad\quad\qquad\qquad\times \big( \sum_{\{\sigma_i\}}\mathcal{O}_1\;e^{-\beta H[\{\sigma_i\}]+\tilde{\gamma}\sum_{\langle ij\rangle}m_{ij}\sigma_i\sigma_j}\big)\nonumber\\
&\quad\quad\quad\quad\qquad\;\;
\cdots\times\big( \sum_{\{\sigma_i\}}\mathcal{O}_N\;e^{-\beta H[\{\sigma_i\}]+\tilde{\gamma}\sum_{\langle ij\rangle}m_{ij}\sigma_i\sigma_j}\big),\label{EqMeasurementAveragedMomentsAnIntermediateStep}
\end{align}
{where the above equation follows from Eq.~\eqref{EqClassicalProbability} and~\eqref{EqClassicalExpValue}. Then}
\begin{align}
&\hspace{-0.5cm}\overline{
\langle { {\cal O}}_1\rangle_{\vec m}
\langle { {\cal O}}_2\rangle_{\vec m}
...
\langle {{\cal O}}_N\rangle_{\vec m}}\propto\nonumber\\
&\hspace{-0.2cm}\propto \sum_{\vec{m}} \big( \sum_{\{\sigma_i\}}e^{-\beta H[\{\sigma_i\}]+\tilde{\gamma}\sum_{\langle ij\rangle}m_{ij}\sigma_i\sigma_j}\big)^{1-N} \times\nonumber\\
&\quad\quad\quad\times\big( \sum_{\{\sigma_i\}}\mathcal{O}_1\;e^{-\beta H[\{\sigma_i\}]+\tilde{\gamma}\sum_{\langle ij\rangle}m_{ij}\sigma_i\sigma_j}\big)\times\cdots\nonumber\\
&\quad\quad\quad\quad
\cdots\times\big( \sum_{\{\sigma_i\}}\mathcal{O}_N\;e^{-\beta H[\{\sigma_i\}]+\tilde{\gamma}\sum_{\langle ij\rangle}m_{ij}\sigma_i\sigma_j}\big)\nonumber\\\nonumber
=
&\lim_{R\rightarrow1} \sum_{\vec{m}} \big( \sum_{\{\sigma_i\}}e^{-\beta H[\{\sigma_i\}]+\tilde{\gamma}\sum_{\langle ij\rangle}m_{ij}\sigma_i\sigma_j}\big)^{R-N} \times\nonumber\\
&\quad\quad\quad\times\big( \sum_{\{\sigma_i\}}\mathcal{O}_1\;e^{-\beta H[\{\sigma_i\}]+\tilde{\gamma}\sum_{\langle ij\rangle}m_{ij}\sigma_i\sigma_j}\big)\times\cdots\nonumber\\
&\quad\quad\quad\quad
\cdots\times\big( \sum_{\{\sigma_i\}}\mathcal{O}_N\;e^{-\beta H[\{\sigma_i\}]+\tilde{\gamma}\sum_{\langle ij\rangle}m_{ij}\sigma_i\sigma_j}\big)\nonumber\\\nonumber
=
&\lim_{R\rightarrow1} \sum_{\vec{m}} \Big(\sum_{\{\sigma_i^{(a)}\}}\mathcal{O}_1^{(1)}\mathcal{O}_2^{(2)}\cdots \mathcal{O}_N^{(N)}\times\\&\quad\quad\quad
\times e^{-\beta \sum_{a=1}^{R} H[\{\sigma_i^{(a)}\}]+\tilde{\gamma}\sum_{\langle ij\rangle}m_{ij}\sum_{a=1}^{R}\sigma_i^{(a)}\sigma_j^{(a)}}\Big)\nonumber\\\label{EqPenultimateStepForMeasurementAveragedExpectationValues}
\end{align}
where in the last step we have used the replica trick, 
which will allow us to
average over the measurement outcomes {$\vec{m}$}. 
In particular, note that
\begin{align}
    &\sum_{m_{ij}}e^{\tilde{\gamma}m_{ij}\sum_{a=1}^{R}\sigma_i^{(a)}\sigma_j^{(a)}}=2\exp{\big\{\ln \cosh(\tilde{\gamma}\sum_{a=1}^{R}\sigma_i^{(a)}\sigma_j^{(a)})\big\}}\label{EqFirstStepAveragingOverOutcomes}\\&=2\exp\Big\{\frac{\tilde{\gamma}^2\big(\sum_{a=1}^{R}\sigma_i^{(a)}\sigma_j^{(a)}\big)^2}{2}+\mathcal{O}(\tilde{\gamma}^4)\Big\}\nonumber\\
    &=2e^{R\tilde{\gamma}^2/2}\exp\Big\{\frac{\tilde{\gamma}^2}{2}\sum_{\substack{a,b=1\\ a\neq b}}^{R} \sigma_i^{(a)}\sigma_j^{(a)}\sigma_i^{(b)}\sigma_j^{(b)}+\mathcal{O}(\tilde{\gamma}^4)\Big\}\label{EqAverageOverMeasuremntsHasBeenPerformed}
\end{align}
{Then using Eq.~\eqref{EqPenultimateStepForMeasurementAveragedExpectationValues} and Eq.~\eqref{EqAverageOverMeasuremntsHasBeenPerformed}, we obtain the replica theory expression in Eq.~\eqref{EqLatticeReplicaTheoryBetaV} for measurement-averaged moments of expectation values.}

\section{Derivation of the Identity for Moments of the Spin Correlation Function
\label{AppDerivationOfIdentityForMoments}}

In this section, we discuss the derivation of the equality in Eq.~\eqref{EqNishimoriEqualitiesReplicatedVersion}, which is just an arbitrary $R$ replica number  version of the famous identity due to Nishimori~\cite{Nishimori1981}. 
This derivation of Nishimori's identity obtained from studying the replica 
theory was 
{demonstrated}
in Ref.~\cite{LeDoussalHarrisI,GRL2001}, {where they ultimately used the $R$-replica theory identity in Eq.~\eqref{EqNishimoriEqualitiesReplicatedVersion}  for the theory in Eq.~\eqref{EqLatticeReplicaTheory} on the line $\beta=\Delta$ in the $R\rightarrow0$ replica limit, which is relevant for the ordinary Nishimori line in the RBIM. In this work (in Sec.~\ref{SubSecExactResultsA}), we use the same $R$-replica theory identity in Eq.~\eqref{EqNishimoriEqualitiesReplicatedVersion} for the theory in Eq.~\eqref{EqLatticeReplicaTheory} on the line $\beta=\Delta$  \textit{but} in the $R\rightarrow1$ replica limit, which is relevant for the {\it higher} Nishimori critical point in the learning phase diagram.}

Let us consider the following
correlation function for the {$R$-replica theory} in Eq.~\eqref{EqLatticeReplicaTheory} on the 
$\beta=\Delta$ {line}, 
\begin{align}
    \langle (\sigma_{i_1}^{(a_1)}\cdots\sigma_{i_k}^{(a_1)})\cdots(\sigma_{i_1}^{(a_{2q-1})}\cdots\sigma_{i_k}^{(a_{2q-1})})\rangle_{R}=\nonumber\\
    \frac{1}{Z_R}\sum_{\{\sigma_{i}^{a}\}_{a=1}^{R}}(\sigma_{i_1}^{(a_1)}\cdots\sigma_{i_k}^{(a_1)})\cdots(\sigma_{i_1}^{(a_{2q-1})}\cdots\sigma_{i_k}^{(a_{2q-1})})\times\nonumber\\
    \times\exp{\big\{\frac{\beta}{2} \sum_{\langle
ij \rangle}\Big(\sum_{a=1}^{R} \sigma_i^{(a)}\sigma_j^{(a)}+1\Big)^2\big\}},\label{EqAppCorrelationFunctionFirstStep}
\end{align}
{where the odd number $(2q-1)$ of replica indices
$1 \leq a_1, a_2, ..., a_{2q-1} \leq R$
are all pairwise unequal, and}
where $Z_{R}$ is the partition function on the 
$\beta=\Delta$ {line} given in Eq.~\eqref{EqPartionFunctionWithoutExtraSpinOnNishimoriStarLine}. Note that until this point we have not introduced the extra spin into the above replica theory to write it as a theory {of $R+1$ replicas} with gauge {invariance}
and 
{enlarged}
{replica}
permutation symmetry. 
We will
do that now. Let us first change the sum over dummy spin variables $\sigma_{i}^{(a)}$ to $\tilde{\sigma}_{i}^{(a)}$ and analogous to Eq.~\eqref{EqVariableChange} define the following subsequent variable change 
\begin{equation}
   \tilde{\sigma}_{i}^{(a)}= \sigma_{i}^{(a)}s_{i}.
\end{equation}
Then analogous to Eq.~\eqref{EqTransformedPartFunExtraSpin} we obtain
\begin{align}
    &\langle (\sigma_{i_1}^{(a_1)}\cdots\sigma_{i_k}^{(a_1)})\cdots(\sigma_{i_1}^{(a_{2q-1})}\cdots\sigma_{i_k}^{(a_{2q-1})})\rangle_{R}=\nonumber\\
    &=\frac{1}{Z_R}\sum_{\{\sigma_{i}^{a}\}_{a=1}^{R}}(\sigma_{i_1}^{(a_1)}\cdots\sigma_{i_k}^{(a_1)})\cdots(\sigma_{i_1}^{(a_{2q-1})}\cdots\sigma_{i_k}^{(a_{2q-1})})\times\nonumber\\
&\qquad\times(s_{i_1}s_{i_2}\dots s_{i_k})\exp{\big\{\frac{\beta}{2} \sum_{\langle
ij \rangle}\Big(\sum_{a=1}^{R} \sigma_i^{(a)}\sigma_j^{(a)}+s_is_j\Big)^2\big\}}.
\end{align}
{(The above equality follows because there is an odd number
$(2q-1)$ of pairwise unequal replica indices $a_1, a_2, ...,
a_{2q-1}$, and $s_i=\pm 1$.)}
Then defining $s_{i}$ as the `extra' $(R+1)^{\text{th}}$ spin $\sigma_{i}^{(R+1)}$ as in Eq.~\eqref{EqExtraSpinDef} and summing over all possible configurations of 
{$\{\sigma_i^{(R+1)}\}$,}
we obtain
\begin{align}
    &\langle (\sigma_{i_1}^{(a_1)}\cdots\sigma_{i_k}^{(a_1)})\cdots(\sigma_{i_1}^{(a_{2q-1})}\cdots\sigma_{i_k}^{(a_{2q-1})})\rangle_{R}=\nonumber\\
    &=\frac{1/2^{N_{S}}}{Z_R}\sum_{\{\sigma_{i}^{a}\}_{a=1}^{R+1}}(\sigma_{i_1}^{(a_1)}\cdots\sigma_{i_k}^{(a_1)})\cdots(\sigma_{i_1}^{(a_{2q-1})}\cdots\sigma_{i_k}^{(a_{2q-1})})\times\nonumber\\
&\qquad\qquad\qquad\times(\sigma^{(R+1)}_{i_1}\sigma^{(R+1)}_{i_2}\dots \sigma^{(R+1)}_{i_k})\times\nonumber\nonumber\\
&\hspace{-0.4cm}\qquad\qquad\qquad\times\exp{\big\{\frac{\beta}{2} \sum_{\langle
ij \rangle}\Big(\sum_{a=1}^{R} \sigma_i^{(a)}\sigma_j^{(a)}+\sigma^{(R+1)}_i\sigma^{(R+1)}_j\Big)^2\big\}}\\
&=\frac{1/2^{N_{S}}}{Z_R}\sum_{\{\sigma_{i}^{a}\}_{a=1}^{R+1}}(\sigma_{i_1}^{(a_1)}\cdots\sigma_{i_k}^{(a_1)})\cdots(\sigma_{i_1}^{(a_{2q-1})}\cdots\sigma_{i_k}^{(a_{2q-1})})\times\nonumber\\
&\qquad\qquad\qquad\qquad\times(\sigma^{(R+1)}_{i_1}\sigma^{(R+1)}_{i_2}\dots \sigma^{(R+1)}_{i_k})\times\nonumber\\
&\qquad\qquad\qquad\qquad\times\exp{\big\{\frac{\beta}{2} \sum_{\langle
ij \rangle}\Big(\sum_{a=1}^{R+1} \sigma_i^{(a)}\sigma_j^{(a)}\Big)^2\big\}}
\end{align}
Then using the 
{enlarged}
permutation symmetry
{$S_{R+1}$}
of
{the $(R+1)$} replica copies, we can exchange
the 
$(R+1)^{\text{th}}$ copy with
{a $({2q})^{\rm th}$ replica copy whose replica index $a_{2q}$ 
(with $1 \leq a_{2q} \leq R$) is unequal to any of the previous $(2q-1)$ replica indices, i.e.\ {it} satisfies}
$a_{2q}\neq a_1,a_2,\cdots,a_{2q-1}$, {and} we obtain
\begin{align}
&\langle (\sigma_{i_1}^{(a_1)}\cdots\sigma_{i_k}^{(a_1)})\cdots(\sigma_{i_1}^{(a_{2q-1})}\cdots\sigma_{i_k}^{(a_{2q-1})})\rangle_{R}=\nonumber\\
&=\frac{1/2^{N_{S}}}{Z_R}\sum_{\{\sigma_{i}^{a}\}_{a=1}^{R+1}}(\sigma_{i_1}^{(a_1)}\cdots\sigma_{i_k}^{(a_1)})\cdots(\sigma_{i_1}^{(a_{2q-1})}\cdots\sigma_{i_k}^{(a_{2q-1})})\times\nonumber\\
&\qquad\qquad\qquad\qquad\times(\sigma^{(a_{2q})}_{i_1}\sigma^{(a_{2q})}_{i_2}\dots \sigma^{(a_{2q})}_{i_k})\times\nonumber\\
&\qquad\qquad\qquad\qquad\times\exp{\big\{\frac{\beta}{2} \sum_{\langle
ij \rangle}\Big(\sum_{a=1}^{R+1} \sigma_i^{(a)}\sigma_j^{(a)}\Big)^2\big\}}.\label{EqOddToEvenMomentsPenultimateStep}    
\end{align}
As noted already, the 
{Boltzmann}
weight in the 
{above}
partition function is gauge invariant under local gauge transformation{:} $\sigma_i^{(a)}\rightarrow-\sigma_{i}^{(a)}$ $\forall a\in \{1,\cdots,R+1\}$. In particular, we can fix the gauge by setting the $(R+1)^{\text{th}}$ spin $\sigma_i^{(R+1)}$ to $+1$ for all $i$. 
{The}
Boltzmann weight {then} does not depend on $\sigma_i^{(R+1)}$ for any $i$ and it becomes the same as in Eq.~\eqref{EqAppCorrelationFunctionFirstStep}. 
{This gauge fixing gets rid of the factor of $1/2^{N_S}$ appearing in the RHS of Eq.~\eqref{EqOddToEvenMomentsPenultimateStep}, and}
we obtain 
the result that on the 
$\beta=\Delta$  line,
\begin{align}
&\langle (\sigma_{i_1}^{(a_1)}\cdots\sigma_{i_k}^{(a_1)})\cdots(\sigma_{i_1}^{(a_{2q-1})}\cdots\sigma_{i_k}^{(a_{2q-1})})\rangle_{R}=\nonumber\\
&=\frac{1}{Z_R}\sum_{\{\sigma_{i}^{a}\}_{a=1}^{R}}(\sigma_{i_1}^{(a_1)}\cdots\sigma_{i_k}^{(a_1)})\cdots(\sigma_{i_1}^{(a_{2q-1})}\cdots\sigma_{i_k}^{(a_{2q-1})})\times\nonumber\\
&\qquad\qquad\qquad\qquad\times(\sigma^{(a_{2q})}_{i_1}\sigma^{(a_{2q})}_{i_2}\dots \sigma^{(a_{2q})}_{i_k})\times\nonumber\\
&\qquad\qquad\qquad\qquad\times\exp{\big\{\frac{\beta}{2} \sum_{\langle
ij \rangle}\Big(\sum_{a=1}^{R} \sigma_i^{(a)}\sigma_j^{(a)}+1\Big)^2\big\}}\nonumber\\
&=\langle (\sigma_{i_1}^{(a_1)}\cdot\cdot\sigma_{i_k}^{(a_1)})\cdots(\sigma_{i_1}^{(a_{2q-1})}\cdot\cdot\sigma_{i_k}^{(a_{2q-1})})(\sigma_{i_1}^{(a_{2q})}\cdot\cdot\sigma_{i_k}^{(a_{2q})})\rangle_{R}.    
\end{align}
Therefore, we have concluded the derivation of Eq.~\eqref{EqNishimoriEqualitiesReplicatedVersion}, which holds for any number of replicas $R$.
The $R\rightarrow0$ replica limit of the above identity corresponds to the 
{identity in our Eq.~\eqref{EqNishimoriEquality}}
obtained by Nishimori~\cite{Nishimori1981} for the bond-randomness averaged moments in the RBIM on the ordinary Nishimori line. 
On the other hand, we use the $R\rightarrow1$ replica limit version of the above identity for the {measurement-averaged moments at the {\it higher} Nishimori critical point, i.e.\ the} 
tricritical point in the {classical} Ising learning phase diagram.

\section{Gaussian Measurements
\label{AppGaussianMeasurements}}

In this Appendix, we will discuss a different
microscopic measurement protocol with continuous measurement outcomes $\mathfrak{m}_{ij}\in \mathbb{R}$ instead of the discrete measurement outcomes $m_{ij}\in \{+1,-1\}$ considered in Sec.~\ref{SecModel}.
We will refer to this measurement protocol as the `Gaussian' measurement protocol.
This protocol is analogous to various measurement protocols discussed in Ref.~\cite{NahumJacobsen}.
{For Gaussian measurements, using Eq.~\eqref{eq:PmsigmaGaussian} and Bayes' theorem, the 
{conditional}
probability
{of}
a 
spin configuration $\{\sigma_{i}\}$ 
{given}
the measurement outcomes $\{\mathfrak{m}_{ij}\}$ is 
[instead of Eq.~\eqref{EqProbabilitySpinGivenMeasBetaVersion}] given by
\begin{equation}
    P(\{\sigma_i\}|\{\mathfrak{m}_{ij}\})=\frac{e^{-\beta H[\{\sigma_i\}]-\frac{\Delta}{2}\sum_{\langle ij\rangle}(\mathfrak{m}_{ij}-\sigma_i\sigma_j)^2}}{Z[\{\mathfrak{m}_{ij}\}]},\label{EqProbabilitySpinGivenMeas}
\end{equation}
where the Ising Hamiltonian $H[\{\sigma_{i}\}]$ is given in Eq.~\eqref{Eq2DIsingHamiltonian} and $Z[\{\mathfrak{m}_{ij}\}]$ is given by
\begin{equation}
     Z[\{\mathfrak{m}_{ij}\}]=\sum_{\{\sigma_i\}}e^{-\beta H[\{\sigma_i\}]-\frac{\Delta}{2} \sum_{\langle ij\rangle}(\mathfrak{m}_{ij}-\sigma_i\sigma_j)^2}.\label{EqFixedMeasPartFunGaussian}
\end{equation}
{Analogous to the discussion in Sec.~\ref{SecModel}, the above setup with Bayesian Gaussian measurements of bond-energies is 
equivalent to the problem of Born-rule Gaussian measurements with the bond-energy operator on the RK wavefunction corresponding to the $2D$ classical Ising model in Eq.~\eqref{EqRKWF}. 
In contrast to binary measurements [Eq.~\eqref{EqKrausBinary}],  the Kraus operator for Gaussian measurements of the bond energy operator on the RK wavefunction is given by
\begin{equation}\label{EqKrausGauss}
    \hat{K}_{\vec{\mathfrak{m}}}
    =
    {e^{-\frac{\Delta}{4}\sum_{\langle ij\rangle}(m_{ij}-\hat{\sigma}^z_i\hat{\sigma}^z_j)^2}},\;\;\;
    {m_{ij}\in \mathbb{R},}
\end{equation}
where $\Delta$ quantifies the strength of the measurements and the Kraus operators satisfy the POVM~\cite{Note1} condition
\begin{equation}
    \int_{\vec{\mathfrak{m}}}\hat{K}_{\vec{\mathfrak{m}}}^{\dagger}\hat{K}_{\vec{\mathfrak{m}}}=\hat{1},
\end{equation}
where we have defined $\int_{\vec{\mathfrak{m}}}$  by
\begin{equation}
    \int_{\vec{\mathfrak{m}}}:=\prod_{<ij>} \int\frac{d\mathfrak{m}_{ij}}{\sqrt{2\pi/\Delta}}\label{EqIntegrationMeasureVecM}.
\end{equation}
See e.g. App.~C of Ref.~\cite{PatilLudwig20251} for
details on the general setup of
performing Born-rule Gaussian measurements on 
quantum RK wavefunction{s} and its equivalence to the
Bayesian inference problem for the corresponding classical stat-mech model.}

\subsection{Replica Theory}

Following the steps
{in App.~\ref{AppDerivationOfReplicaTheory},}
the measurement-averaged moments of expectation values for the case of Gaussian measurements are given by
\begin{align}
   &\overline{\left [
\langle { {\cal O}}_1\rangle_{\vec{\mathfrak{m}}}
\langle { {\cal O}}_2\rangle_{\vec{\mathfrak{m}}}
...
\langle {{\cal O}}_N\rangle_{\vec{\mathfrak{m}}}
\right]} \propto
\lim_{R\rightarrow1} \int_{\vec{\mathfrak{m}}} \Big(\sum_{\{\sigma_i^{(a)}\}}\mathcal{O}_1^{(1)}\mathcal{O}_2^{(2)}\cdots \mathcal{O}_N^{(N)}\nonumber\\&
\;\;\;\;\;\;\;\;\;\;\;\;\;\;\;\times e^{-\beta \sum_{a=1}^{R} H[\{\sigma_i^{(a)}\}]-\frac{\Delta}{2}\sum_{\langle ij\rangle}\sum_{a=1}^{R}(\mathfrak{m}_{ij}-\sigma_i^{(a)}\sigma_j^{(a)})^2}\Big).\nonumber\\
\end{align}
Then performing the integration over $\mathfrak{m}_{ij}$ and using Eq.~\eqref{Eq2DIsingHamiltonian} we obtain
\begin{align}
&\overline{\left [
\langle { {\cal O}}_1\rangle_{\vec{\mathfrak{m}}}
\langle { {\cal O}}_2\rangle_{\vec{\mathfrak{m}}}
...
\langle {{\cal O}}_N\rangle_{\vec{\mathfrak{m}}}
\right]} \propto
\lim_{R\rightarrow1} 
\sum_{\{\sigma_i^{(a)}\}}\mathcal{O}_1^{(1)}\mathcal{O}_2^{(2)}\cdots \mathcal{O}_N^{(N)}\nonumber\\&
\;\;\;\;\;\;\times \exp{\bigg\{\sum_{\langle ij\rangle} \big(\beta \sum_{a=1}^{R} \sigma_i^{(a)}\sigma_j^{(a)}+\frac{\Delta}{2R}\sum_{a\neq b}^{R}\sigma_i^{(a)}\sigma_j^{(a)}\sigma_i^{(b)}\sigma_j^{(b)}\big)\bigg\}}.\nonumber\\
\end{align}
So with Gaussian measurements we have to study the replica theory
\begin{align}
    -{\tilde{\mathcal{H}}}&=\beta \sum_{\langle
ij \rangle}\sum_{a=1}^{R} \sigma_i^{(a)}\sigma_j^{(a)}+\frac{\Delta}{2R}\sum_{\langle ij\rangle}\sum_{\substack{a,b=1\\ a\neq b}}^{R}\sigma_i^{(a)}\sigma_j^{(a)}\sigma_i^{(b)}\sigma_j^{(b)},\label{EqGassianReplicaTheory}\\
&=\beta \sum_{\langle
ij \rangle}\sum_{a=1}^{R} \sigma_i^{(a)}\sigma_j^{(a)}+{\frac{\Delta}{2}}\sum_{\langle ij\rangle}\sum_{\substack{a,b=1\\ a\neq b}}^{R}\sigma_i^{(a)}\sigma_j^{(a)}\sigma_i^{(b)}\sigma_j^{(b)}\nonumber\\
&\qquad\;\;\;\;\;\;\;\;\;\;\;\;+\frac{\Delta ({1-R})}
{2R}
\sum_{\langle ij\rangle}\sum_{\substack{a,b=1\\ a\neq b}}^{R}\sigma_i^{(a)}\sigma_j^{(a)}\sigma_i^{(b)}\sigma_j^{(b)}.
\end{align}
Clearly, the last term in the above equation vanishes in the $R\rightarrow1$ replica limit, and thus in the $R\rightarrow1$ replica limit, we obtain the replica theory in Eq.~\eqref{EqLatticeReplicaTheory}.
So in the case of Gaussian measurements, we see that the physics of the $2D$ 
{classical}
Ising learning phase diagram is 
{exactly}
captured by the $R\rightarrow1$ limit of the replica theory in Eq.~\eqref{EqLatticeReplicaTheory}.
Note that here we do not have to make the assumption of ignoring the $\mathcal{O}(\Delta^2)$ [$\Delta={\tilde{\gamma}}^2$] terms that was made in the case of binary measurements {of strength $\tilde{\gamma}$} in going from Eq.~\eqref{EqLatticeReplicaTheoryBetaV} to Eq.~\eqref{EqLatticeReplicaTheory}.
{That is, the replica theory in Eq.~\eqref{EqLatticeReplicaTheory} captures the physics of the $2D$ classical Ising model under Gaussian bond-energy measurements at all length scales, unlike the case of binary measurements where it is expected to be a valid description of the problem only at long-distances.}

\subsection{Simplification of Edwards-Anderson Correlation Function Without Replicas}

In this appendix, we will demonstrate {that} with
Gaussian measurements
various features of the {{\it higher} Nishimori line} $\beta=\Delta$ and the 
{{\it higher} Nishimori critical point Eq.~\eqref{EqLocationFixedPoint}, which is the tricritical point in the $2D$ Ising learning phase diagram,}
are readily seen without the use of replicas.
{Although, for brevity, we 
demonstrate this
for the $2D$ {classical} Ising model under Gaussian bond-energy measurements, it 
holds
for any $D$-dimensional {classical} Ising model ($D>1$) under bond-energy measurements.}
To illustrate this let us consider the Edwards-Anderson correlator $\overline{\langle \sigma_{i}\sigma_{j}\rangle_{\vec{\mathfrak{m}}}^2}$ with Gaussian measurements
\begin{equation}
    \overline{\langle \sigma_{i}\sigma_{j}\rangle_{\vec{\mathfrak{m}}}^2}=\int_{\vec{\mathfrak{m}}} P(\vec{\mathfrak{m}})\langle \sigma_{i}\sigma_{j}\rangle_{\vec{\mathfrak{m}}}
\langle \sigma_{i}\sigma_{j}\rangle_{\vec{\mathfrak{m}}}.
\end{equation}
Then analogous to Eq.~\eqref{EqMeasurementAveragedMomentsAnIntermediateStep} 
{with $N=2$ and $\mathcal{O}_{1}=\mathcal{O}_{2}=\sigma_{i}\sigma_{j}$,}
but now with Gaussian measurements, we obtain that
\begin{align}
&\overline{\langle \sigma_{i}\sigma_{j}\rangle_{\vec{\mathfrak{m}}}^2}=
\nonumber\\&=\frac{1}{\int_{\vec{\mathfrak{m}}}Z[\vec{\mathfrak{m}}]}\int_{\vec{\mathfrak{m}}} \frac{\big(\sum_{\{\sigma_i\}}\sigma_{i}\sigma_{j}\;e^{\sum_{\langle ij\rangle} (\beta \sigma_i\sigma_j-\frac{\Delta}{2}(\mathfrak{m}_{ij}-\sigma_i\sigma_j)^2)}\big)^2}{\sum_{\{\sigma_i\}}e^{\sum_{\langle ij\rangle} (\beta \sigma_i\sigma_j-\frac{\Delta}{2}\sum_{\langle ij\rangle}(\mathfrak{m}_{ij}-\sigma_i\sigma_j)^2)}}\nonumber\\
&=\frac{e^{-\frac{\Delta}{2} N_b}}{Z}\int_{\vec{\mathfrak{m}}}e^{-\frac{\Delta}{2} \sum_{\langle ij\rangle}\mathfrak{m}_{ij}^2}\times\nonumber\\&\;\;\;\;
\qquad\qquad\qquad\times\frac{\big(\sum_{\{\sigma_i\}}\sigma_{i}\sigma_{j}\;e^{\sum_{\langle ij\rangle} (\beta \sigma_i\sigma_j+\Delta \mathfrak{m}_{ij}\sigma_i\sigma_j)}\big)^2}{\sum_{\{\sigma_i\}}e^{\sum_{\langle ij\rangle} (\beta \sigma_i\sigma_j+\Delta \mathfrak{m}_{ij}\sigma_i\sigma_j)}},\nonumber\\\label{EqIntermediateStepInSImplificationOfEAOPinGaussianMeaus}
\end{align}
where $N_b$ is the number of bonds in the square lattice and $Z$ is the partition function of the unmeasured Ising model{,} which is equal to $\int_{\vec{\mathfrak{m}}}Z[\vec{\mathfrak{m}}]$ after performing the Gaussian integral over $\vec{\mathfrak{m}}$ in Eq.~\eqref{EqFixedMeasPartFunGaussian} by using Eq.~\eqref{EqIntegrationMeasureVecM}.
We will now make the following variable change from $\mathfrak{m}_{ij}$ to $\mathfrak{n}_{ij}$ in Eq.~\eqref{EqIntermediateStepInSImplificationOfEAOPinGaussianMeaus} 
\begin{equation}\label{EqVariableChangeToGothicN}
    \mathfrak{n}_{ij}=\beta+\Delta \mathfrak{m}_{ij},
\end{equation}
and we obtain~\footnote{{Note that the expression in Eq.~\eqref{EqEACorrelationFunctionGaussianMeasurementsAfterChangeOfVariables} \textit{almost} resembles the definition of the EA correlation function for the RBIM with Gaussian bond disorder, where $\beta$ is the mean
bond strength
and $\Delta$ quantifies the variance in bond strengths.
The only difference is that, unlike the EA correlation function for RBIM, 
we do not have a square of $\sum_{\{\sigma_i\}}e^{\sum_{\langle ij\rangle} \mathfrak{n}_{ij} \sigma_i\sigma_j}$ in the denominator on the RHS of Eq.~\eqref{EqEACorrelationFunctionGaussianMeasurementsAfterChangeOfVariables}. This 
{distinction}
is, of course, crucial for the difference 
in corresponding replica limits $R\rightarrow0$ and $R\rightarrow1$ for the RBIM and 
{the}
$2D$ Ising measurement/learning problem, respectively.}}
\begin{align}
&\overline{\langle \sigma_{i}\sigma_j\rangle_{\vec{\mathfrak{m}}}^2}\nonumber\\
    &=\frac{e^{-\frac{\Delta}{2} N_b}}{Z}\int_{\vec{\mathfrak{n}}}e^{-\sum_{\langle ij\rangle}\frac{(\mathfrak{n}_{ij}-\beta)^2}{2\Delta}}\frac{\big(\sum_{\{\sigma_i\}}\sigma_{i}\sigma_{j}\;e^{\sum_{\langle ij\rangle} \mathfrak{n}_{ij} \sigma_i\sigma_j}\big)^2}{\sum_{\{\sigma_i\}}e^{\sum_{\langle ij\rangle} \mathfrak{n}_{ij} \sigma_i\sigma_j}},\label{EqEACorrelationFunctionGaussianMeasurementsAfterChangeOfVariables}
\end{align}
where the integral $\vec{\mathfrak{n}}$, following the measure of integration $\int_{\vec{\mathfrak{m}}}$  {in} Eq.~\eqref{EqIntegrationMeasureVecM}, is defined as
\begin{equation}
    \int_{\vec{\mathfrak{n}}}:=\prod_{<ij>}\int\frac{d\mathfrak{n}_{ij}}{\sqrt{2\pi \Delta}}.
\end{equation}
Let us now consider the theory on the  {{\it higher} Nishimori line} $\beta=\Delta$ and
analogous to Nishimori~\cite{Nishimori_1980}, we will consider making the following transformations
\begin{equation}
\label{EqGaugeTransform}    \sigma_i\rightarrow\sigma_is_i,\;\;\,\mathfrak{n}_{ij}\rightarrow \mathfrak{n}_{ij}s_is_j,\;\;\;s_i=\pm 1,
\end{equation}
{where we choose a fixed but arbitrary value $s_i=\pm 1$
at each site $i$ on the square lattice. Then}
\begin{align}
&\overline{\langle \sigma_{i}\sigma_j\rangle_{\vec{\mathfrak{m}}}^2}\nonumber=\\
    &\frac{e^{-\frac{\Delta}{2} N_b}}{Z}\int_{\vec{\mathfrak{n}}}e^{-\sum_{\langle ij\rangle}\frac{(\mathfrak{n}_{ij}s_is_j-\Delta)^2}{2\Delta}}\frac{\big(\sum_{\{\sigma_i\}}\sigma_{i}\sigma_{j}\;e^{\sum_{\langle ij\rangle} \mathfrak{n}_{ij} \sigma_i\sigma_j}\big)^2}{\sum_{\{\sigma_i\}}e^{\sum_{\langle ij\rangle} \mathfrak{n}_{ij} \sigma_i\sigma_j}}\nonumber\\
    &=\frac{e^{-{\Delta} N_{b}}}{Z}\int_{\vec{\mathfrak{n}}}e^{-\sum_{\langle ij\rangle}\frac{\mathfrak{n}_{ij}^2}{2\Delta}} e^{\sum_{\langle ij\rangle}\mathfrak{n}_{ij}s_is_j}\times\nonumber
    \\&\qquad\qquad\qquad\qquad\qquad\times\frac{\big(\sum_{\{\sigma_i\}}\sigma_{i}\sigma_{j}\;e^{\sum_{\langle ij\rangle} \mathfrak{n}_{ij} \sigma_i\sigma_j}\big)^2}{\sum_{\{\sigma_i\}}e^{\sum_{\langle ij\rangle} \mathfrak{n}_{ij} \sigma_i\sigma_j}}\nonumber\\
    &=\frac{e^{-\Delta N_{b}}}{Z}\int_{\vec{\mathfrak{n}}}e^{-\sum_{\langle ij\rangle}\frac{\mathfrak{n}_{ij}^2}{2\Delta}}\frac{ \big(\sum_{\{s_{i}\}}e^{\sum_{\langle ij\rangle}\mathfrak{n}_{ij}s_is_j}\big)}{2^{N_s}}\times\nonumber
    \\&\qquad\qquad\qquad\qquad\qquad\times\frac{\big(\sum_{\{\sigma_i\}}\sigma_{i}\sigma_{j}\;e^{\sum_{\langle ij\rangle} \mathfrak{n}_{ij} \sigma_i\sigma_j}\big)^2}{\sum_{\{\sigma_i\}}e^{\sum_{\langle ij\rangle} \mathfrak{n}_{ij} \sigma_i\sigma_j}},\label{EqEASimplificationForGaussianMeasIntermediateStepII}
\end{align}
where in the last
equation above we have summed over all possible $2^{N_{s}}$ ($N_s$ is the number of sites) choices for `gauge' spins $\{s_{i}\}$ in Eq.~\eqref{EqGaugeTransform}.
Now we see that the factor of $\sum_{\{s_i\}}\exp{\{{\sum_{\langle ij\rangle} \mathfrak{n}_{ij} s_is_j}\}}$ in the numerator and the factor of $\sum_{\{\sigma_i\}}\exp{\{{\sum_{\langle ij\rangle} \mathfrak{n}_{ij} \sigma_i\sigma_j}\}}$ in the denominator
{of}
Eq.~\eqref{EqEASimplificationForGaussianMeasIntermediateStepII} cancel exactly.
Therefore{,} we obtain
\begin{align}
&\overline{\langle \sigma_{i}\sigma_j\rangle_{\vec{\mathfrak{m}}}^2}
=\nonumber\\
&=\frac{e^{-\Delta N_b}}{Z} \int_{\vec{\mathfrak{n}}}e^{-\sum_{\langle ij\rangle}\frac{\mathfrak{n}_{ij}^2}{2\Delta}}\frac{\big(\sum_{\{\sigma_i\}}\sigma_{i}\sigma_{j}\;e^{\sum_{\langle ij\rangle} \mathfrak{n}_{ij} \sigma_i\sigma_j}\big)^2}{2^{N_{s}}}\nonumber\\
&=\frac{e^{-\Delta N_b}}{Z\cdot 2^{N_s}} \int_{\vec{\mathfrak{n}}}e^{-\sum_{\langle ij\rangle}\frac{\mathfrak{n}_{ij}^2}{2\Delta}}\big(\sum_{\{\sigma_i^{(1)}\}}\sigma^{(1)}_{i}\sigma^{(1)}_{j}\;e^{\sum_{\langle ij\rangle} \mathfrak{n}_{ij} \sigma^{(1)}_i\sigma^{(1)}_j}\big)\times\nonumber\\
&\qquad\qquad\qquad\qquad\qquad\times\big(\sum_{\{\sigma_i^{(2)}\}}\sigma^{(2)}_{i}\sigma^{(2)}_{j}\;e^{\sum_{\langle ij\rangle} \mathfrak{n}_{ij} \sigma^{(2)}_i\sigma^{(2)}_j}\big)\nonumber\\
&=\frac{e^{-\Delta N_b}}{Z\cdot 2^{N_s}} \sum_{\{\sigma_i^{(a)}\}_{a=1}^2}\sigma^{(1)}_{i}\sigma^{(1)}_{j}\sigma^{(2)}_{i}\sigma^{(2)}_{j}\;e^{\frac{\Delta}{2} \sum_{\langle ij\rangle} (\sigma^{(1)}_i\sigma^{(1)}_j+\sigma^{(2)}_i\sigma^{(2)}_j)^2}\nonumber\\
&=\frac{1}{Z\cdot 2^{N_s}} \sum_{\{\sigma_i^{(a)}\}_{a=1}^2}\sigma^{(1)}_{i}\sigma^{(1)}_{j}\sigma^{(2)}_{i}\sigma^{(2)}_{j}\;e^{\Delta \sum_{\langle ij\rangle} \sigma^{(1)}_i\sigma^{(1)}_j\sigma^{(2)}_i\sigma^{(2)}_j}
\end{align}
Note that the Boltzmann weight in the above equation is invariant under the local transformation $\sigma_i^{(1)}\rightarrow-\sigma_i^{(1)}$ and $\sigma_i^{(2)}\rightarrow-\sigma_i^{(2)}$ at any given site $i$. This 
`gauge'
invariance can be fixed by 
{setting}
$\sigma_{i}^{(2)}=+1$ 
{at}
all sites {$i$} and this
`gauge'
fixing gets rid of the factor of $\frac{1}{2^{N_s}}$ in the above equation, and we obtain
\begin{align}
    \overline{\langle \sigma_{i}\sigma_j\rangle_{\vec{\mathfrak{m}}}^2} &=\frac{1}{Z} \sum_{\{\sigma_i^{(1)}\}}\sigma^{(1)}_{i}\sigma^{(1)}_{j}\;e^{\Delta \sum_{\langle ij\rangle} \sigma^{(1)}_i\sigma^{(1)}_j}.\nonumber
\end{align}
Thus, on the 
{{\it higher} Nishimori line} 
$\beta=\Delta$, the Edwards-Anderson correlator, i.e.\ the measurement-averaged second moment of the spin-spin correlation function is given by
\begin{equation}\label{EqEqSecondMomentIsUnaffectedOnTheNishimoriStarLine}
     \overline{\langle \sigma_{i}\sigma_j\rangle_{\vec{\mathfrak{m}}}^2}=\langle \sigma_{i}\sigma_j\rangle,
\end{equation}
where the correlation function on the RHS, given by $\langle \sigma_{i}\sigma_j\rangle${,}  is evaluated in the unmeasured $2D$ Ising model at inverse temperature $\beta=\Delta$.
As we already know, 
the measurement-averaged first moment of the spin-spin correlation function{, for any measurement strength $\Delta$, is}
equal to the spin-spin correlation function in the unmeasured $2D$ Ising model [see {e.g.} Eq.~\eqref{EqFirstMomentIsEqualToTheUnmeasured}], i.e.
\begin{equation}
    \label{EqFirstMomentIsUnaffected}\overline{\langle\sigma_i\sigma_j\rangle_{\vec{\mathfrak{m}}}} =\langle\sigma_i\sigma_j\rangle.
\end{equation}
Therefore,
{from Eq.~\eqref{EqEqSecondMomentIsUnaffectedOnTheNishimoriStarLine} and~\eqref{EqFirstMomentIsUnaffected}, on the {\it higher} Nishimori line $\beta=\Delta$}
\begin{equation}\label{EqFinal}
    \overline{\langle\sigma_i\sigma_j\rangle_{\vec{\mathfrak{m}}}^2} =\overline{\langle\sigma_i\sigma_j\rangle_{\vec{\mathfrak{m}}}} =\langle\sigma_i\sigma_j\rangle{.}
\end{equation}
{In particular,
at $\beta=\Delta=\beta_c$, where $\beta_c=\ln(\sqrt{1+\sqrt{2}})$ is the critical point of the (unmeasured) $2D$ {classical} Ising model,
all of the correlation functions in the above equation will be power law decaying, and this determines the location of the {\it higher} Nishimori critical point, just as in Eq.~\eqref{EqLocationFixedPoint}.
{This location of the {\it higher} Nishimori critical point exactly gives
the location of the tricritical point in the 
learning phase diagram {of the $2D$ classical Ising model} for the case of \textit{Gaussian} bond-energy measurements. 
This is to be contrasted with the case of binary measurements, where unlike the case of Gaussian measurements, we expect the replica theory in Eq.~\eqref{EqLatticeReplicaTheory} to be only a valid long-distance description of the problem, and hence we could not determine the exact location of the tricritical point in the learning phase diagram for the case of binary measurements.
However, we expect the universal long-distance behavior at the tricritical point to be the same, i.e.\ that of the {\it higher} Nishimori critical point, regardless of whether we use Gaussian or binary measurements.}
In particular, from Eq.~\eqref{EqFinal} and using the known power-law exponent for the spin-spin correlation function at the (unmeasured) $2D$ 
{classical}
Ising critical point, we obtain that at the tricritical point $\beta=\Delta=\beta_c$}
\begin{equation}\label{EqFinalFinal}
    \overline{\langle\sigma_i\sigma_j\rangle_{\vec{\mathfrak{m}}}^2} =\overline{\langle\sigma_i\sigma_j\rangle_{\vec{\mathfrak{m}}}} =\langle\sigma_i\sigma_j\rangle\sim\frac{1}{|i-j|^{1/4}}{.}
\end{equation}
{Finally, for the case of binary measurements, as discussed in Sec.~\ref{SecEmergentGaugeAndEnlargedPermutationSymmetry}, we expect the replica theory in Eq.~\eqref{EqLatticeReplicaTheory} to be a valid description of the measurement problem only at large distances. 
This is in contrast to Gaussian measurements discussed in this appendix, where the replica theory in Eq.~\eqref{EqLatticeReplicaTheory} is an exact description valid at all length scales. Consequently, for the tricritical point in the case of binary measurements, the first equality in Eq.~\eqref{EqFinalFinal} is spoiled at short distances. 
However, even in the case of binary measurements,
at long distances,  we expect 
both the measurement-averaged first and 
second moment of the spin-spin correlation function 
at the tricritical point
{to} exhibit 
power-law 
decay 
{with the exact same exponent as that for Gaussian measurements in Eq.~\eqref{EqFinalFinal}.}}
\vspace{1em}
\subsection{Dual Spin Correlation Function at the Tricritical Point
\label{AppDualSpinCorrFun}}

In this section of the appendix, we will follow both Kadanoff and Ceva~\cite{KadanoffCeva} in the (unmeasured) $2D$ Ising 
model and 
Read and Ludwig~\cite{ReadLudwig2001}
in the $2D$ RBIM to define the 
{\textit{dual spin correlation function}}
in a given measurement trajectory $\{\mathfrak{m}_{ij}\}$
and also its measurement-averaged moments.
Following Ref.~\cite{KadanoffCeva} and~\cite{ReadLudwig2001},
in a given measurement trajectory $\vec{\mathfrak{m}}=\{\mathfrak{m}_{ij}\}$, a natural way to define the dual spin correlation function 
{$\langle \mu_{\tilde{x}}\mu_{\tilde{y}} \rangle_{\vec{m}}$ for plaquettes at locations $\tilde{x}$ and $\tilde{y}$ is the following}:
\begin{widetext}
\begin{equation}
    \langle \mu_{\tilde{x}}\mu_{\tilde{y}} \rangle_{\vec{m}}=\frac{\sum_{\{\sigma_i\}}e^{\sum_{\langle ij\rangle\notin\mathcal{L}} [\beta \sigma_i\sigma_j-\frac{\Delta}{2}(\mathfrak{m}_{ij}-\sigma_i\sigma_j)^2]+\sum_{\langle ij\rangle\in\mathcal{L}} [-\beta \sigma_i\sigma_j-\frac{\Delta}{2}(-\mathfrak{m}_{ij}-\sigma_i\sigma_j)^2]}}{\sum_{\{\sigma_i\}}e^{\sum_{\langle ij\rangle} (\beta \sigma_i\sigma_j-\frac{\Delta}{2}(\mathfrak{m}_{ij}-\sigma_i\sigma_j)^2)}},\label{EqDualCorrFunInAGivenMeasTrajectory}
\end{equation}
{where we have considered the ratio of the partition function obtained by flipping the sign of \textit{both} the inverse temperature $\beta$ \textit{and} 
the measurement outcomes $\{\mathfrak{m}_{ij}\}$ on a string $\mathcal{L}$ of bonds connecting two plaquettes $\tilde{x}$ and $\tilde{y}$ in the partition function $Z[\{\mathfrak{m}_{ij}\}]$ in Eq.~\eqref{EqFixedMeasPartFunGaussian} {with that of the
unchanged partition function $Z[\{\mathfrak{m}_{ij}\}]$}.}
Analogous to Ref.~\cite{KadanoffCeva,ReadLudwig2001}, it is easy to check that on the infinite plane the above definition of the dual spin correlation function is independent of the actual string $\mathcal{L}$ of bonds
where the inverse temperature and measurement outcomes
are flipped in their sign and it only depends on the locations $\tilde{x}$ and $\tilde{y}$ of the plaquettes where the string ends.
We can now consider the measurement-averaged $q^{\text{th}}$ moments of this dual spin correlation function, 
{which}
are given by
\begin{equation}
    \overline{\langle \mu_{\tilde{x}}\mu_{\tilde{y}}\rangle_{\vec{\mathfrak{m}}}^q}=\int_{\vec{\mathfrak{m}}} P(\vec{\mathfrak{m}})\langle \mu_{\tilde{x}}\mu_{\tilde{y}}\rangle_{\vec{\mathfrak{m}}}^q.
\end{equation}
Then {using Eq.~\eqref{EqDualCorrFunInAGivenMeasTrajectory},} 
\begin{align}
    \overline{\langle \mu_{\tilde{x}}\mu_{\tilde{y}}\rangle_{\vec{\mathfrak{m}}}^q}=&
\frac{1}{Z}\int_{\vec{\mathfrak{m}}} \frac{\Big(\sum_{\{\sigma_i\}}e^{\sum_{\langle ij\rangle\notin\mathcal{L}} [\beta \sigma_i\sigma_j-\frac{\Delta}{2}(\mathfrak{m}_{ij}-\sigma_i\sigma_j)^2]+\sum_{\langle ij\rangle\in\mathcal{L}} [-\beta \sigma_i\sigma_j-\frac{\Delta}{2}(-\mathfrak{m}_{ij}-\sigma_i\sigma_j)^2]}\Big)^{q}}{\Big(\sum_{\{\sigma_i\}}e^{\sum_{\langle ij\rangle} (\beta \sigma_i\sigma_j-\frac{\Delta}{2}(\mathfrak{m}_{ij}-\sigma_i\sigma_j)^2)}\Big)^{q-1}}\label{EqDefMomentsOFDualSpinCorrFun}\\
=&\frac{e^{-\frac{\Delta}{2} N_b}}{Z}\int_{\vec{\mathfrak{m}}} e^{-\frac{\Delta}{2} \sum_{\langle ij\rangle}m_{ij}^2} \frac{\Big(\sum_{\{\sigma_i\}}e^{\sum_{\langle ij\rangle\notin\mathcal{L}} (\beta +\Delta\mathfrak{m}_{ij})\sigma_i\sigma_j-\sum_{\langle ij\rangle\in\mathcal{L}} (\beta +\Delta\mathfrak{m}_{ij})\sigma_i\sigma_j}\Big)^{q}}{\Big(\sum_{\{\sigma_i\}}e^{\sum_{\langle ij\rangle}(\beta +\Delta\mathfrak{m}_{ij})\sigma_i\sigma_j}\Big)^{q-1}},
\end{align}
and changing the variables to $\mathfrak{n}_{ij}=\beta+\Delta m_{ij}$ [Eq.~\eqref{EqVariableChangeToGothicN}],
\begin{equation}
    \overline{\langle \mu_{\tilde{x}}\mu_{\tilde{y}}\rangle_{\vec{\mathfrak{m}}}^q}=\frac{e^{-\frac{\Delta}{2} N_b}}{Z}\int_{\vec{\mathfrak{m}}} e^{-\sum_{\langle ij\rangle}\frac{(\mathfrak{n}_{ij}-\beta)^2}{2\Delta}} \frac{\Big(\sum_{\{\sigma_i\}}e^{\sum_{\langle ij\rangle\notin\mathcal{L}} \mathfrak{n}_{ij}\sigma_i\sigma_j-\sum_{\langle ij\rangle\in\mathcal{L}} \mathfrak{n}_{ij}\sigma_i\sigma_j}\Big)^{q}}{\Big(\sum_{\{\sigma_i\}}e^{\sum_{\langle ij\rangle}\mathfrak{n}_{ij}\sigma_i\sigma_j}\Big)^{q-1}}.
\end{equation}
Then, {analogous to Merz and Chalker~\cite{MerzChalker},} on the {\it higher} Nishimori line $\beta=\Delta$ [Fig.~\ref{fig:2DGaussianMeasurements}], we can consider
performing the `gauge' transformation in Eq.~\eqref{EqGaugeTransform} and 
sum
over all `gauge' spins $\{s_i\}$, to obtain [analogous to Eq.~\eqref{EqEASimplificationForGaussianMeasIntermediateStepII}]
\begin{align}
    \overline{\langle \mu_{\tilde{x}}\mu_{\tilde{y}}\rangle_{\vec{\mathfrak{m}}}^q}&=\frac{e^{-\Delta N_b}}{Z}\int_{\vec{\mathfrak{n}}}e^{-\sum_{\langle ij\rangle}\frac{\mathfrak{n}_{ij}^2}{2\Delta}}  \frac{ \big(\sum_{\{s_{i}\}}e^{\sum_{\langle ij\rangle}\mathfrak{n}_{ij}s_is_j}\big)}{2^{N_s}} \frac{\Big(\sum_{\{\sigma_i\}}e^{\sum_{\langle ij\rangle\notin\mathcal{L}} \mathfrak{n}_{ij}\sigma_i\sigma_j-\sum_{\langle ij\rangle\in\mathcal{L}} \mathfrak{n}_{ij}\sigma_i\sigma_j}\Big)^{q}}{\Big(\sum_{\{\sigma_i\}}e^{\sum_{\langle ij\rangle}\mathfrak{n}_{ij}\sigma_i\sigma_j}\Big)^{q-1}}\nonumber\\    &=\frac{e^{-\Delta N_b}}{Z\times 2^{N_s}}\int_{\vec{\mathfrak{n}}}e^{-\sum_{\langle ij\rangle}\frac{\mathfrak{n}_{ij}^2}{2\Delta}} \frac{\Big(\sum_{\{\sigma_i\}}e^{\sum_{\langle ij\rangle\notin\mathcal{L}} \mathfrak{n}_{ij}\sigma_i\sigma_j-\sum_{\langle ij\rangle\in\mathcal{L}} \mathfrak{n}_{ij}\sigma_i\sigma_j}\Big)^{q}}{\Big(\sum_{\{\sigma_i\}}e^{\sum_{\langle ij\rangle}\mathfrak{n}_{ij}\sigma_i\sigma_j}\Big)^{q-2}}\label{EqSimipliedDualSpinCorrFun}.
\end{align}
\end{widetext}
It is clear that the integrand in Eq.~\eqref{EqSimipliedDualSpinCorrFun} is invariant under the following simultaneous change of variables:
\begin{align}
    &q\rightarrow (2-q), \nonumber\\
    &\mathfrak{n_{ij}}\rightarrow-\mathfrak{n_{ij}} \text{ if } \langle ij\rangle \in \mathcal{L},\nonumber\\
    &\mathfrak{n_{ij}}\rightarrow \mathfrak{n_{ij}} \text{ if } \langle ij\rangle \notin \mathcal{L}\nonumber.
\end{align}
 Since we integrate over all $\mathfrak{n}_{ij}$, this 
 implies
 that
\begin{equation}
    \overline{\langle \mu_{\tilde{x}}\mu_{\tilde{y}}\rangle_{\vec{\mathfrak{m}}}^q}=\overline{\langle \mu_{\tilde{x}}\mu_{\tilde{y}}\rangle_{\vec{\mathfrak{m}}}^{2-q}}.\label{EqReflectionUnderMomentOfTheDualSpinCorrFun}
\end{equation}
The above equality between moments holds exactly on the {\it higher} Nishimori line $\beta=\Delta$ in the learning phase diagram for Gaussian measurements. 
For binary measurements, even though the exact equality in Eq.~\eqref{EqReflectionUnderMomentOfTheDualSpinCorrFun} might be spoiled at short distances, since the universal properties of the
{novel}
tricritical point in the learning phase diagram for binary measurements are 
also
expected to be governed by the {\it higher} Nishimori critical point, 
this implies that
the above equality 
holds as an equivalence at long-distances
{at the novel \textit{learning} tricritical point for binary measurements}, i.e.
\begin{equation}
    \overline{\langle \mu_{\tilde{x}}\mu_{\tilde{y}}\rangle_{\vec{\mathfrak{m}}}^q} \sim  \overline{\langle \mu_{\tilde{x}}\mu_{\tilde{y}}\rangle_{\vec{\mathfrak{m}}}^{2-q}}, \;\;\;\;\;\;|\tilde{x}-\tilde{y}| \gg 1 \,.
    \label{EqDualCorrelationsReflectionEquivalenceAtLongDistances}
\end{equation}
This means that the exponents $\eta_q$ of the measurement-averaged moments of the dual spin correlation function at the tricritical point in the learning phase diagram satisfy
\begin{equation}
    \eta_q=\eta_{2-q}.\label{EqDualExponentsEquality}
\end{equation}
This, in particular, implies that the measurement-averaged \textit{second} moment ($q=2$) of the dual spin correlation function has \textit{zero} scaling dimension~\cite{Note9}
at the tricritical point, i.e.\ the {\it higher} Nishimori critical point, in the learning phase diagram, i.e.
\begin{equation}
    \eta_2=0.\label{EqAppExponentSecondMomentDualIsZero}
\end{equation}
The 
{results in Eqs.~\eqref{EqDualExponentsEquality} and~\eqref{EqAppExponentSecondMomentDualIsZero}}
should be compared 
with those
of Merz and Chalker~\cite{MerzChalker}, who obtained the equality between the scaling dimensions for bond-randomness averaged $q^{\text{th}}$ and $(1-q)^{\text{th}}$ moments for the dual spin correlation function at the ordinary Nishimori critical point in the $2D$ RBIM, and 
consequently also
the result that the bond-randomness averaged \textit{first} moment of the dual spin correlation function has \textit{zero} scaling dimension
at the ordinary Nishimori critical point in the $2D$ RBIM.
Analogous to the argument in Ref.~\cite{MerzChalker}, we see that due to Jensen's inequality  
$ \overline{\langle \mu_{\tilde{x}}\mu_{\tilde{y}}\rangle_{\vec{\mathfrak{m}}}^q}\geq \Big(\overline{\langle \mu_{\tilde{x}}\mu_{\tilde{y}}\rangle_{\vec{\mathfrak{m}}}^2}\Big)^{q/2}$~\footnote{Note that by its definition in Eq.~\eqref{EqDualCorrFunInAGivenMeasTrajectory}, $\langle \mu_{\tilde{x}}\mu_{\tilde{y}}\rangle_{\vec{\mathfrak{m}}}$ is non-negative in every {measurement} trajectory $\vec{\mathfrak{m}}$.},
and 
{hence}
the scaling dimension $\eta_q$ for the measurement-averaged $q^{\text{th}}$ 
{moment}
of the dual spin correlation function should be \textit{non-positive} for $q>2$.
Finally, analogous to
the
other correlation functions that we discussed 
{[see e.g. Eq.~\eqref{EqFirstMomentIsEqualToTheUnmeasured} or Eq.~\eqref{EqFirstMomentIsUnaffected}],}
it follows from Eq.~\eqref{EqDefMomentsOFDualSpinCorrFun} that the measurement-averaged \textit{first} moment {($q=1$)} of the dual spin correlation function at the {\it higher} Nishimori critical {point} is just given by the dual spin correlation function at the unmeasured $2D$ Ising critical point. Due to {the} KW duality~\cite{KramersWannier1941}, the scaling dimension of the dual spin operator at the unmeasured $2D$ Ising critical point is
known
to be equal to that of the spin operator~\cite{KadanoffCeva}, {and} 
therefore
\begin{equation}
    \eta_1=\frac{1}{8}.
\end{equation}

{In summary, we have obtained that the first and the second moment of the dual spin correlation function at the learning tricritical point, i.e.\  the {\it higher} Nishimori critical point, have $1/8$ and zero scaling dimensions, respectively, and the rest of the higher moments ($q>2$) should have \textit{non-positive} scaling dimensions.}

\section{Translating observables between the ordinary Nishimori line in the RBIM and in the learning phase diagram
\label{AppTranslatingBetweenRBIMandBayesianInference}}

It is important to distinguish between the measurement-averaged observables in the Bayesian inference formulation of the ordinary Nishimori line, the line $\beta=0$ in the learning phase diagram~[Fig.~\ref{fig:2DGaussianMeasurements}] governed by the replica $R\rightarrow1$ limit, and the bond-randomness averaged observables on 
the ordinary Nishimori line in the RBIM, i.e.\ its
$R\rightarrow0$ replica limit formulation. 
{The main point is that the \textit{measurement-averaged} $(2q)^{\text{th}}$ even moment of a multipoint spin correlation function in the Bayesian inference formulation of the ordinary Nishimori line as the $\beta=0$ line in Fig.~\ref{fig:2DGaussianMeasurements}, which is given by a  replica limit $R\rightarrow1$ theory with gauge invariance, is equal to both the \textit{bond-randomness} averaged $(2q)^{\text{th}}$ even 
and the $(2q-1)^{\text{th}}$ odd moment of the same correlation function in the RBIM formulation of the ordinary Nishimori line, which corresponds to the `gauge fixed' replica theory in the replica $R\rightarrow0$ limit. That is,
\begin{equation}
    \overline{\langle \sigma_{i_1}\sigma_{i_2}\cdots\sigma_{i_k}\rangle^{2q}}= \left[\left\langle \sigma_{\boldsymbol{i}_1} \ldots \sigma_{\boldsymbol{i}_k}\right\rangle^{2 q-1}\right]=\left[\left\langle \sigma_{\boldsymbol{i}_1} \ldots \sigma_{\boldsymbol{i}_k}\right\rangle^{2 q}\right],\label{EqRBIMtoLearning}
\end{equation}
where the correlation function $\overline{\langle \sigma_{i_1}\sigma_{i_2}\cdots\sigma_{i_k}\rangle^{2q}}$ is the $(2q)^{\text{th}}$ measurement-averaged moment on the $\beta=0$ line in the learning phase diagram of the Ising model, and the $[\left\langle \sigma_{\boldsymbol{i}_1} \ldots \sigma_{\boldsymbol{i}_k}\right\rangle^{2 q-1}]$ and $[\left\langle \sigma_{\boldsymbol{i}_1} \ldots \sigma_{\boldsymbol{i}_k}\right\rangle^{2q}]$ are respectively the $(2q-1)^{\text{th}}$ and $(2q)^{\text{th}}$ bond-randomness averaged moments on the ordinary Nishimori line in the RBIM, where the latter two are known to be equal due to Nishimori~\cite{Nishimori1981}.}
{This 
{relation
in Eq.~\eqref{EqRBIMtoLearning}} for the ordinary Nishimori line
has been emphasized in Ref.~\cite{NahumJacobsen} and in a related context of
{a}
monitored quantum circuit in Ref.~\cite{ZhuTantivasadakarnVishwanthTrebstVerresen}, and we discuss it here for completeness.}

The line $\beta=0$ for the replica theory in Eq.~\eqref{EqLatticeReplicaTheory} 
has local symmetry: $\sigma_i^{(a)}\rightarrow-\sigma_i^{(a)}, \; \forall a=1, \cdots ,R$. 
Due to this local symmetry,
the measurement-averaged odd moment
of any multipoint spin correlation function anywhere on the $\beta=0$ line in the learning phase diagram~[Fig.~\ref{fig:2DGaussianMeasurements}], which is governed by the $R\rightarrow1$ replica limit of the replica theory in Eq.~\eqref{EqLatticeReplicaTheory} {with $\beta=0$,} vanishes identically. 
In particular, the measurement-averaged first moment of the spin-spin correlation function vanishes identically everywhere on the line $\beta=0$ in the learning phase diagram.
{For $\beta=0$,} 
the replica theory {in} Eq.~\eqref{EqLatticeReplicaTheory} in the $R\rightarrow1$ replica limit
can be written as 
{a replica $R\rightarrow0$}
theory {in}  Eq.~\eqref{EqLatticeReplicaTheory} with $\beta$ 
equal to $\Delta$,
where the latter is precisely the formulation of the ordinary Nishimori line in the RBIM with {Gaussian bond-randomness~\cite{Nishimori1981,LeDoussalHarrisI}.}
[This proceeds exactly analogous to our discussion in Sec.~\ref{SecEmergentGaugeAndEnlargedPermutationSymmetry}, where we demonstrated that
the line $\beta=\Delta$ for the replica theory Eq.~\eqref{EqLatticeReplicaTheory} {with $R$ replicas can be written}
as the $\beta=0$ line for the gauge-invariant and enlarged replica-symmetric theory with $R+1$ replicas in Eq.~\eqref{EqGaugeSymmetricPartionFunction}.]
However, the measurement-averaged moments of the correlation functions on the $\beta=0$ {line} in the learning phase diagram {[Fig.~\ref{fig:2DGaussianMeasurements}]}, 
which has local symmetry as a replica theory 
{in the $R\rightarrow1$ replica limit,}
and the bond-randomness averaged moments in the equivalent problem of the ordinary Nishimori line in the RBIM, governed by the $R\rightarrow0$ replica limit, are \textit{not} in one-to-one correspondence.
For example, as we just discussed above, the measurement-averaged first moment of the spin-spin correlation function 
vanishes identically on the $\beta=0$ line in the learning phase diagram but that is not true for the bond-randomness averaged first moment of the spin-spin  correlation function
on the ordinary Nishimori line in the RBIM. 
{In particular, considering the replica $R\rightarrow0$ limit of}
Eq.~\eqref{EqHamiltonianOnNishimoriStarLineUngauged}, \eqref{EqGaugeSymmetricPartionFunction} {from Sec.~\ref{SecEmergentGaugeAndEnlargedPermutationSymmetry}} and 
{Eq.~\eqref{EqAppCorrelationFunctionFirstStep}, 
\eqref{EqOddToEvenMomentsPenultimateStep} in App.~\ref{AppDerivationOfIdentityForMoments},}
it 
{is}
easily verified that the measurement-averaged $(2q)^{\text{th}}$ moment of a {spin} correlation function on the $\beta=0$ line in the learning phase diagram, described by the gauge-invariant replica theory in the $R\rightarrow1$ replica limit, is equal to the bond-randomness  averaged $(2q)^{\text{th}}$ and $(2q-1)^{\text{th}}$ moment of the same correlation function on the {ordinary} Nishimori line in the RBIM, governed by the $R\rightarrow0$ replica limit. 
{[As noted in our Eq.~\eqref{EqNishimoriEquality}, the bond-randomness  averaged $(2q)^{\text{th}}$ and $(2q-1)^{\text{th}}$ moment of any multipoint spin correlation function on the Nishimori line in the RBIM are 
well-known
to be equal to each other~\cite{Nishimori1981}.] }
{That is, we obtain the relation in Eq.~\eqref{EqRBIMtoLearning} above.}

\section{Derivation of \texorpdfstring{$c_{\text{eff}}> c_{\rm Ising}=1/2$}{Lg}}

\subsection{The Derivation
\label{AppDerivationOfCeffBound}}
In this appendix, we will obtain the bound $c_{\text{eff}}> c_{\rm Ising}=1/2$
discussed in Sec.~\ref{SecCeffThmBound}. 
To obtain this bound, apart from the usual assumptions about 
translational
{invariance}, rotational (Lorentz) {invariance}, conformal invariance, and (unbroken) replica permutation symmetry of the discussed replica CFTs, we will make the following fairly standard assumptions:
\begin{itemize}
    \item The amplitudes and the scaling dimensions of discussed replica correlation functions [see below] are analytic functions of the number of replicas $R$ around the number of replicas $R=1$ of interest.
    (This assumption is intimately connected to the validity of
replica trick.)
    \item The scaling limit of the lattice replica theory [defined in Eq.~\eqref{EqScalingLimit}] and the $R\rightarrow$ replica limit commute with each other.
\end{itemize}

To obtain 
the bound discussed in Sec.~\ref{SecCeffThmBound}, 
we will consider the $R$-replica 
Nishimori critical point which occurs on the line $\beta=\Delta$ of the $R$-replica theory in Eq.~\eqref{EqLatticeReplicaTheory}. In particular, we will consider the following replica theory
\begin{align}
&Z_R=\sum_{\{\sigma_i^{(a)}\}_{a=1}^{R}}\exp{\bigg\{\beta_c(R) \sum_{\langle
ij \rangle}\sum_{a=1}^{R} \sigma_i^{(a)}\sigma_j^{(a)}}\nonumber\\
&\qquad\qquad\qquad\qquad
+\frac{\beta_c(R)}{2}\sum_{\langle ij\rangle}\sum_{\substack{a,b=1\\ a\neq b}}^{R}\sigma_i^{(a)}\sigma_j^{(a)}\sigma_i^{(b)}\sigma_j^{(b)}\bigg\},\label{EqRReplicaNishimoriCriticalPoint}
\end{align}
where 
$(\beta=\beta_c(R),{\Delta=\beta_c(R)})$
denotes the location of the
$R$-replica 
Nishimori critical point on the line $\beta=\Delta$ for the $R$-replica theory [see Sec.~\ref{SecEmergentGaugeAndEnlargedPermutationSymmetry}]. 
For example, the $R\rightarrow1$ 
Nishimori critical point is 
what we referred to as the
{\it higher} Nishimori critical point in this work, 
which
governs the
tricritical point in the learning phase diagram (Gaussian measurements) in Fig.~\ref{fig:2DGaussianMeasurements}, and the $R\rightarrow0$ Nishimori critical point corresponds to the ordinary Nishimori critical point in the RBIM (Gaussian disorder).
We expect that the $R$-replica Nishimori critical point exists for an appropriate finite value of 
$(\beta=\beta_c(R),{\Delta=\beta_c(R)})$
for sufficiently small values of number $R$ of replicas comprising the interval $0\leq R\leq 1$.

We will now consider perturbing the $R$-replica Nishimori critical point in Eq.~\eqref{EqRReplicaNishimoriCriticalPoint} 
with
a particular operator $O_{ij}^{(R)}$
[after subtracting its expectation value $\langle \cdots\rangle_*$ in Eq.~\eqref{EqRReplicaNishimoriCriticalPoint}]
given by
\begin{align}
    &O^{(R)}_{ij}=\bigg({-}\sum_{\substack{a,b=1\\ a\neq b}}^{R}\varphi_{ij}^{(a)}\varphi_{ij}^{(b)}+\alpha(R)\sum_{a=1}^{R} \varphi_{ij}^{(a)}\bigg)\nonumber\\
    &\qquad\qquad\big\langle\bigg({-}\sum_{\substack{a,b=1\\ a\neq b}}^{R}\varphi_{ij}^{(a)}\varphi_{ij}^{(b)}+\alpha(R)\sum_{a=1}^{R} \varphi_{ij}^{(a)}\bigg)\big\rangle_*,\label{EqOperatorCausingFlowToTheIsingCriticalPoint}
\end{align}
where we have defined $\varphi_{ij}$ as the bond-energy given by
$$\varphi_{ij}=\sigma_i\sigma_j,$$
and
$\alpha(R)$ is chosen s.t. the perturbation
{$(\delta \Delta)\sum_{ij}O_{ij}^{(R)}$ [$(\delta\Delta)>0$]}
induces the RG flow to the `clean' $2D$ Ising critical point with decoupled $R$-replica copies.
{[The negative sign in front of the inter-replica coupling in Eq.~\eqref{EqOperatorCausingFlowToTheIsingCriticalPoint} is chosen because at the `clean' Ising critical point the replicas are decoupled and the inter-replica coupling vanishes.]}
Indeed, such RG flow is known to exist from the ordinary Nishimori critical point{, the replica $R\rightarrow0$ Nishimori critical point,} to the clean $2D$ Ising critical point in the {$2D$} RBIM, 
where the perturbation is of the form in Eq.~\eqref{EqOperatorCausingFlowToTheIsingCriticalPoint} with some $\alpha(R=0)\neq 0$ (see e.g.~\cite{PiccoPujolHonecher2006}). 
An analogous RG flow is seen from the tricritical point in the learning phase diagram in Fig.~\ref{fig:2DGaussianMeasurements}, which is the replica $R\rightarrow1$ Nishimori critical point, to the unmeasured $2D$ Ising critical point.
In the latter case of the replica $R\rightarrow1$ limit, the operator in Eq.~\ref{EqOperatorCausingFlowToTheIsingCriticalPoint} that induces the RG flow to the $2D$ Ising critical point has $\alpha(R=1)=0$, but in general $\alpha(R\neq 1)\neq 0$, and assuming analyticity in replica index $R$ around $R=1$, this implies that
\begin{equation}
    \alpha(R)=\mathcal{O}(R-1).\label{EqAlphaR}
\end{equation}
The lattice operator in Eq.~\eqref{EqOperatorCausingFlowToTheIsingCriticalPoint} at the 
$R$-replica Nishimori critical point
can be written 
as a
superposition of scaling fields of the fixed point $2D$ CFT 
that governs
the $R$-replica Nishimori critical point (see, e.g.,~\cite{CardyLesHouches2008,LUKYANOVTERRAS}), 
\begin{equation}\label{EqContinuumFieldExpansionofO}
    O^{(R)}_{ij}\sim \epsilon^{\mathcal{X}_0(R)}C_0O^{(R)}_0(x)+\sum_{\mathfrak{p}\neq 0} \epsilon^{\mathcal{X}_\mathfrak{p}(R)}C_\mathfrak{p}O^{(R)}_\mathfrak{p}(x)
\end{equation}
where $\epsilon$ is the lattice spacing, $C_\mathfrak{p}$ are non-universal constants, and
 $O^{(R)}_\mathfrak{p}(x)$ [$x$ 
 {is}
 the continuum label] is the scaling field with scaling dimension $\mathcal{X}_\mathfrak{p}(R)$ at the fixed point $2D$ CFT corresponding to the
 {$R$-replica}
Nishimori critical point.
We note that the scaling fields $O^{(R)}_\mathfrak{p}(x)$ in the expansion in Eq.~\eqref{EqContinuumFieldExpansionofO} are \textit{not} normalized to have amplitude $=1$ in the two-point correlation function at the CFT fixed point. 
Rather,
assuming the validity of 
{the}
replica trick, since the two-point correlation function of the lattice operator $O_{ij}^{(R)}$ in Eq.~\eqref{EqOperatorCausingFlowToTheIsingCriticalPoint}
is expected to be analytic in the replica index $R$ around $R=1$, the 
{amplitudes}
of 
{the}
two-point correlation functions for 
{the}
scaling fields $O_\mathfrak{p}^{(R)}(x)$ on the RHS of Eq.~\eqref{EqContinuumFieldExpansionofO} are also expected to be
{some}
analytic functions of 
{the}
replica index $R$.
Moreover, we also expect that in the expansion in Eq.~\eqref{EqContinuumFieldExpansionofO} only the leading scaling field $O^{(R)}_0(x)$ is relevant in the RG sense, i.e.\ $\mathcal{X}_0(R)<2$, and all the other fields $O^{(R)}_\mathfrak{p}(x)$ with $\mathfrak{p}\neq 0$ are irrelevant corrections, i.e.\ $\mathcal{X}_\mathfrak{p}(R)>2$ for $\mathfrak{p}\neq 0$.

Let us now consider the following partition function $Z_R'$ of the
$R$-replica Nishimori critical point in Eq.~\eqref{EqRReplicaNishimoriCriticalPoint} perturbed by
the operator $O^{(R)}_{ij}$ 
\begin{align}
    &Z'_R=\sum_{\{\sigma_i^{(a)}\}_{a=1}^{R}}\exp{\bigg\{\beta_c(R) \sum_{\langle
ij \rangle}\sum_{a=1}^{R} \sigma_i^{(a)}\sigma_j^{(a)}}\nonumber\\
&\qquad\qquad\qquad\qquad
+\frac{\beta_c(R)}{2}\sum_{\langle ij\rangle}\sum_{\substack{a,b=1\\ a\neq b}}^{R}\sigma_i^{(a)}\sigma_j^{(a)}\sigma_i^{(b)}\sigma_j^{(b)}\nonumber\\&\qquad\qquad\qquad\qquad
+
\delta \Delta\sum_{\langle ij\rangle}O^{(R)}_{ij}\bigg\}.\label{EqPerturbedPartitionFunction}
\end{align}
We will
consider taking the scaling limit of the perturbed lattice replica theory in Eq.~\eqref{EqPerturbedPartitionFunction}, where we will send the lattice spacing $\epsilon\rightarrow0$ and the perturbation strength $\delta\Delta\rightarrow0$ while holding the `renormalized coupling' $g=(\delta \Delta) \epsilon^{\mathcal{X}_0(R)-2}$ to a fixed value:
\begin{align}\label{EqScalingLimit}
&\text{Scaling Limit}: \epsilon\rightarrow0\;\;\;\delta\Delta\rightarrow0\nonumber
\\
&(\delta \Delta) \epsilon^{\mathcal{X}_0(R)-2}=g=\text{constant}. 
\end{align}
In this scaling limit, the perturbed partition function in Eq.~\eqref{EqPerturbedPartitionFunction} is described by the following continuum action
\begin{equation}
    \mathcal{S}=S_*^{(R)}
    +
    g\int d^2x \;\; C_0O_0^{(R)}(x),\label{EqPerturbedFieldTheory}
\end{equation}
where $S_*^{(R)}$ is the fixed point $2D$ CFT action for 
the $R$-replica Nishimori critical point and $O_0^{(R)}$ is the leading relevant scaling field in the perturbation $O^{(R)}_{ij}$ in Eq.~\eqref{EqContinuumFieldExpansionofO}. 
[Note that in the scaling limit in Eq.~\eqref{EqScalingLimit} all the irrelevant scaling fields $O^{(R)}_\mathfrak{p}(x)$ ($\mathfrak{p}\neq 0$) in Eq.~\eqref{EqContinuumFieldExpansionofO} drop out of Eq.~\eqref{EqPerturbedFieldTheory}.]
Then we can consider applying Zamolodchikov's $c$-theorem sum-rule~\cite{Zamolodchikovctheorem,Cardy1988}, analogous to the proof of {the} $c$-effective theorem in Ref.~\cite{PatilLudwig20251}, to the field theory in Eq.~\eqref{EqPerturbedFieldTheory} to obtain
\begin{align}
    &\frac{R}{2}-c(R)=-6\pi^2{g^2}{}(2-\mathcal{X}_0(R))^2\times\nonumber\\
    &\qquad\qquad\qquad\qquad\times\int_0^{\infty} dr\, r^3 C_0^2 \langle O_0^{(R)}(r)O_0^{(R)}(0) \rangle_c,\label{EqSumRule}
\end{align}
where the \textit{connected} correlation function $\langle O_0^{(R)}(r)O_0^{(R)}(0) \rangle_c$ is evaluated with the field theory action in Eq.~\eqref{EqPerturbedFieldTheory},
and $c(R)$ and $\frac{R}{2}$ are the replica dependent central charges of the $R$-replica 
Nishimori critical point (the `UV' central charge) and the decoupled $R$-copies of the $2D$ Ising critical point (the `IR' central charge), respectively. 
We are interested in the Casimir effective central charge in the $R\rightarrow1$ replica limit
\begin{equation}
    c_{\text{eff}}=\frac{dc(R)}{dR}\big|_{R=1}.
\end{equation}
{The above Casimir effective central charge $c_{\text{eff}}$, together with the central charge of the unmeasured Ising critical point ($=1/2$),}
characterizes the finite-size scaling of the Shannon entropy of the measurement record~\cite{NahumJacobsen} at the tricritical point in the 
learning phase diagram [Fig.~\ref{fig:2DGaussianMeasurements}].
To this end, let us divide both sides of Eq.~\eqref{EqSumRule} by $R-1$ and take the replica $R\rightarrow1$ limit. 
Thus, we obtain
\begin{align}
    &\frac{1}{2}-c_{\text{eff}}=-6\pi^2g^2(2-\mathcal{X}_0(R=1))^2\times \nonumber\\
    &\qquad\qquad\qquad\times\int_{0}^{\infty} dr\,r^3\Big[C_0^2\lim_{R\rightarrow1}\frac{\langle O_0^{(R)}(r)O_0^{(R)}(0) \rangle_c}{R-1}\Big],\label{EqSumRuleSimplifiedForREqualTo1ReplicaLimit}
\end{align}
where we have used the assumption that around $R=1$ $\mathcal{X}_0(R)<2$ and that $\mathcal{X}_0(R)$ is an analytic function of $R$. 

The correlation function $\langle O_0^{(R)}(r)O_0^{(R)}(0) \rangle_c$ in the field theory in Eq.~\eqref{EqPerturbedFieldTheory} is precisely given by the scaling limit [Eq.~\eqref{EqScalingLimit}] of the correlation function of the 
lattice {perturbation} operator $O_{ij}^{(R)}$ [Eq.~\eqref{EqOperatorCausingFlowToTheIsingCriticalPoint}] in the perturbed lattice replica theory in Eq.~\eqref{EqPerturbedPartitionFunction}. In particular, from Eq.~\eqref{EqContinuumFieldExpansionofO},
\begin{align}
    &C_0^2\langle O_0^{(R)}(r)O_0^{(R)}(0) \rangle_c\nonumber\\&=\lim_{\substack{\epsilon,\delta \Delta \rightarrow0\\(\delta \Delta)\epsilon^{\mathcal{X}_0(R)-2}=g}}\epsilon^{-2\mathcal{X}_0(R)} \langle O^{(R)}_{ij}O^{(R)}_{kl}\rangle_{c},\label{EqLatticeToContinuumCorrelationFunction}
\end{align}
where we have set the origin of the continuum space at the bond $kl$ in the lattice model and
the lattice distance of bond $ij$ from the origin is given by $r/\epsilon$, where $r$ is the continuum
distance from the origin and $\epsilon$ is the lattice spacing.
In Eq.~\eqref{EqLatticeToContinuumCorrelationFunction}, the correlation function $\langle O_0^{(R)}(r)O_0^{(R)}(0) \rangle_c$ is evaluated in the field theory in Eq.~\eqref{EqPerturbedFieldTheory} and $\langle O^{(R)}_{ij}O^{(R)}_{kl}\rangle_{c}$ is evaluated on the lattice using the lattice $R$-replica theory in Eq.~\eqref{EqPerturbedPartitionFunction}. 
Analogous to Ref.~\cite{PatilLudwig20251}, using the replica permutation symmetry of the replica theory in Eq.~\eqref{EqPerturbedPartitionFunction}, one can easily verify that
\begin{align}
    &\langle O^{(R)}_{ij}O^{(R)}_{kl}\rangle_c=\nonumber\\
    &2(R-1)\bigg(\langle \varphi_{ij}^{(1)}\varphi^{(1)}_{kl}\varphi^{(2)}_{ij}\varphi^{(2)}_{kl}\rangle_{R=1}+\langle\varphi_{ij}^{(1)}\varphi_{ij}^{(2)}\varphi^{(3)}_{kl}\varphi^{(4)}_{kl}\rangle_{R=1}\nonumber\\&\;\;\;\;\;\;\;\;\;\;\;\;\;\;\;\;\;\;\;-2\langle \varphi^{(1)}_{ij}\varphi^{(1)}_{kl}\varphi^{(2)}_{ij}\varphi^{(3)}_{kl}\rangle_{R=1}\bigg)+ \mathcal{O}((R-1)^2),\label{EqSimplifiedOOCorrelationFunction}
\end{align}
where we have also used that $\alpha(R)$
in Eq.~\eqref{EqOperatorCausingFlowToTheIsingCriticalPoint} is $\mathcal{O}(R-1)$ [Eq.~\eqref{EqAlphaR}]. 
We note that the correlation functions $\langle \cdots\rangle_{R=1}$ in Eq.~\eqref{EqSimplifiedOOCorrelationFunction} are evaluated with the lattice replica theory in Eq.~\eqref{EqPerturbedPartitionFunction} in the replica $R\rightarrow1$ limit. 
Since the lattice replica theory in Eq.~\eqref{EqPerturbedPartitionFunction} in the replica $R\rightarrow1$ limit governs the 
problem of measurements of bond-energies on the $2D$ classical Ising model, correlation functions $\langle \cdots\rangle_{R=1}$ in Eq.~\eqref{EqSimplifiedOOCorrelationFunction} can be written in terms of measurement-averaged moments of correlation functions of the bond-energy $\varphi_{ij}:=\sigma_i\sigma_j$.
In particular, from Eq.~\eqref{EqSimplifiedOOCorrelationFunction},
\begin{align}
    &\langle O^{(R)}_{ij}O^{(R)}_{kl}\rangle_c=
    2(R-1)\overline{\big(\langle \varphi_{ij}\varphi_{kl}\rangle_{\vec{m}}-\langle \varphi_{ij}\rangle_{\vec{m}} \langle\varphi_{kl}\rangle_{\vec{m}}\big)^2}\nonumber\\&\qquad\qquad\qquad
    + \mathcal{O}((R-1)^2)\label{EqFinalOOCorrelationFunction},
\end{align}
where the correlation function $\langle \cdots \rangle_{\vec{m}}$ is calculated in a given measurement trajectory $\vec{m}=\{m_{ij}\}$ and the overbar `$\overline{(\cdots)}$' denotes the average over Gaussian measurement outcomes.
[Recall that,
as discussed in App.~\ref{AppGaussianMeasurements}, for the case of Gaussian measurements the replica theory in Eq.~\eqref{EqPerturbedPartitionFunction} is an exact description of the measurement problem in the $R\rightarrow1$ replica limit.]
Then,
\begin{align}
    &\frac{\epsilon^{-2\mathcal{X}_0(R)}\langle O^{(R)}_{ij}O^{(R)}_{kl}\rangle_c}{R-1}=\nonumber\\
    & =2\epsilon^{-2\mathcal{X}_0(R)}\overline{\big(\langle \varphi_{ij}\varphi_{kl}\rangle_{\vec{m}}-\langle \varphi_{ij}\rangle_{\vec{m}} \langle\varphi_{kl}\rangle_{\vec{m}}\big)^2}\nonumber\\&\qquad\qquad\qquad\qquad\qquad
    + \epsilon^{-2\mathcal{X}_0(R)} \mathcal{O}((R-1))\label{EqFinalOOCorrelationFunctionModified},
\end{align}
and
assuming the commutativity of the replica and the scaling limits, we obtain
\begin{align}
    &\lim_{R\rightarrow1}\bigg[\lim_{\substack{\epsilon,\delta \Delta \rightarrow0\\(\delta \Delta)\epsilon^{\mathcal{X}_0(R)-2}=g}}\frac{\epsilon^{-2\mathcal{X}_0(R)}\langle O^{(R)}_{ij}O^{(R)}_{kl}\rangle_c}{R-1}\bigg]=\nonumber\\
    &2\lim_{\substack{\epsilon,\delta \Delta \rightarrow0\\(\delta \Delta)\epsilon^{\mathcal{X}_0(R=1)-2}=g}}\epsilon^{-2\mathcal{X}_0(R=1)}\overline{\big(\langle \varphi_{ij}\varphi_{kl}\rangle_{\vec{m}}-\langle \varphi_{ij}\rangle_{\vec{m}} \langle\varphi_{kl}\rangle_{\vec{m}}\big)^2}.\label{EqIntermediateStepCeffThm}
\end{align}
Then, using 
{Eq.~\eqref{EqIntermediateStepCeffThm}}
and Eq.~\eqref{EqLatticeToContinuumCorrelationFunction}, we obtain that
\begin{align}
    &C_0^2\times\lim_{R\rightarrow1}\frac{\langle O_0^{(R)}(r)O_0^{(R)}(0) \rangle_c}{R-1}=\nonumber\\&2\lim_{\substack{\epsilon,\delta \Delta \rightarrow0\\(\delta \Delta)\epsilon^{\mathcal{X}_0(R=1)-2}=g}}\epsilon^{-2\mathcal{X}_0(R=1)} \overline{\big(\langle \varphi_{ij}\varphi_{kl}\rangle_{\vec{m}}-\langle \varphi_{ij}\rangle_{\vec{m}} \langle\varphi_{kl}\rangle_{\vec{m}}\big)^2},\label{EqFinalCorrelatorExpressionInTermsOfMeasurementAvgSquare}
\end{align}
where since the RHS of the above equation is manifestly non-negative this implies non-negativity of the quantity on the LHS as well.
Finally, using Eq.~\eqref{EqSumRuleSimplifiedForREqualTo1ReplicaLimit} and the non-negativity of the LHS of Eq.~\eqref{EqFinalCorrelatorExpressionInTermsOfMeasurementAvgSquare}, due to the overall negative sign on the RHS of Eq.~\eqref{EqSumRuleSimplifiedForREqualTo1ReplicaLimit}, we obtain that
\begin{equation}
   c_{\text{eff}}> \frac{1}{2}, \label{EqCEffIsGreaterThanHalf}
\end{equation}
which is the desired result 
discussed in Sec.~\ref{SecCeffThmBound}.
}

\subsection{A comment on the effective central charge at the ordinary Nishimori critical point
\label{AppCommentOnTheCeffAtOrdinaryNishimori}}

Lastly, an interesting observation 
can be made about the $R\rightarrow0$ replica theory in Eq.~\eqref{EqRReplicaNishimoriCriticalPoint} and the corresponding replica $R\rightarrow0$ Casimir effective central charge $c_{\text{eff}}^{(R\rightarrow0)}$ defined as
\begin{equation}
    c_{\text{eff}}^{(R\rightarrow0)}:=\frac{dc(R)}{dR}\Big|_{R=0}.
\end{equation}
This replica $R\rightarrow0$ Casimir effective central charge 
characterizes
the ordinary Nishimori critical point in the $2D$ RBIM, which also appears on the $\beta=0$ line of the learning phase diagram [Fig.~\ref{fig:2DGaussianMeasurements}] and also at the threshold of the optimal decoder 
{for}
toric code under incoherent errors~\cite{DennisKitaevLandahlPreskill}. 
This replica $R\rightarrow0$ Casimir effective central charge we have numerically estimated (see 
Sec.~\ref{SecNumericalResults}, Fig.~\ref{fig:FreeEnergy}) to be 
\begin{equation}
    c_{\text{eff}}^{(R\rightarrow0)} = {0.462(1)} \,,
    \label{EqCeffZeroForNishimoriOrdinary} 
\end{equation}
which is consistent with previously obtained estimates in the literature~\cite{HoneckerPiccoPujol2001,PiccoPujolHonecher2006}.
This value is clearly less than $1/2$.
This is in contrast to the replica $R\rightarrow1$ Casimir effective central charge corresponding to the \textit{higher} Nishimori critical point, which is rigorously demonstrated to be greater than $1/2$ by the $c$-effective theorem result in Eq.~\eqref{EqCEffIsGreaterThanHalf} and whose value we have determined to be 
\[
 	c_{\text{eff}}^{(R\rightarrow1)} = 0.522(1) \,, 
\]
see Sec.~\ref{SecNumericalResults} [Fig.~\ref{fig:FreeEnergy}].
The observation of the $R\rightarrow0$ Casimir effective central charge in Eq.~\eqref{EqCeffZeroForNishimoriOrdinary} being less than $1/2$
can be understood by considering
the following function $f(R)$
\begin{equation}
	\label{EqDefF(R)}
   	 f(R)=\frac{R}{2}-c(R),
\end{equation}
which represents the difference between the IR and the UV central charge for the replica theory in 
Eq.~\eqref{EqPerturbedPartitionFunction}, which
under RG flows from the $R$-replica Nishimori critical point to the `clean' decoupled replica $2D$ Ising critical point.
Following Ref.~\cite{PatilLudwig20251}, 
note that the function $f(R)$ has two `natural' zeroes 
\footnote{See also Ref.~\cite{NambiKhannaAllocaIadecolaHickeyVasseurWilson}, where an analogous argument was very recently made in the case of $(1+1)D$ measurement-induced phase transitions.} 
-- one at $R=0$ and one at $R=1$, 
as illustrated in Fig.~\ref{fig:Reversal-of-Central-Charge-Change}. 
\begin{figure}
    \centering
    \includegraphics[width=\linewidth]{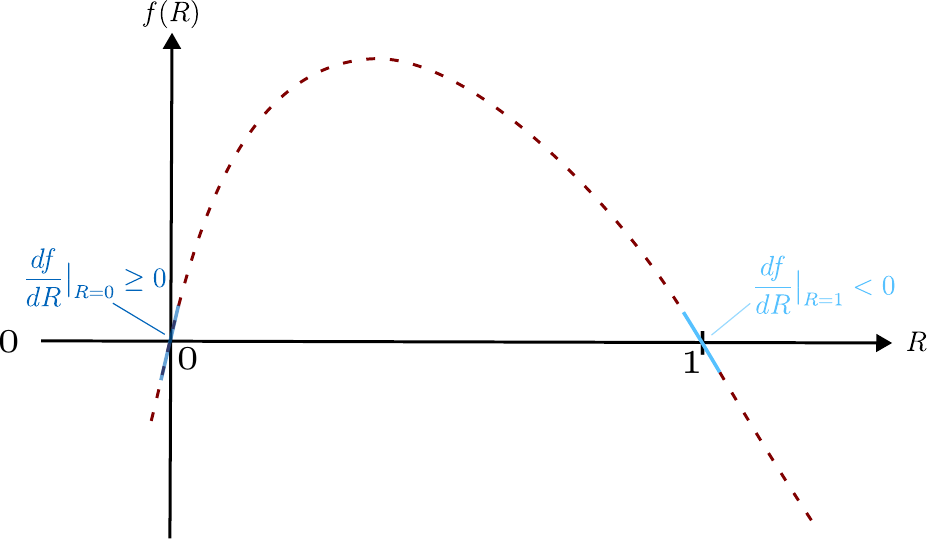}
    \caption{{\bf Difference in the UV and IR replica central charge.}
    Shown is a schematic illustration of function $f(R)=R/2-c(R)$ in Eq.~\eqref{EqDefF(R)},
    which is known to have zeros at $R=0$ and $R=1$, and 
    the $c$-effective theorem result in Eq.~\eqref{EqCEffIsGreaterThanHalf} implies a negative slope at $R=1$ for the function.
    The dashed line then schematically connects these two roots assuming that there is no other root in the interval $0<R<1$,
    and it implies that the slope at $R=0$ is a non-negative number as illustrated in the figure.
     \label{fig:Reversal-of-Central-Charge-Change}}
\end{figure}
The latter zero at $R=1$ follows because, exactly at $R=1$, we only have a single copy of the $2D$ Ising model at $\beta_{c}(R=1)=\beta_c=\frac{1}{2}\ln(1+\sqrt{2})$ [Eq.~\eqref{EqLocationFixedPoint}], which corresponds to the critical point of the $2D$ clean Ising model and therefore has central charge $1/2$.
The former zero at $R=0$ results trivially since $f(R=0)=0\times (1/2)-c(R=0)=c(R=0)$,
which vanishes because the partition function is unity in this limit by the definition of the replica 
trick~\footnote{Equivalently, note that there are simply \textit{no} replicas at $R=0$.}.
The result we obtained in Eq.~\eqref{EqCEffIsGreaterThanHalf} for 
the $R\rightarrow1$ replica limit implies that the slope of $f(R)$ at $R=1$ is
{negative
i.e.\ 
$(df/dR)|_{R=1}<0$;}
therefore, when $R<1$ and near $R=1$,
{the function $f(R)$ is positive, i.e.} $f(R)>0$. 
{{Let us now}
assume that  $f(R)=R/2-c(R)$
 has zeroes {\it only} at $R=0$ and $R=1$ in the interval $0\leq R\leq 1$, since these are the only zeroes
 which are forced to appear due to  the very robust and concrete physical 
 reasoning mentioned above.
{If we assume that $R=0$ and $R=1$ are the only zeroes 
{of}
$f(R)$ in the interval $0\leq R\leq1$, 
as schematically illustrated in Fig.~\ref{fig:Reversal-of-Central-Charge-Change}, 
our result in Eq.~\eqref{EqCEffIsGreaterThanHalf}, which is equivalent to $(df(R)/dR)|_{R=1}<0$, implies that~\footnote{
Since $f(R)$ is zero at $R=0$ and $R=1$, and \textit{if} it does not have any other zeroes in $R\in (0,1)$, 
then $f(R)$ does not change sign in the interval $R\in (0,1)$. Since the slope of $f(R)$ defined 
$(df(R)/dR)|_{R=1}=\lim_{\delta\rightarrow0^{+}}\frac{(f(1)-f(1-\delta))}{\delta}=-\lim_{\delta\rightarrow0^{+}}\frac{(f(1-\delta))}{\delta}$ 
is negative at $R=1$ (See Eqs. \eqref{EqCEffIsGreaterThanHalf} and \eqref{EqDefF(R)}), the slope 
$(df(R)/dR)|_{R=0}=\lim_{\delta\rightarrow0^{+}}\frac{(f(0+\delta)-f(0))}{\delta}=\lim_{\delta\rightarrow0^{+}}\frac{(f(0+\delta))}{\delta}$ at $R=0$ must be non-negative. {This is because $f(1-\delta)$ and $f(0+\delta)$ should 
be both positive for $\delta>0$} [We are assuming that $f(R)$ is differentiable in the closed interval $R\in [0,1]$.]}, 
}
\begin{equation}\label{EqCeffR=0}
    \frac{df(R)}{dR}\bigg|_{R=0}\geq0 
     \ \Rightarrow \ 
    \frac{1}{2}\geq \frac{dc(R)}{dR}\bigg|_{R=0}
    \ \Rightarrow \ 
    \frac{1}{2}\geq c_{{\rm eff}}^{(R\rightarrow0)}.
    \quad
\end{equation}
{Therefore, given the assumption that $f(R)=\frac{R}{2}-c(R)$ does not have zeroes in the 
{open interval
$0 < R < 1$},
{our result in
Eq.~\eqref{EqCEffIsGreaterThanHalf}} implies that the
Casimir
effective central charge $c_{\text{eff}}^{(R\rightarrow0)}$
{in the $R\to 0$ replica limit}, which characterizes
the ordinary Nishimori critical point, should be less than $1/2$, which is consistent 
with
numerical 
{estimates}
for it.

\section{Numerical simulations}
\label{App:Numerics}

\subsubsection*{Majorana mapping}

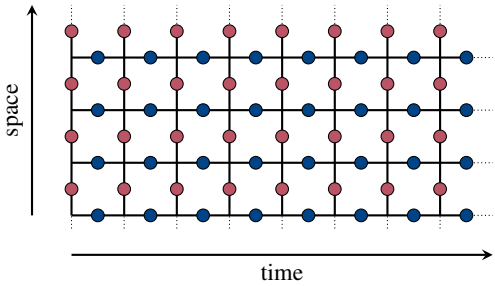
\begin{figure}[b]
    \centering
    \begin{tikzpicture}[>=stealth, scale=0.7]

        \def\nx{8}   
        \def\ny{4}   

        \foreach \x in {1,...,\nx} {
                \draw[black, thick] (\x,1) -- (\x,\ny + 0.5);
            }

        \foreach \y in {1, ..., \ny} {
                \draw[black, thick] (1,\y) -- (\nx + 0.5,\y);
            }

        \foreach \x in {1,...,\nx} {
                \draw[densely dotted, black] (\x,\ny + 0.5) -- (\x,\ny + 1);
            }

        \foreach \x in {1,...,\nx} {
                \draw[densely dotted, black] (\x,0.75) -- (\x,1);
            }
        \foreach \y in {1,...,\ny} {
                \draw[densely dotted, black] (\nx + 0.5,\y) -- (\nx + 1,\y);
            }
        \foreach \x in {1,...,\nx} {
                \foreach \y in {1, ..., \ny} {
                        \node[circle, draw=black, fill=darkred_malte, inner sep=1.7pt] at (\x,\y + 0.5) {};
                    }
            }

        \foreach \x in {1,...,\nx} {
                \foreach \y in {1, ..., \ny} {
                        \node[circle, draw=black, fill=darkblue_malte, inner sep=1.7pt] at (\x + 0.5,\y) {};
                    }
            }

        \draw[->, thick] (0.25,1) -- (0.25,\ny + 1) node[midway, above, rotate=90] {space};
        \draw[->, thick] (1,0.25) -- (\nx + 1,0.25) node[midway, below] {time};

    \end{tikzpicture}
    \caption{{\bf One-dimensional imaginary time evolution.}
    We interpret the $2+0$D system as a $1+1$D imaginary time evolution, where the vertical direction is space and the horizontal direction is imaginary time. The red nodes represent $ZZ$ interactions and the blue nodes represent $X$ fields. The dotted lines indicate, that we have periodic boundary conditions in space and the system extends to very large imaginary time.
    }
    \label{fig:imaginary_time_evolution}
\end{figure}

In order to calculate the free energy density in our numerical simulations, we interpret the $(2+0)$D system as a $(1+1)$D imaginary time evolution (see Fig.~\ref{fig:imaginary_time_evolution}). The vertical couplings in this picture are $ZZ$ interactions, while the horizontal couplings are $X$ fields. We can write them as
\begin{align}
    \tikz[baseline=-2.5pt]{
        \draw[white, thick] (-0.25, 0) -- (0.25, 0);
        \draw[black, thick] (0, -0.25) -- (0, 0.25);
        \node (d) [circle, draw=black, fill=darkred_malte, inner sep=1.7pt] {};
    } &= e^{(\beta +\tilde \gamma s) ZZ} \\
    \tikz[baseline=-2.5pt]{
        \draw[black, thick] (-0.25, 0) -- (0.25, 0);
        \node (d) [circle, draw=black, fill=darkblue_malte, inner sep=1.7pt] {};
    } &= e^{\beta +\tilde \gamma s} \mathbf{1} + e^{-(\beta +\tilde \gamma s)} X.
\end{align}
Performing a Jordan-Wigner-Transformation, we can then fermionize the system: $Z_j Z_{j+1} = ic_{2j}c_{2j+1}\ ,\ X_j = ic_{2j-1}c_{2j}$, where $c_k$ denotes the Majorana fermion at site $k$}. The vertical couplings are straightforward to fermionize, resulting in
\begin{equation}
    \tikz[baseline=-2.5pt]{
        \draw[white, thick] (-0.25, 0) -- (0.25, 0);
        \draw[black, thick] (0, -0.25) -- (0, 0.25);
        \node (d) [circle, draw=black, fill=darkred_malte, inner sep=1.7pt] {};
    } 
    =e^{(\beta+\tilde{\gamma} s)ic_{2i}c_{2i+1}}
    \longrightarrow e^{2(\beta +\tilde \gamma s) \sigma^x_{2i, 2i+1}} \, ,
\end{equation}
where the last matrix is the corresponding transfer matrix in the first quantization single-particle basis: $ic_{2i}c_{2i+1} = \frac{1}{4} (ic_{2i}, c_{2i+1}) (2\sigma^x)(-ic_{2i}, c_{2i+1})^T$. 
The Jordan-Wigner transformation of the horizontal couplings is more subtle, since we have to distinguish between two cases: $\beta +\tilde \gamma s > 0$ and $\beta +\tilde \gamma s < 0$. In the first case, we rewrite the horizontal couplings as
\begin{align}
    \tikz[baseline=-2.5pt]{
        \draw[black, thick] (-0.25, 0) -- (0.25, 0);
        \node (d) [circle, draw=black, fill=darkblue_malte, inner sep=1.7pt] {};
    } &= e^{\beta +\tilde \gamma s} \left( \mathbf{1} + e^{-2(\beta +\tilde \gamma s)} X \right) \nonumber \\
    &= e^{\beta +\tilde \gamma s} \left( \mathbf{1} + \tanh(\mathrm{KW}(\beta +\tilde \gamma s)) X \right) \nonumber\\
    &= \frac{e^{\beta +\tilde \gamma s}}{\cosh(\mathrm{KW}(\beta +\tilde \gamma s))} e^{\mathrm{KW}(\beta +\tilde \gamma s) X},
\end{align}
where $\mathrm{KW}(x) = \tanh^{-1}(e^{-2x})$ is the Kramers-Wannier dual of $x$. In this form the Jordan-Wigner transformation is straightforward, resulting in
\begin{equation}
    \tikz[baseline=-2.5pt]{
        \draw[black, thick] (-0.25, 0) -- (0.25, 0);
        \node (d) [circle, draw=black, fill=darkblue_malte, inner sep=1.7pt] {};
    } 
    \xrightarrow{\beta +\tilde \gamma s > 0}
    \frac{e^{\beta +\tilde \gamma s}}{\cosh(\mathrm{KW}(\beta +\tilde \gamma s))} e^{2\mathrm{KW}(\beta +\tilde \gamma s) \sigma^x_{2i-1, 2i}} \,.
\end{equation}
In the case $\beta +\tilde \gamma s < 0$, we rewrite the horizontal couplings as
\begin{align}
    \tikz[baseline=-2.5pt]{
        \draw[black, thick] (-0.25, 0) -- (0.25, 0);
        \node (d) [circle, draw=black, fill=darkblue_malte, inner sep=1.7pt] {};
    } &= e^{-(\beta +\tilde \gamma s)} \left( e^{2(\beta +\tilde \gamma s)} \mathbf{1} + X \right) \nonumber \\
    &= e^{-(\beta +\tilde \gamma s)} X \left( \mathbf{1} +  e^{2(\beta +\tilde \gamma s)} X \right) \nonumber \\
    &= e^{-(\beta +\tilde \gamma s)} X \left( \mathbf{1} + \tanh(\mathrm{KW}(-(\beta +\tilde \gamma s))) X \right) \nonumber \\
    &= \frac{e^{-(\beta +\tilde \gamma s) }X}{\cosh(\mathrm{KW}(-(\beta +\tilde \gamma s)))} e^{\mathrm{KW}(-(\beta +\tilde \gamma s)) X} \,.
\end{align}
Now the Jordan-Wigner transformation results in
\begin{equation}
    \tikz[baseline=-2.5pt]{
        \draw[black, thick] (-0.25, 0) -- (0.25, 0);
        \node (d) [circle, draw=black, fill=darkblue_malte, inner sep=1.7pt] {};
    } 
    \xrightarrow{\beta +\tilde \gamma s < 0}
    \frac{- \mathbf{1}_{2i-1, 2i} \, e^{-(\beta +\tilde \gamma s)}}{\cosh(\mathrm{KW}(-(\beta +\tilde \gamma s)))} e^{2\mathrm{KW}(-(\beta +\tilde \gamma s)) \sigma^x_{2i-1, 2i}} \,,
\end{equation}
where the $-\mathbf{1}_{2i-1, 2i}$ comes due to the additional $X$ operator in the original expression
\begin{equation}
    X = e^{\frac{i \pi}{2} X} \longrightarrow e^{i \pi \sigma^x_{2i-1, 2i}} = -\mathbf{1}_{2i-1, 2i} \,.
\end{equation}
We treat this factor, by `pulling' the Pauli $X$ through the imaginary time axis of the system from the past to the future, which results in a sign change for all vertical couplings adjacent to the horizontal coupling in question. This is illustrated in Fig.~\ref{fig:minus-one-propagation}. Using this method, we get rid of the $-\mathbf{1}_{2i-1, 2i}$ factors. 

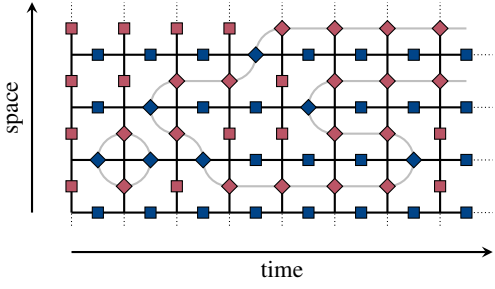
\begin{figure}[t]
    \centering
    \begin{tikzpicture}[>=stealth, scale=0.7]

        \def\nx{8}   
        \def\ny{4}   

        \foreach \x in {1,...,\nx} {
                \draw[black, thick] (\x,1) -- (\x,\ny + 0.5);
            }

        \foreach \y in {1, ..., \ny} {
                \draw[black, thick] (1,\y) -- (\nx + 0.5,\y);
            }

        \foreach \x in {1,...,\nx} {
                \draw[densely dotted, black] (\x,\ny + 0.5) -- (\x,\ny + 1);
            }

        \foreach \x in {1,...,\nx} {
                \draw[densely dotted, black] (\x,0.75) -- (\x,1);
            }
        \foreach \y in {1,...,\ny} {
                \draw[densely dotted, black] (\nx + 0.5,\y) -- (\nx + 1,\y);
            }
        \def\myxlist{{1, 3, 8, 1, 4, 5, 8, 1, 2, 5, 1, 2, 3, 4}}
        \def\myylist{{1, 1, 1, 2, 2, 2, 2, 3, 3, 3, 4, 4, 4, 4}}
        \foreach \i in {1,...,14} {
                \def\x{\myxlist[\i-1]}
                \def\y{\myylist[\i-1]}
                \node[rectangle, draw=black, fill=darkred_malte, inner sep=2.1pt] at (\x,\y + 0.5) {};
            }

        \def\myxlist{{1, 2, 3, 4, 5, 6, 7, 8, 4, 5, 6, 8, 1, 3, 4, 6, 7, 8, 1, 2, 3, 5, 6, 7, 8}}
        \def\myylist{{1, 1, 1, 1, 1, 1, 1, 1, 2, 2, 2, 2, 3, 3, 3, 3, 3, 3, 4, 4, 4, 4, 4, 4, 4}}
        \foreach \i in {1,...,25} {
                \def\x{\myxlist[\i-1]}
                \def\y{\myylist[\i-1]}
                \node[rectangle, draw=black, fill=darkblue_malte, inner sep=2.1pt] at (\x + 0.5,\y) {};
            }
        
        \draw[thick, color=lightgray, style=solid] (2 - 0.5, 2) 
        to[out=270, in=180] (2, 2 - 0.5) 
        to[out=0, in=270] (2 + 0.5, 2)
        to[out=90, in=0] (2, 2 + 0.5)
        to[out=180, in=90] (2 - 0.5, 2);

        \draw[thick, color=lightgray, style=solid] (8.5, 4.5)
        to[out=180, in=0] (5, 4.5)
        to[out=180, in=90] (4.5, 4)
        to[out=270, in=0] (4, 3.5)
        to[out=180, in=0] (3, 3.5)
        to[out=180, in=90](3 - 0.5, 3) 
        to[out=270, in=180] (3, 3 - 0.5) 
        to[out=0, in=90] (3 + 0.5, 2)
        to[out=270, in=180] (4, 1.5)
        to[out=0, in=180] (7, 1.5)
        to[out=0, in=270] (7.5, 2)
        to[out=90, in=0] (7, 2.5)
        to[out=180, in=0] (6, 2.5)
        to[out=180, in=270] (5.5, 3)
        to[out=90, in=180] (6, 3.5)
        to[out=0, in=180] (8.5, 3.5);

        \def\myxlist{{2, 4, 5, 6, 7, 2, 3, 6, 7, 3, 4, 6, 7, 8, 5, 6, 7, 8}}
        \def\myylist{{1, 1, 1, 1, 1, 2, 2, 2, 2, 3, 3, 3, 3, 3, 4, 4, 4, 4}}
        \foreach \i in {1,...,18} {
                \def\xx{\myxlist[\i-1]}
                \def\yy{\myylist[\i-1]}
                \node[rectangle, draw=black, fill=darkred_malte, inner sep=2.1pt, rotate=45] at (\xx, \yy + 0.5) {};  
        }

        \def\myxlist{{2, 3, 4, 8, 3, 6, 5}}
        \def\myylist{{2, 2, 2, 2, 3, 3, 4}}
        \foreach \i in {1,...,7} {
                \def\xx{\myxlist[\i-1]}
                \def\yy{\myylist[\i-1]}
                \node[rectangle, draw=black, fill=darkblue_malte, inner sep=2.1pt, rotate=45] at (\xx - 0.5, \yy) {};  
        }
        
        \draw[->, thick] (0.25,1) -- (0.25,\ny + 1) node[midway, above, rotate=90] {space};
        \draw[->, thick] (1,0.25) -- (\nx + 1,0.25) node[midway, below] {time};

    \end{tikzpicture}
    \caption{{\bf Propagating Pauli $X$ through the Ising circuit.}
    This figure illustrates how we treat the $X$ operator which appears in the Jordan-Wigner transformation of the horizontal couplings when $\beta +\tilde \gamma s < 0$ (blue diamonds). We `pull' the $X$ through the system until it annihilates with another $X$ or reaches the boundary of the system. This propagation of the $X_j$ flips the sign of all the vertical couplings, as $(Z_j Z_{j+1}) X_j = - X_j (Z_j Z_{j+1})$, such that the red squares along the path of the $X$ are flipped to red diamonds.
    }
    \label{fig:minus-one-propagation}
\end{figure}

The vertical coupling matrices are either
\begin{equation}
    \tikz[baseline=-2.5pt]{
        \draw[white, thick] (-0.25, 0) -- (0.25, 0);
        \draw[black, thick] (0, -0.25) -- (0, 0.25);
        \node (d) [rectangle, draw=black, fill=darkred_malte, inner sep=2.1pt] {};
    } = \begin{pmatrix}
        \cosh\left(2(\beta +\tilde \gamma s)\right) & \sinh\left(2(\beta +\tilde \gamma s)\right) \\
        \sinh\left(2(\beta +\tilde \gamma s)\right) & \cosh\left(2(\beta +\tilde \gamma s)\right)
    \end{pmatrix}
\end{equation}
or
\begin{equation}
    \tikz[baseline=-2.5pt]{
        \draw[white, thick] (-0.25, 0) -- (0.25, 0);
        \draw[black, thick] (0, -0.25) -- (0, 0.25);
        \node (d) [rectangle, draw=black, fill=darkred_malte, inner sep=2.1pt, rotate=45] {};
    } = \begin{pmatrix}
        \cosh\left(-2(\beta +\tilde \gamma s)\right) & \sinh\left(-2(\beta +\tilde \gamma s)\right) \\
        \sinh\left(-2(\beta +\tilde \gamma s)\right) & \cosh\left(-2(\beta +\tilde \gamma s)\right)
    \end{pmatrix},
\end{equation}
depending on the propagation of the $-\mathbf{1}_{2i-1, 2i}$ factors (if any).
The horizontal coupling matrices are
\begin{align}
    \tikz[baseline=-2.5pt]{
        \draw[black, thick] (-0.25, 0) -- (0.25, 0);
        \node (d) [rectangle, draw=black, fill=darkblue_malte, inner sep=2.1pt] {};
    } & =     \tikz[baseline=-2.5pt]{
        \draw[black, thick] (-0.25, 0) -- (0.25, 0);
        \node (d) [rectangle, draw=black, fill=darkblue_malte, inner sep=2.1pt, rotate=45] {};
    }                               \notag \\
      & =
    \begin{pmatrix}
        \cosh\left(2\mathrm{KW}|\beta +\tilde \gamma s|\right) & \sinh\left(2\mathrm{KW}|\beta +\tilde \gamma s|\right) \\
        \sinh\left(2\mathrm{KW}|\beta +\tilde \gamma s|\right) & \cosh\left(2\mathrm{KW}|\beta +\tilde \gamma s|\right)
    \end{pmatrix},
\end{align}
where we dropped the common prefactor $\frac{e^{|\beta +\tilde \gamma s|}}{\cosh(\mathrm{KW}(|\beta +\tilde \gamma s|))}$.

We sample the measurement outcomes $s$, using classical Monte Carlo methods, as explained in \cite{PutzGarrattNishimoriTrebstZhu}. Evolving the system in imaginary time, we iteratively re-orthogonalize the eigenvectors of the fermionized system after each time step using a QR decomposition~\footnote{factorization of a general matrix into a product of an orthogonal (or unitary) and an upper triangular matrix.}. 
The minus logarithm of the diagonal entries $-\ln|r_{ii}|$ of the $R$ matrix can be viewed as the single particle energy spectrum~\cite{DecoherenceWangEtAl}. By summing them over, we can then calculate the free energy density of the respective timestep $y$ as
\begin{equation}
    f_y = - \frac{\sum_i \log |r_{ii}|}{2 L_x} - \frac{f_y^\text{correction}}{L_x},
\end{equation}
where 
\begin{equation}
    f_y^\text{correction} = \left(\sum_{x=1}^{L_x} \log \cosh \left(\mathrm{KW} |\beta +\tilde \gamma s_{xy}^\text{h}|\right) - |\beta +\tilde \gamma s_{xy}^\text{h}| \right)
\end{equation}
is a correction term which we have to subtract from the free energy density, since we dropped the common prefactor $\frac{e^{|\beta +\tilde \gamma s|}}{\cosh(\mathrm{KW}(|\beta +\tilde \gamma s|))}$ in the horizontal coupling matrices. $s_{xy}^\text{h}$ is the measurement outcome of the horizontal coupling at position $(x,y)$.

\subsubsection*{Additional numerical data for the higher Nishimori line}

Let us close with providing supplementary numerical data, which we used for the identification of the emergent {\it higher} Nishimori line 
in the numerical phase diagram of Fig.~\ref{fig:numerical_higher_nishimori_line} with binary measurement outcomes.\\[2mm]

\begin{figure}[h!]
    \centering
    \includegraphics[width=0.9\linewidth]{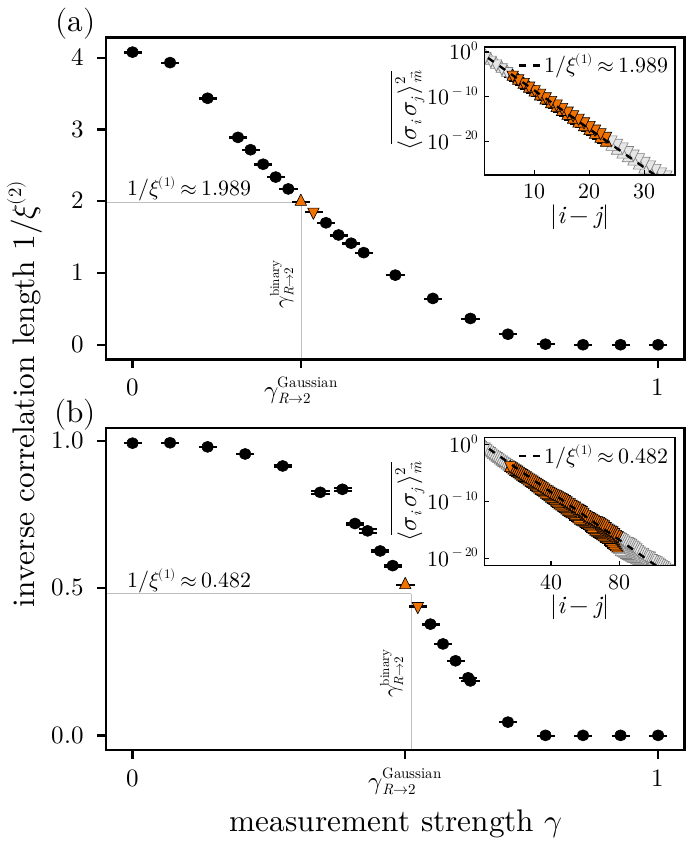}
    \caption{{\bf {\textit{Higher}} Nishimori condition for binary measurement outcomes.}
    	This figure is analogous to Fig.~\ref{fig:xi-betac-half}, but for cuts at $\beta =\beta_c/4$ (a) and $\beta =3\beta_c/4$ (b).
     }
    \label{fig:xi-additional-beta-cuts}
\end{figure}

\newpage

\bibliography{HigherNishimori}

\begin{thebibliography}{101}%
\makeatletter
\providecommand \@ifxundefined [1]{%
 \@ifx{#1\undefined}
}%
\providecommand \@ifnum [1]{%
 \ifnum #1\expandafter \@firstoftwo
 \else \expandafter \@secondoftwo
 \fi
}%
\providecommand \@ifx [1]{%
 \ifx #1\expandafter \@firstoftwo
 \else \expandafter \@secondoftwo
 \fi
}%
\providecommand \natexlab [1]{#1}%
\providecommand \enquote  [1]{``#1''}%
\providecommand \bibnamefont  [1]{#1}%
\providecommand \bibfnamefont [1]{#1}%
\providecommand \citenamefont [1]{#1}%
\providecommand \href@noop [0]{\@secondoftwo}%
\providecommand \href [0]{\begingroup \@sanitize@url \@href}%
\providecommand \@href[1]{\@@startlink{#1}\@@href}%
\providecommand \@@href[1]{\endgroup#1\@@endlink}%
\providecommand \@sanitize@url [0]{\catcode `\\12\catcode `\$12\catcode `\&12\catcode `\#12\catcode `\^12\catcode `\_12\catcode `\%12\relax}%
\providecommand \@@startlink[1]{}%
\providecommand \@@endlink[0]{}%
\providecommand \url  [0]{\begingroup\@sanitize@url \@url }%
\providecommand \@url [1]{\endgroup\@href {#1}{\urlprefix }}%
\providecommand \urlprefix  [0]{URL }%
\providecommand \Eprint [0]{\href }%
\providecommand \doibase [0]{https://doi.org/}%
\providecommand \selectlanguage [0]{\@gobble}%
\providecommand \bibinfo  [0]{\@secondoftwo}%
\providecommand \bibfield  [0]{\@secondoftwo}%
\providecommand \translation [1]{[#1]}%
\providecommand \BibitemOpen [0]{}%
\providecommand \bibitemStop [0]{}%
\providecommand \bibitemNoStop [0]{.\EOS\space}%
\providecommand \EOS [0]{\spacefactor3000\relax}%
\providecommand \BibitemShut  [1]{\csname bibitem#1\endcsname}%
\let\auto@bib@innerbib\@empty
\bibitem [{\citenamefont {Kitaev}(2003)}]{KITAEV20032}%
  \BibitemOpen
  \bibfield  {author} {\bibinfo {author} {\bibfnamefont {A.}~\bibnamefont {Kitaev}},\ }\bibfield  {title} {\bibinfo {title} {{Fault-tolerant quantum computation by anyons}},\ }\href {https://doi.org/https://doi.org/10.1016/S0003-4916(02)00018-0} {\bibfield  {journal} {\bibinfo  {journal} {Annals of Physics}\ }\textbf {\bibinfo {volume} {303}},\ \bibinfo {pages} {2} (\bibinfo {year} {2003})}\BibitemShut {NoStop}%
\bibitem [{\citenamefont {Dennis}\ \emph {et~al.}(2002)\citenamefont {Dennis}, \citenamefont {Kitaev}, \citenamefont {Landahl},\ and\ \citenamefont {Preskill}}]{DennisKitaevLandahlPreskill}%
  \BibitemOpen
  \bibfield  {author} {\bibinfo {author} {\bibfnamefont {E.}~\bibnamefont {Dennis}}, \bibinfo {author} {\bibfnamefont {A.}~\bibnamefont {Kitaev}}, \bibinfo {author} {\bibfnamefont {A.}~\bibnamefont {Landahl}},\ and\ \bibinfo {author} {\bibfnamefont {J.}~\bibnamefont {Preskill}},\ }\bibfield  {title} {\bibinfo {title} {{Topological quantum memory}},\ }\href {https://doi.org/10.1063/1.1499754} {\bibfield  {journal} {\bibinfo  {journal} {Journal of Mathematical Physics}\ }\textbf {\bibinfo {volume} {43}},\ \bibinfo {pages} {4452} (\bibinfo {year} {2002})}\BibitemShut {NoStop}%
\bibitem [{\citenamefont {Nishimori}(1980)}]{Nishimori_1980}%
  \BibitemOpen
  \bibfield  {author} {\bibinfo {author} {\bibfnamefont {H.}~\bibnamefont {Nishimori}},\ }\bibfield  {title} {\bibinfo {title} {{Exact results and critical properties of the Ising model with competing interactions}},\ }\href {https://doi.org/10.1088/0022-3719/13/21/012} {\bibfield  {journal} {\bibinfo  {journal} {Journal of Physics C: Solid State Physics}\ }\textbf {\bibinfo {volume} {13}},\ \bibinfo {pages} {4071} (\bibinfo {year} {1980})}\BibitemShut {NoStop}%
\bibitem [{\citenamefont {Nishimori}(1981)}]{Nishimori1981}%
  \BibitemOpen
  \bibfield  {author} {\bibinfo {author} {\bibfnamefont {H.}~\bibnamefont {Nishimori}},\ }\bibfield  {title} {\bibinfo {title} {{Internal Energy, Specific Heat and Correlation Function of the Bond-Random Ising Model}},\ }\href {https://doi.org/10.1143/PTP.66.1169} {\bibfield  {journal} {\bibinfo  {journal} {Progress of Theoretical Physics}\ }\textbf {\bibinfo {volume} {66}},\ \bibinfo {pages} {1169} (\bibinfo {year} {1981})}\BibitemShut {NoStop}%
\bibitem [{\citenamefont {Nishimori}(2001)}]{NishimoriBook}%
  \BibitemOpen
  \bibfield  {author} {\bibinfo {author} {\bibfnamefont {H.}~\bibnamefont {Nishimori}},\ }\href {https://doi.org/10.1093/acprof:oso/9780198509417.001.0001} {\emph {\bibinfo {title} {{Statistical Physics of Spin Glasses and Information Processing: An Introduction}}}}\ (\bibinfo  {publisher} {Oxford University Press},\ \bibinfo {year} {2001})\BibitemShut {NoStop}%
\bibitem [{\citenamefont {Zhu}\ \emph {et~al.}(2023)\citenamefont {Zhu}, \citenamefont {Tantivasadakarn}, \citenamefont {Vishwanath}, \citenamefont {Trebst},\ and\ \citenamefont {Verresen}}]{ZhuTantivasadakarnVishwanthTrebstVerresen}%
  \BibitemOpen
  \bibfield  {author} {\bibinfo {author} {\bibfnamefont {G.-Y.}\ \bibnamefont {Zhu}}, \bibinfo {author} {\bibfnamefont {N.}~\bibnamefont {Tantivasadakarn}}, \bibinfo {author} {\bibfnamefont {A.}~\bibnamefont {Vishwanath}}, \bibinfo {author} {\bibfnamefont {S.}~\bibnamefont {Trebst}},\ and\ \bibinfo {author} {\bibfnamefont {R.}~\bibnamefont {Verresen}},\ }\bibfield  {title} {\bibinfo {title} {{Nishimori's Cat: Stable Long-Range Entanglement from Finite-Depth Unitaries and Weak Measurements}},\ }\href {https://doi.org/10.1103/PhysRevLett.131.200201} {\bibfield  {journal} {\bibinfo  {journal} {Phys. Rev. Lett.}\ }\textbf {\bibinfo {volume} {131}},\ \bibinfo {pages} {200201} (\bibinfo {year} {2023})}\BibitemShut {NoStop}%
\bibitem [{\citenamefont {Lee}\ \emph {et~al.}(2022)\citenamefont {Lee}, \citenamefont {Ji}, \citenamefont {Bi},\ and\ \citenamefont {Fisher}}]{LeeJiFisherBi}%
  \BibitemOpen
  \bibfield  {author} {\bibinfo {author} {\bibfnamefont {J.~Y.}\ \bibnamefont {Lee}}, \bibinfo {author} {\bibfnamefont {W.}~\bibnamefont {Ji}}, \bibinfo {author} {\bibfnamefont {Z.}~\bibnamefont {Bi}},\ and\ \bibinfo {author} {\bibfnamefont {M.~P.~A.}\ \bibnamefont {Fisher}},\ }\href {https://arxiv.org/abs/2208.11699} {\bibinfo {title} {{Decoding Measurement-Prepared Quantum Phases and Transitions: from Ising model to gauge theory, and beyond}}} (\bibinfo {year} {2022}),\ \Eprint {https://arxiv.org/abs/2208.11699} {arXiv:2208.11699 [cond-mat.str-el]} \BibitemShut {NoStop}%
\bibitem [{\citenamefont {Chen}\ \emph {et~al.}(2025)\citenamefont {Chen}, \citenamefont {Zhu}, \citenamefont {Verresen}, \citenamefont {Seif}, \citenamefont {B{\"a}umer}, \citenamefont {Layden}, \citenamefont {Tantivasadakarn}, \citenamefont {Zhu}, \citenamefont {Sheldon}, \citenamefont {Vishwanath}, \citenamefont {Trebst},\ and\ \citenamefont {Kandala}}]{Chen2025}%
  \BibitemOpen
  \bibfield  {author} {\bibinfo {author} {\bibfnamefont {E.~H.}\ \bibnamefont {Chen}}, \bibinfo {author} {\bibfnamefont {G.-Y.}\ \bibnamefont {Zhu}}, \bibinfo {author} {\bibfnamefont {R.}~\bibnamefont {Verresen}}, \bibinfo {author} {\bibfnamefont {A.}~\bibnamefont {Seif}}, \bibinfo {author} {\bibfnamefont {E.}~\bibnamefont {B{\"a}umer}}, \bibinfo {author} {\bibfnamefont {D.}~\bibnamefont {Layden}}, \bibinfo {author} {\bibfnamefont {N.}~\bibnamefont {Tantivasadakarn}}, \bibinfo {author} {\bibfnamefont {G.}~\bibnamefont {Zhu}}, \bibinfo {author} {\bibfnamefont {S.}~\bibnamefont {Sheldon}}, \bibinfo {author} {\bibfnamefont {A.}~\bibnamefont {Vishwanath}}, \bibinfo {author} {\bibfnamefont {S.}~\bibnamefont {Trebst}},\ and\ \bibinfo {author} {\bibfnamefont {A.}~\bibnamefont {Kandala}},\ }\bibfield  {title} {\bibinfo {title} {{Nishimori transition across the error threshold for constant-depth quantum circuits}},\ }\href {https://doi.org/10.1038/s41567-024-02696-6} {\bibfield  {journal} {\bibinfo  {journal} {Nature
  Physics}\ }\textbf {\bibinfo {volume} {21}},\ \bibinfo {pages} {161} (\bibinfo {year} {2025})}\BibitemShut {NoStop}%
\bibitem [{\citenamefont {Fan}\ \emph {et~al.}(2024)\citenamefont {Fan}, \citenamefont {Bao}, \citenamefont {Altman},\ and\ \citenamefont {Vishwanath}}]{FanBaoAltmanVishwanath}%
  \BibitemOpen
  \bibfield  {author} {\bibinfo {author} {\bibfnamefont {R.}~\bibnamefont {Fan}}, \bibinfo {author} {\bibfnamefont {Y.}~\bibnamefont {Bao}}, \bibinfo {author} {\bibfnamefont {E.}~\bibnamefont {Altman}},\ and\ \bibinfo {author} {\bibfnamefont {A.}~\bibnamefont {Vishwanath}},\ }\bibfield  {title} {\bibinfo {title} {{Diagnostics of Mixed-State Topological Order and Breakdown of Quantum Memory}},\ }\href {https://doi.org/10.1103/PRXQuantum.5.020343} {\bibfield  {journal} {\bibinfo  {journal} {PRX Quantum}\ }\textbf {\bibinfo {volume} {5}},\ \bibinfo {pages} {020343} (\bibinfo {year} {2024})}\BibitemShut {NoStop}%
\bibitem [{\citenamefont {Bao}\ \emph {et~al.}(2023)\citenamefont {Bao}, \citenamefont {Fan}, \citenamefont {Vishwanath},\ and\ \citenamefont {Altman}}]{BaoFanVishwanathAltman}%
  \BibitemOpen
  \bibfield  {author} {\bibinfo {author} {\bibfnamefont {Y.}~\bibnamefont {Bao}}, \bibinfo {author} {\bibfnamefont {R.}~\bibnamefont {Fan}}, \bibinfo {author} {\bibfnamefont {A.}~\bibnamefont {Vishwanath}},\ and\ \bibinfo {author} {\bibfnamefont {E.}~\bibnamefont {Altman}},\ }\href {https://arxiv.org/abs/2301.05687} {\bibinfo {title} {{Mixed-state topological order and the errorfield double formulation of decoherence-induced transitions}}} (\bibinfo {year} {2023}),\ \Eprint {https://arxiv.org/abs/2301.05687} {arXiv:2301.05687 [quant-ph]} \BibitemShut {NoStop}%
\bibitem [{\citenamefont {Lee}\ \emph {et~al.}(2023)\citenamefont {Lee}, \citenamefont {Jian},\ and\ \citenamefont {Xu}}]{LeeJianXu}%
  \BibitemOpen
  \bibfield  {author} {\bibinfo {author} {\bibfnamefont {J.~Y.}\ \bibnamefont {Lee}}, \bibinfo {author} {\bibfnamefont {C.-M.}\ \bibnamefont {Jian}},\ and\ \bibinfo {author} {\bibfnamefont {C.}~\bibnamefont {Xu}},\ }\bibfield  {title} {\bibinfo {title} {{Quantum Criticality Under Decoherence or Weak Measurement}},\ }\href {https://doi.org/10.1103/PRXQuantum.4.030317} {\bibfield  {journal} {\bibinfo  {journal} {PRX Quantum}\ }\textbf {\bibinfo {volume} {4}},\ \bibinfo {pages} {030317} (\bibinfo {year} {2023})}\BibitemShut {NoStop}%
\bibitem [{\citenamefont {Sala}\ \emph {et~al.}(2024)\citenamefont {Sala}, \citenamefont {Gopalakrishnan}, \citenamefont {Oshikawa},\ and\ \citenamefont {You}}]{SalaGopalakrishnanOshikawaYou}%
  \BibitemOpen
  \bibfield  {author} {\bibinfo {author} {\bibfnamefont {P.}~\bibnamefont {Sala}}, \bibinfo {author} {\bibfnamefont {S.}~\bibnamefont {Gopalakrishnan}}, \bibinfo {author} {\bibfnamefont {M.}~\bibnamefont {Oshikawa}},\ and\ \bibinfo {author} {\bibfnamefont {Y.}~\bibnamefont {You}},\ }\bibfield  {title} {\bibinfo {title} {{Spontaneous strong symmetry breaking in open systems: Purification perspective}},\ }\href {https://doi.org/10.1103/PhysRevB.110.155150} {\bibfield  {journal} {\bibinfo  {journal} {Phys. Rev. B}\ }\textbf {\bibinfo {volume} {110}},\ \bibinfo {pages} {155150} (\bibinfo {year} {2024})}\BibitemShut {NoStop}%
\bibitem [{\citenamefont {Ellison}\ and\ \citenamefont {Cheng}(2025)}]{EllisonCheng}%
  \BibitemOpen
  \bibfield  {author} {\bibinfo {author} {\bibfnamefont {T.~D.}\ \bibnamefont {Ellison}}\ and\ \bibinfo {author} {\bibfnamefont {M.}~\bibnamefont {Cheng}},\ }\bibfield  {title} {\bibinfo {title} {{Toward a Classification of Mixed-State Topological Orders in Two Dimensions}},\ }\href {https://doi.org/10.1103/PRXQuantum.6.010315} {\bibfield  {journal} {\bibinfo  {journal} {PRX Quantum}\ }\textbf {\bibinfo {volume} {6}},\ \bibinfo {pages} {010315} (\bibinfo {year} {2025})}\BibitemShut {NoStop}%
\bibitem [{\citenamefont {Wang}\ \emph {et~al.}(2025{\natexlab{a}})\citenamefont {Wang}, \citenamefont {Wu},\ and\ \citenamefont {Wang}}]{WangWuWang}%
  \BibitemOpen
  \bibfield  {author} {\bibinfo {author} {\bibfnamefont {Z.}~\bibnamefont {Wang}}, \bibinfo {author} {\bibfnamefont {Z.}~\bibnamefont {Wu}},\ and\ \bibinfo {author} {\bibfnamefont {Z.}~\bibnamefont {Wang}},\ }\bibfield  {title} {\bibinfo {title} {{Intrinsic Mixed-State Topological Order}},\ }\href {https://doi.org/10.1103/PRXQuantum.6.010314} {\bibfield  {journal} {\bibinfo  {journal} {PRX Quantum}\ }\textbf {\bibinfo {volume} {6}},\ \bibinfo {pages} {010314} (\bibinfo {year} {2025}{\natexlab{a}})}\BibitemShut {NoStop}%
\bibitem [{\citenamefont {Chen}\ and\ \citenamefont {Grover}(2024{\natexlab{a}})}]{YHChenGrover}%
  \BibitemOpen
  \bibfield  {author} {\bibinfo {author} {\bibfnamefont {Y.-H.}\ \bibnamefont {Chen}}\ and\ \bibinfo {author} {\bibfnamefont {T.}~\bibnamefont {Grover}},\ }\bibfield  {title} {\bibinfo {title} {{Unconventional topological mixed-state transition and critical phase induced by self-dual coherent errors}},\ }\href {https://doi.org/10.1103/PhysRevB.110.125152} {\bibfield  {journal} {\bibinfo  {journal} {Phys. Rev. B}\ }\textbf {\bibinfo {volume} {110}},\ \bibinfo {pages} {125152} (\bibinfo {year} {2024}{\natexlab{a}})}\BibitemShut {NoStop}%
\bibitem [{\citenamefont {Chen}\ and\ \citenamefont {Grover}(2024{\natexlab{b}})}]{YHChenGroverSeparability}%
  \BibitemOpen
  \bibfield  {author} {\bibinfo {author} {\bibfnamefont {Y.-H.}\ \bibnamefont {Chen}}\ and\ \bibinfo {author} {\bibfnamefont {T.}~\bibnamefont {Grover}},\ }\bibfield  {title} {\bibinfo {title} {{Separability Transitions in Topological States Induced by Local Decoherence}},\ }\href {https://doi.org/10.1103/PhysRevLett.132.170602} {\bibfield  {journal} {\bibinfo  {journal} {Phys. Rev. Lett.}\ }\textbf {\bibinfo {volume} {132}},\ \bibinfo {pages} {170602} (\bibinfo {year} {2024}{\natexlab{b}})}\BibitemShut {NoStop}%
\bibitem [{\citenamefont {Lessa}\ \emph {et~al.}(2025)\citenamefont {Lessa}, \citenamefont {Ma}, \citenamefont {Zhang}, \citenamefont {Bi}, \citenamefont {Cheng},\ and\ \citenamefont {Wang}}]{LessaStrongToWeak}%
  \BibitemOpen
  \bibfield  {author} {\bibinfo {author} {\bibfnamefont {L.~A.}\ \bibnamefont {Lessa}}, \bibinfo {author} {\bibfnamefont {R.}~\bibnamefont {Ma}}, \bibinfo {author} {\bibfnamefont {J.-H.}\ \bibnamefont {Zhang}}, \bibinfo {author} {\bibfnamefont {Z.}~\bibnamefont {Bi}}, \bibinfo {author} {\bibfnamefont {M.}~\bibnamefont {Cheng}},\ and\ \bibinfo {author} {\bibfnamefont {C.}~\bibnamefont {Wang}},\ }\bibfield  {title} {\bibinfo {title} {{Strong-to-Weak Spontaneous Symmetry Breaking in Mixed Quantum States}},\ }\href {https://doi.org/10.1103/PRXQuantum.6.010344} {\bibfield  {journal} {\bibinfo  {journal} {PRX Quantum}\ }\textbf {\bibinfo {volume} {6}},\ \bibinfo {pages} {010344} (\bibinfo {year} {2025})}\BibitemShut {NoStop}%
\bibitem [{\citenamefont {Sohal}\ and\ \citenamefont {Prem}(2025)}]{SohalPrem}%
  \BibitemOpen
  \bibfield  {author} {\bibinfo {author} {\bibfnamefont {R.}~\bibnamefont {Sohal}}\ and\ \bibinfo {author} {\bibfnamefont {A.}~\bibnamefont {Prem}},\ }\bibfield  {title} {\bibinfo {title} {{Noisy Approach to Intrinsically Mixed-State Topological Order}},\ }\href {https://doi.org/10.1103/PRXQuantum.6.010313} {\bibfield  {journal} {\bibinfo  {journal} {PRX Quantum}\ }\textbf {\bibinfo {volume} {6}},\ \bibinfo {pages} {010313} (\bibinfo {year} {2025})}\BibitemShut {NoStop}%
\bibitem [{\citenamefont {Su}\ \emph {et~al.}(2024)\citenamefont {Su}, \citenamefont {Yang},\ and\ \citenamefont {Jian}}]{SuYangJian}%
  \BibitemOpen
  \bibfield  {author} {\bibinfo {author} {\bibfnamefont {K.}~\bibnamefont {Su}}, \bibinfo {author} {\bibfnamefont {Z.}~\bibnamefont {Yang}},\ and\ \bibinfo {author} {\bibfnamefont {C.-M.}\ \bibnamefont {Jian}},\ }\bibfield  {title} {\bibinfo {title} {{Tapestry of dualities in decohered quantum error correction codes}},\ }\href {https://doi.org/10.1103/PhysRevB.110.085158} {\bibfield  {journal} {\bibinfo  {journal} {Phys. Rev. B}\ }\textbf {\bibinfo {volume} {110}},\ \bibinfo {pages} {085158} (\bibinfo {year} {2024})}\BibitemShut {NoStop}%
\bibitem [{\citenamefont {Lee}(2025)}]{LeeExactCalculation}%
  \BibitemOpen
  \bibfield  {author} {\bibinfo {author} {\bibfnamefont {J.~Y.}\ \bibnamefont {Lee}},\ }\bibfield  {title} {\bibinfo {title} {{Exact Calculations of Coherent Information for Toric Codes under Decoherence: Identifying the Fundamental Error Threshold}},\ }\href {https://doi.org/10.1103/hlfh-86yz} {\bibfield  {journal} {\bibinfo  {journal} {Phys. Rev. Lett.}\ }\textbf {\bibinfo {volume} {134}},\ \bibinfo {pages} {250601} (\bibinfo {year} {2025})}\BibitemShut {NoStop}%
\bibitem [{\citenamefont {Eckstein}\ \emph {et~al.}(2024)\citenamefont {Eckstein}, \citenamefont {Han}, \citenamefont {Trebst},\ and\ \citenamefont {Zhu}}]{EcksteinPRX}%
  \BibitemOpen
  \bibfield  {author} {\bibinfo {author} {\bibfnamefont {F.}~\bibnamefont {Eckstein}}, \bibinfo {author} {\bibfnamefont {B.}~\bibnamefont {Han}}, \bibinfo {author} {\bibfnamefont {S.}~\bibnamefont {Trebst}},\ and\ \bibinfo {author} {\bibfnamefont {G.-Y.}\ \bibnamefont {Zhu}},\ }\bibfield  {title} {\bibinfo {title} {{Robust Teleportation of a Surface Code and Cascade of Topological Quantum Phase Transitions}},\ }\href {https://doi.org/10.1103/PRXQuantum.5.040313} {\bibfield  {journal} {\bibinfo  {journal} {PRX Quantum}\ }\textbf {\bibinfo {volume} {5}},\ \bibinfo {pages} {040313} (\bibinfo {year} {2024})}\BibitemShut {NoStop}%
\bibitem [{\citenamefont {Wang}\ \emph {et~al.}(2025{\natexlab{b}})\citenamefont {Wang}, \citenamefont {Vasseur}, \citenamefont {Trebst}, \citenamefont {Ludwig},\ and\ \citenamefont {Zhu}}]{DecoherenceWangEtAl}%
  \BibitemOpen
  \bibfield  {author} {\bibinfo {author} {\bibfnamefont {Q.}~\bibnamefont {Wang}}, \bibinfo {author} {\bibfnamefont {R.}~\bibnamefont {Vasseur}}, \bibinfo {author} {\bibfnamefont {S.}~\bibnamefont {Trebst}}, \bibinfo {author} {\bibfnamefont {A.~W.~W.}\ \bibnamefont {Ludwig}},\ and\ \bibinfo {author} {\bibfnamefont {G.-Y.}\ \bibnamefont {Zhu}},\ }\href@noop {} {\bibinfo {title} {{Decoherence-induced self-dual criticality in topological states of matter}}} (\bibinfo {year} {2025}{\natexlab{b}}),\ \Eprint {https://arxiv.org/abs/2502.14034} {arXiv:2502.14034 [quant-ph]} \BibitemShut {NoStop}%
\bibitem [{\citenamefont {Eckstein}\ \emph {et~al.}(2025)\citenamefont {Eckstein}, \citenamefont {Han}, \citenamefont {Trebst},\ and\ \citenamefont {Zhu}}]{eckstein2025learningtransitionstopologicalsurface}%
  \BibitemOpen
  \bibfield  {author} {\bibinfo {author} {\bibfnamefont {F.}~\bibnamefont {Eckstein}}, \bibinfo {author} {\bibfnamefont {B.}~\bibnamefont {Han}}, \bibinfo {author} {\bibfnamefont {S.}~\bibnamefont {Trebst}},\ and\ \bibinfo {author} {\bibfnamefont {G.-Y.}\ \bibnamefont {Zhu}},\ }\href {https://arxiv.org/abs/2512.19786} {\bibinfo {title} {{Learning transitions of topological surface codes}}} (\bibinfo {year} {2025}),\ \Eprint {https://arxiv.org/abs/2512.19786} {arXiv:2512.19786 [quant-ph]} \BibitemShut {NoStop}%
\bibitem [{\citenamefont {Iba}(1999)}]{Iba_1999}%
  \BibitemOpen
  \bibfield  {author} {\bibinfo {author} {\bibfnamefont {Y.}~\bibnamefont {Iba}},\ }\bibfield  {title} {\bibinfo {title} {{The Nishimori line and Bayesian statistics}},\ }\href {https://doi.org/10.1088/0305-4470/32/21/302} {\bibfield  {journal} {\bibinfo  {journal} {Journal of Physics A: Mathematical and General}\ }\textbf {\bibinfo {volume} {32}},\ \bibinfo {pages} {3875} (\bibinfo {year} {1999})}\BibitemShut {NoStop}%
\bibitem [{\citenamefont {P\"utz}\ \emph {et~al.}(2025{\natexlab{a}})\citenamefont {P\"utz}, \citenamefont {Garratt}, \citenamefont {Nishimori}, \citenamefont {Trebst},\ and\ \citenamefont {Zhu}}]{PutzGarrattNishimoriTrebstZhu}%
  \BibitemOpen
  \bibfield  {author} {\bibinfo {author} {\bibfnamefont {M.}~\bibnamefont {P\"utz}}, \bibinfo {author} {\bibfnamefont {S.~J.}\ \bibnamefont {Garratt}}, \bibinfo {author} {\bibfnamefont {H.}~\bibnamefont {Nishimori}}, \bibinfo {author} {\bibfnamefont {S.}~\bibnamefont {Trebst}},\ and\ \bibinfo {author} {\bibfnamefont {G.-Y.}\ \bibnamefont {Zhu}},\ }\href@noop {} {\bibinfo {title} {{Learning transitions in classical Ising models and deformed toric codes}}} (\bibinfo {year} {2025}{\natexlab{a}}),\ \Eprint {https://arxiv.org/abs/2504.12385} {arXiv:2504.12385 [cond-mat.stat-mech]} \BibitemShut {NoStop}%
\bibitem [{\citenamefont {Castelnovo}\ and\ \citenamefont {Chamon}(2008)}]{CastelnovoChamon}%
  \BibitemOpen
  \bibfield  {author} {\bibinfo {author} {\bibfnamefont {C.}~\bibnamefont {Castelnovo}}\ and\ \bibinfo {author} {\bibfnamefont {C.}~\bibnamefont {Chamon}},\ }\bibfield  {title} {\bibinfo {title} {{Quantum topological phase transition at the microscopic level}},\ }\href {https://doi.org/10.1103/PhysRevB.77.054433} {\bibfield  {journal} {\bibinfo  {journal} {Phys. Rev. B}\ }\textbf {\bibinfo {volume} {77}},\ \bibinfo {pages} {054433} (\bibinfo {year} {2008})}\BibitemShut {NoStop}%
\bibitem [{\citenamefont {Papanikolaou}\ \emph {et~al.}(2007)\citenamefont {Papanikolaou}, \citenamefont {Raman},\ and\ \citenamefont {Fradkin}}]{PapanikolaouRamanFradkin}%
  \BibitemOpen
  \bibfield  {author} {\bibinfo {author} {\bibfnamefont {S.}~\bibnamefont {Papanikolaou}}, \bibinfo {author} {\bibfnamefont {K.~S.}\ \bibnamefont {Raman}},\ and\ \bibinfo {author} {\bibfnamefont {E.}~\bibnamefont {Fradkin}},\ }\bibfield  {title} {\bibinfo {title} {{Topological phases and topological entropy of two-dimensional systems with finite correlation length}},\ }\href {https://doi.org/10.1103/PhysRevB.76.224421} {\bibfield  {journal} {\bibinfo  {journal} {Phys. Rev. B}\ }\textbf {\bibinfo {volume} {76}},\ \bibinfo {pages} {224421} (\bibinfo {year} {2007})}\BibitemShut {NoStop}%
\bibitem [{\citenamefont {Ardonne}\ \emph {et~al.}(2004)\citenamefont {Ardonne}, \citenamefont {Fendley},\ and\ \citenamefont {Fradkin}}]{ArdonneFedleyFradkin}%
  \BibitemOpen
  \bibfield  {author} {\bibinfo {author} {\bibfnamefont {E.}~\bibnamefont {Ardonne}}, \bibinfo {author} {\bibfnamefont {P.}~\bibnamefont {Fendley}},\ and\ \bibinfo {author} {\bibfnamefont {E.}~\bibnamefont {Fradkin}},\ }\bibfield  {title} {\bibinfo {title} {{Topological order and conformal quantum critical points}},\ }\href {https://doi.org/https://doi.org/10.1016/j.aop.2004.01.004} {\bibfield  {journal} {\bibinfo  {journal} {Annals of Physics}\ }\textbf {\bibinfo {volume} {310}},\ \bibinfo {pages} {493} (\bibinfo {year} {2004})}\BibitemShut {NoStop}%
\bibitem [{\citenamefont {Isakov}\ \emph {et~al.}(2011)\citenamefont {Isakov}, \citenamefont {Fendley}, \citenamefont {Ludwig}, \citenamefont {Trebst},\ and\ \citenamefont {Troyer}}]{IsakovFendleyLudwigTrebstTroyer}%
  \BibitemOpen
  \bibfield  {author} {\bibinfo {author} {\bibfnamefont {S.~V.}\ \bibnamefont {Isakov}}, \bibinfo {author} {\bibfnamefont {P.}~\bibnamefont {Fendley}}, \bibinfo {author} {\bibfnamefont {A.~W.~W.}\ \bibnamefont {Ludwig}}, \bibinfo {author} {\bibfnamefont {S.}~\bibnamefont {Trebst}},\ and\ \bibinfo {author} {\bibfnamefont {M.}~\bibnamefont {Troyer}},\ }\bibfield  {title} {\bibinfo {title} {{Dynamics at and near conformal quantum critical points}},\ }\href {https://doi.org/10.1103/PhysRevB.83.125114} {\bibfield  {journal} {\bibinfo  {journal} {Phys. Rev. B}\ }\textbf {\bibinfo {volume} {83}},\ \bibinfo {pages} {125114} (\bibinfo {year} {2011})}\BibitemShut {NoStop}%
\bibitem [{\citenamefont {Zhu}\ and\ \citenamefont {Zhang}(2019)}]{Zhu19deform}%
  \BibitemOpen
  \bibfield  {author} {\bibinfo {author} {\bibfnamefont {G.-Y.}\ \bibnamefont {Zhu}}\ and\ \bibinfo {author} {\bibfnamefont {G.-M.}\ \bibnamefont {Zhang}},\ }\bibfield  {title} {\bibinfo {title} {{Gapless Coulomb State Emerging from a Self-Dual Topological Tensor-Network State}},\ }\href {https://doi.org/10.1103/PhysRevLett.122.176401} {\bibfield  {journal} {\bibinfo  {journal} {Phys. Rev. Lett.}\ }\textbf {\bibinfo {volume} {122}},\ \bibinfo {pages} {176401} (\bibinfo {year} {2019})}\BibitemShut {NoStop}%
\bibitem [{\citenamefont {Huxford}\ \emph {et~al.}(2023)\citenamefont {Huxford}, \citenamefont {Nguyen},\ and\ \citenamefont {Kim}}]{Kim23deform}%
  \BibitemOpen
  \bibfield  {author} {\bibinfo {author} {\bibfnamefont {J.}~\bibnamefont {Huxford}}, \bibinfo {author} {\bibfnamefont {D.~X.}\ \bibnamefont {Nguyen}},\ and\ \bibinfo {author} {\bibfnamefont {Y.~B.}\ \bibnamefont {Kim}},\ }\bibfield  {title} {\bibinfo {title} {{Gaining insights on anyon condensation and 1-form symmetry breaking across a topological phase transition in a deformed toric code model}},\ }\href {https://doi.org/10.21468/SciPostPhys.15.6.253} {\bibfield  {journal} {\bibinfo  {journal} {SciPost Phys.}\ }\textbf {\bibinfo {volume} {15}},\ \bibinfo {pages} {253} (\bibinfo {year} {2023})}\BibitemShut {NoStop}%
\bibitem [{\citenamefont {{Sahay}}\ \emph {et~al.}()\citenamefont {{Sahay}}, \citenamefont {{von Keyserlingk}}, \citenamefont {{Verresen}},\ and\ \citenamefont {{Zhang}}}]{Verresen25deform}%
  \BibitemOpen
  \bibfield  {author} {\bibinfo {author} {\bibfnamefont {R.}~\bibnamefont {{Sahay}}}, \bibinfo {author} {\bibfnamefont {C.}~\bibnamefont {{von Keyserlingk}}}, \bibinfo {author} {\bibfnamefont {R.}~\bibnamefont {{Verresen}}},\ and\ \bibinfo {author} {\bibfnamefont {C.}~\bibnamefont {{Zhang}}},\ }\href@noop {} {\bibinfo {title} {{Enforced Gaplessness from States with Exponentially Decaying Correlations}}},\ \Eprint {https://arxiv.org/abs/2503.01977} {2503.01977} \BibitemShut {NoStop}%
\bibitem [{\citenamefont {Nahum}\ and\ \citenamefont {Jacobsen}(2025)}]{NahumJacobsen}%
  \BibitemOpen
  \bibfield  {author} {\bibinfo {author} {\bibfnamefont {A.}~\bibnamefont {Nahum}}\ and\ \bibinfo {author} {\bibfnamefont {J.~L.}\ \bibnamefont {Jacobsen}},\ }\href {https://arxiv.org/abs/2504.01264} {\bibinfo {title} {{Bayesian critical points in classical lattice models}}} (\bibinfo {year} {2025}),\ \Eprint {https://arxiv.org/abs/2504.01264} {arXiv:2504.01264 [cond-mat.stat-mech]} \BibitemShut {NoStop}%
\bibitem [{\citenamefont {Hauser}\ \emph {et~al.}(2026)\citenamefont {Hauser}, \citenamefont {Bao}, \citenamefont {Sang}, \citenamefont {Lavasani}, \citenamefont {Agrawal},\ and\ \citenamefont {Fisher}}]{hauser2024informationdynamicsdecoheredquantum}%
  \BibitemOpen
  \bibfield  {author} {\bibinfo {author} {\bibfnamefont {J.}~\bibnamefont {Hauser}}, \bibinfo {author} {\bibfnamefont {Y.}~\bibnamefont {Bao}}, \bibinfo {author} {\bibfnamefont {S.}~\bibnamefont {Sang}}, \bibinfo {author} {\bibfnamefont {A.}~\bibnamefont {Lavasani}}, \bibinfo {author} {\bibfnamefont {U.}~\bibnamefont {Agrawal}},\ and\ \bibinfo {author} {\bibfnamefont {M.~P.~A.}\ \bibnamefont {Fisher}},\ }\bibfield  {title} {\bibinfo {title} {{Information dynamics in decohered quantum memory with repeated syndrome measurements}},\ }\href {https://doi.org/10.1103/v5kq-7mn7} {\bibfield  {journal} {\bibinfo  {journal} {Phys. Rev. B}\ }\textbf {\bibinfo {volume} {113}},\ \bibinfo {pages} {054303} (\bibinfo {year} {2026})}\BibitemShut {NoStop}%
\bibitem [{\citenamefont {Hauser}\ \emph {et~al.}(2025)\citenamefont {Hauser}, \citenamefont {Lavasani}, \citenamefont {Vijay},\ and\ \citenamefont {Fisher}}]{hauser2025informationdynamicssymmetrybreaking}%
  \BibitemOpen
  \bibfield  {author} {\bibinfo {author} {\bibfnamefont {J.}~\bibnamefont {Hauser}}, \bibinfo {author} {\bibfnamefont {A.}~\bibnamefont {Lavasani}}, \bibinfo {author} {\bibfnamefont {S.}~\bibnamefont {Vijay}},\ and\ \bibinfo {author} {\bibfnamefont {M.~P.~A.}\ \bibnamefont {Fisher}},\ }\href {https://arxiv.org/abs/2512.03031} {\bibinfo {title} {{Information dynamics and symmetry breaking in generic monitored $\mathbb{Z}_2$-symmetric open quantum systems}}} (\bibinfo {year} {2025}),\ \Eprint {https://arxiv.org/abs/2512.03031} {arXiv:2512.03031 [quant-ph]} \BibitemShut {NoStop}%
\bibitem [{\citenamefont {Wan}\ \emph {et~al.}(2025)\citenamefont {Wan}, \citenamefont {Dai},\ and\ \citenamefont {Zhu}}]{wan2025revisitingnishimorimulticriticalitylens}%
  \BibitemOpen
  \bibfield  {author} {\bibinfo {author} {\bibfnamefont {Z.-Q.}\ \bibnamefont {Wan}}, \bibinfo {author} {\bibfnamefont {X.-D.}\ \bibnamefont {Dai}},\ and\ \bibinfo {author} {\bibfnamefont {G.-Y.}\ \bibnamefont {Zhu}},\ }\href {https://arxiv.org/abs/2511.02907} {\bibinfo {title} {{Revisiting Nishimori multicriticality through the lens of information measures}}} (\bibinfo {year} {2025}),\ \Eprint {https://arxiv.org/abs/2511.02907} {arXiv:2511.02907 [cond-mat.stat-mech]} \BibitemShut {NoStop}%
\bibitem [{\citenamefont {Colmenarez}\ \emph {et~al.}(2024)\citenamefont {Colmenarez}, \citenamefont {Huang}, \citenamefont {Diehl},\ and\ \citenamefont {M\"uller}}]{MullerEtAll2024}%
  \BibitemOpen
  \bibfield  {author} {\bibinfo {author} {\bibfnamefont {L.}~\bibnamefont {Colmenarez}}, \bibinfo {author} {\bibfnamefont {Z.-M.}\ \bibnamefont {Huang}}, \bibinfo {author} {\bibfnamefont {S.}~\bibnamefont {Diehl}},\ and\ \bibinfo {author} {\bibfnamefont {M.}~\bibnamefont {M\"uller}},\ }\bibfield  {title} {\bibinfo {title} {{Accurate optimal quantum error correction thresholds from coherent information}},\ }\href {https://doi.org/10.1103/PhysRevResearch.6.L042014} {\bibfield  {journal} {\bibinfo  {journal} {Phys. Rev. Res.}\ }\textbf {\bibinfo {volume} {6}},\ \bibinfo {pages} {L042014} (\bibinfo {year} {2024})}\BibitemShut {NoStop}%
\bibitem [{\citenamefont {Huang}\ \emph {et~al.}(2025)\citenamefont {Huang}, \citenamefont {Colmenarez}, \citenamefont {M\"uller},\ and\ \citenamefont {Diehl}}]{DiehlEtAl2025}%
  \BibitemOpen
  \bibfield  {author} {\bibinfo {author} {\bibfnamefont {Z.-M.}\ \bibnamefont {Huang}}, \bibinfo {author} {\bibfnamefont {L.}~\bibnamefont {Colmenarez}}, \bibinfo {author} {\bibfnamefont {M.}~\bibnamefont {M\"uller}},\ and\ \bibinfo {author} {\bibfnamefont {S.}~\bibnamefont {Diehl}},\ }\bibfield  {title} {\bibinfo {title} {{Coherent information as a mixed-state topological order parameter of fermions}},\ }\href {https://doi.org/10.1103/fx56-8nvy} {\bibfield  {journal} {\bibinfo  {journal} {Phys. Rev. Res.}\ }\textbf {\bibinfo {volume} {7}},\ \bibinfo {pages} {043009} (\bibinfo {year} {2025})}\BibitemShut {NoStop}%
\bibitem [{\citenamefont {Kim}\ \emph {et~al.}(2025)\citenamefont {Kim}, \citenamefont {von Keyserlingk},\ and\ \citenamefont {Lamacraft}}]{KimKeyserlingkLamacraft}%
  \BibitemOpen
  \bibfield  {author} {\bibinfo {author} {\bibfnamefont {S.~W.~P.}\ \bibnamefont {Kim}}, \bibinfo {author} {\bibfnamefont {C.}~\bibnamefont {von Keyserlingk}},\ and\ \bibinfo {author} {\bibfnamefont {A.}~\bibnamefont {Lamacraft}},\ }\href@noop {} {\bibinfo {title} {{Measurement-induced phase transitions in quantum inference problems and quantum hidden Markov models}}} (\bibinfo {year} {2025}),\ \Eprint {https://arxiv.org/abs/2504.08888} {arXiv:2504.08888 [cond-mat.stat-mech]} \BibitemShut {NoStop}%
\bibitem [{\citenamefont {Le~Doussal}\ and\ \citenamefont {Harris}(1988)}]{LeDoussalHarrisI}%
  \BibitemOpen
  \bibfield  {author} {\bibinfo {author} {\bibfnamefont {P.}~\bibnamefont {Le~Doussal}}\ and\ \bibinfo {author} {\bibfnamefont {A.~B.}\ \bibnamefont {Harris}},\ }\bibfield  {title} {\bibinfo {title} {{Location of the Ising Spin-Glass Multicritical Point on Nishimori's Line}},\ }\href {https://doi.org/10.1103/PhysRevLett.61.625} {\bibfield  {journal} {\bibinfo  {journal} {Phys. Rev. Lett.}\ }\textbf {\bibinfo {volume} {61}},\ \bibinfo {pages} {625} (\bibinfo {year} {1988})}\BibitemShut {NoStop}%
\bibitem [{\citenamefont {Le~Doussal}\ and\ \citenamefont {Harris}(1989)}]{LeDoussalHarrisII}%
  \BibitemOpen
  \bibfield  {author} {\bibinfo {author} {\bibfnamefont {P.}~\bibnamefont {Le~Doussal}}\ and\ \bibinfo {author} {\bibfnamefont {A.~B.}\ \bibnamefont {Harris}},\ }\bibfield  {title} {\bibinfo {title} {{\ensuremath{\epsilon} expansion for the Nishimori multicritical point of spin glasses}},\ }\href {https://doi.org/10.1103/PhysRevB.40.9249} {\bibfield  {journal} {\bibinfo  {journal} {Phys. Rev. B}\ }\textbf {\bibinfo {volume} {40}},\ \bibinfo {pages} {9249} (\bibinfo {year} {1989})}\BibitemShut {NoStop}%
\bibitem [{\citenamefont {{Georges, A.}}\ \emph {et~al.}(1987)\citenamefont {{Georges, A.}}, \citenamefont {{Hansel, D.}}, \citenamefont {{Le Doussal, P.}},\ and\ \citenamefont {{Maillard, J.M.}}}]{GeorgesHanselLeDoussalMaillard}%
  \BibitemOpen
  \bibfield  {author} {\bibinfo {author} {\bibnamefont {{Georges, A.}}}, \bibinfo {author} {\bibnamefont {{Hansel, D.}}}, \bibinfo {author} {\bibnamefont {{Le Doussal, P.}}},\ and\ \bibinfo {author} {\bibnamefont {{Maillard, J.M.}}},\ }\bibfield  {title} {\bibinfo {title} {{The replica momenta of a spin-glass and the phase diagram of n-colour Ashkin-Teller models}},\ }\href {https://doi.org/10.1051/jphys:019870048010100} {\bibfield  {journal} {\bibinfo  {journal} {J. Phys. France}\ }\textbf {\bibinfo {volume} {48}},\ \bibinfo {pages} {1} (\bibinfo {year} {1987})}\BibitemShut {NoStop}%
\bibitem [{\citenamefont {Gruzberg}\ \emph {et~al.}(2001)\citenamefont {Gruzberg}, \citenamefont {Read},\ and\ \citenamefont {Ludwig}}]{GRL2001}%
  \BibitemOpen
  \bibfield  {author} {\bibinfo {author} {\bibfnamefont {I.~A.}\ \bibnamefont {Gruzberg}}, \bibinfo {author} {\bibfnamefont {N.}~\bibnamefont {Read}},\ and\ \bibinfo {author} {\bibfnamefont {A.~W.~W.}\ \bibnamefont {Ludwig}},\ }\bibfield  {title} {\bibinfo {title} {{Random-bond Ising model in two dimensions: The Nishimori line and supersymmetry}},\ }\href {https://doi.org/10.1103/PhysRevB.63.104422} {\bibfield  {journal} {\bibinfo  {journal} {Phys. Rev. B}\ }\textbf {\bibinfo {volume} {63}},\ \bibinfo {pages} {104422} (\bibinfo {year} {2001})}\BibitemShut {NoStop}%
\bibitem [{\citenamefont {Jian}\ \emph {et~al.}(2020)\citenamefont {Jian}, \citenamefont {You}, \citenamefont {Vasseur},\ and\ \citenamefont {Ludwig}}]{JianYouVasseurLudwig2019}%
  \BibitemOpen
  \bibfield  {author} {\bibinfo {author} {\bibfnamefont {C.-M.}\ \bibnamefont {Jian}}, \bibinfo {author} {\bibfnamefont {Y.-Z.}\ \bibnamefont {You}}, \bibinfo {author} {\bibfnamefont {R.}~\bibnamefont {Vasseur}},\ and\ \bibinfo {author} {\bibfnamefont {A.~W.~W.}\ \bibnamefont {Ludwig}},\ }\bibfield  {title} {\bibinfo {title} {{Measurement-induced criticality in random quantum circuits}},\ }\href {https://doi.org/10.1103/PhysRevB.101.104302} {\bibfield  {journal} {\bibinfo  {journal} {Phys. Rev. B}\ }\textbf {\bibinfo {volume} {101}},\ \bibinfo {pages} {104302} (\bibinfo {year} {2020})},\ \Eprint {https://arxiv.org/abs/1908.08051} {arXiv:1908.08051 [cond-mat.stat-mech]} \BibitemShut {NoStop}%
\bibitem [{\citenamefont {Bao}\ \emph {et~al.}(2020)\citenamefont {Bao}, \citenamefont {Choi},\ and\ \citenamefont {Altman}}]{BaoChoiAltman2019}%
  \BibitemOpen
  \bibfield  {author} {\bibinfo {author} {\bibfnamefont {Y.}~\bibnamefont {Bao}}, \bibinfo {author} {\bibfnamefont {S.}~\bibnamefont {Choi}},\ and\ \bibinfo {author} {\bibfnamefont {E.}~\bibnamefont {Altman}},\ }\bibfield  {title} {\bibinfo {title} {Theory of the phase transition in random unitary circuits with measurements},\ }\href {https://doi.org/10.1103/PhysRevB.101.104301} {\bibfield  {journal} {\bibinfo  {journal} {Phys. Rev. B}\ }\textbf {\bibinfo {volume} {101}},\ \bibinfo {pages} {104301} (\bibinfo {year} {2020})}\BibitemShut {NoStop}%
\bibitem [{Note1()}]{Note1}%
  \BibitemOpen
  \bibinfo {note} {Standing for `Positive Operator Valued Measure'}\BibitemShut {NoStop}%
\bibitem [{\citenamefont {Merz}\ and\ \citenamefont {Chalker}(2002)}]{MerzChalker}%
  \BibitemOpen
  \bibfield  {author} {\bibinfo {author} {\bibfnamefont {F.}~\bibnamefont {Merz}}\ and\ \bibinfo {author} {\bibfnamefont {J.~T.}\ \bibnamefont {Chalker}},\ }\bibfield  {title} {\bibinfo {title} {{Negative scaling dimensions and conformal invariance at the Nishimori point in the $\ifmmode\pm\else\textpm\fi{}J$ random-bond Ising model}},\ }\href {https://doi.org/10.1103/PhysRevB.66.054413} {\bibfield  {journal} {\bibinfo  {journal} {Phys. Rev. B}\ }\textbf {\bibinfo {volume} {66}},\ \bibinfo {pages} {054413} (\bibinfo {year} {2002})}\BibitemShut {NoStop}%
\bibitem [{\citenamefont {Kadanoff}\ and\ \citenamefont {Ceva}(1971)}]{KadanoffCeva}%
  \BibitemOpen
  \bibfield  {author} {\bibinfo {author} {\bibfnamefont {L.~P.}\ \bibnamefont {Kadanoff}}\ and\ \bibinfo {author} {\bibfnamefont {H.}~\bibnamefont {Ceva}},\ }\bibfield  {title} {\bibinfo {title} {{Determination of an Operator Algebra for the Two-Dimensional Ising Model}},\ }\href {https://doi.org/10.1103/PhysRevB.3.3918} {\bibfield  {journal} {\bibinfo  {journal} {Phys. Rev. B}\ }\textbf {\bibinfo {volume} {3}},\ \bibinfo {pages} {3918} (\bibinfo {year} {1971})}\BibitemShut {NoStop}%
\bibitem [{\citenamefont {Patil}\ and\ \citenamefont {Ludwig}(2025)}]{PatilLudwig20251}%
  \BibitemOpen
  \bibfield  {author} {\bibinfo {author} {\bibfnamefont {R.~A.}\ \bibnamefont {Patil}}\ and\ \bibinfo {author} {\bibfnamefont {A.~W.~W.}\ \bibnamefont {Ludwig}},\ }\href {https://arxiv.org/abs/2507.07959} {\bibinfo {title} {Shannon entropy of the measurement record at measurement-dominated criticality and rg flow: A c-theorem for effective central charge and a g-theorem for effective boundary entropy}} (\bibinfo {year} {2025}),\ \Eprint {https://arxiv.org/abs/2507.07959} {arXiv:2507.07959 [cond-mat.stat-mech]} \BibitemShut {NoStop}%
\bibitem [{\citenamefont {Zabalo}\ \emph {et~al.}(2022)\citenamefont {Zabalo}, \citenamefont {Gullans}, \citenamefont {Wilson}, \citenamefont {Vasseur}, \citenamefont {Ludwig}, \citenamefont {Gopalakrishnan}, \citenamefont {Huse},\ and\ \citenamefont {Pixley}}]{ZabaloGullansWilsonVasseurLudwigGopalakrishnanHusePixley}%
  \BibitemOpen
  \bibfield  {author} {\bibinfo {author} {\bibfnamefont {A.}~\bibnamefont {Zabalo}}, \bibinfo {author} {\bibfnamefont {M.~J.}\ \bibnamefont {Gullans}}, \bibinfo {author} {\bibfnamefont {J.~H.}\ \bibnamefont {Wilson}}, \bibinfo {author} {\bibfnamefont {R.}~\bibnamefont {Vasseur}}, \bibinfo {author} {\bibfnamefont {A.~W.~W.}\ \bibnamefont {Ludwig}}, \bibinfo {author} {\bibfnamefont {S.}~\bibnamefont {Gopalakrishnan}}, \bibinfo {author} {\bibfnamefont {D.~A.}\ \bibnamefont {Huse}},\ and\ \bibinfo {author} {\bibfnamefont {J.~H.}\ \bibnamefont {Pixley}},\ }\bibfield  {title} {\bibinfo {title} {{Operator Scaling Dimensions and Multifractality at Measurement-Induced Transitions}},\ }\href {https://doi.org/10.1103/PhysRevLett.128.050602} {\bibfield  {journal} {\bibinfo  {journal} {Phys. Rev. Lett.}\ }\textbf {\bibinfo {volume} {128}},\ \bibinfo {pages} {050602} (\bibinfo {year} {2022})}\BibitemShut {NoStop}%
\bibitem [{\citenamefont {Kumar}\ \emph {et~al.}(2024)\citenamefont {Kumar}, \citenamefont {Aziz}, \citenamefont {Chakraborty}, \citenamefont {Ludwig}, \citenamefont {Gopalakrishnan}, \citenamefont {Pixley},\ and\ \citenamefont {Vasseur}}]{KumarKemalChakrabortyLudwigGopalakrishnanPixleyVasseur}%
  \BibitemOpen
  \bibfield  {author} {\bibinfo {author} {\bibfnamefont {A.}~\bibnamefont {Kumar}}, \bibinfo {author} {\bibfnamefont {K.}~\bibnamefont {Aziz}}, \bibinfo {author} {\bibfnamefont {A.}~\bibnamefont {Chakraborty}}, \bibinfo {author} {\bibfnamefont {A.~W.~W.}\ \bibnamefont {Ludwig}}, \bibinfo {author} {\bibfnamefont {S.}~\bibnamefont {Gopalakrishnan}}, \bibinfo {author} {\bibfnamefont {J.~H.}\ \bibnamefont {Pixley}},\ and\ \bibinfo {author} {\bibfnamefont {R.}~\bibnamefont {Vasseur}},\ }\bibfield  {title} {\bibinfo {title} {{Boundary transfer matrix spectrum of measurement-induced transitions}},\ }\href {https://doi.org/10.1103/PhysRevB.109.014303} {\bibfield  {journal} {\bibinfo  {journal} {Phys. Rev. B}\ }\textbf {\bibinfo {volume} {109}},\ \bibinfo {pages} {014303} (\bibinfo {year} {2024})}\BibitemShut {NoStop}%
\bibitem [{\citenamefont {Honecker}\ \emph {et~al.}(2001)\citenamefont {Honecker}, \citenamefont {Picco},\ and\ \citenamefont {Pujol}}]{HoneckerPiccoPujol2001}%
  \BibitemOpen
  \bibfield  {author} {\bibinfo {author} {\bibfnamefont {A.}~\bibnamefont {Honecker}}, \bibinfo {author} {\bibfnamefont {M.}~\bibnamefont {Picco}},\ and\ \bibinfo {author} {\bibfnamefont {P.}~\bibnamefont {Pujol}},\ }\bibfield  {title} {\bibinfo {title} {{Universality Class of the Nishimori Point in the 2D $\ifmmode\pm\else\textpm\fi{}\mathit{J}$ Random-Bond Ising Model}},\ }\href {https://doi.org/10.1103/PhysRevLett.87.047201} {\bibfield  {journal} {\bibinfo  {journal} {Phys. Rev. Lett.}\ }\textbf {\bibinfo {volume} {87}},\ \bibinfo {pages} {047201} (\bibinfo {year} {2001})}\BibitemShut {NoStop}%
\bibitem [{\citenamefont {Picco}\ \emph {et~al.}(2006)\citenamefont {Picco}, \citenamefont {Honecker},\ and\ \citenamefont {Pujol}}]{PiccoPujolHonecher2006}%
  \BibitemOpen
  \bibfield  {author} {\bibinfo {author} {\bibfnamefont {M.}~\bibnamefont {Picco}}, \bibinfo {author} {\bibfnamefont {A.}~\bibnamefont {Honecker}},\ and\ \bibinfo {author} {\bibfnamefont {P.}~\bibnamefont {Pujol}},\ }\bibfield  {title} {\bibinfo {title} {{Strong disorder fixed points in the two-dimensional random-bond Ising model}},\ }\href {https://doi.org/10.1088/1742-5468/2006/09/P09006} {\bibfield  {journal} {\bibinfo  {journal} {Journal of Statistical Mechanics: Theory and Experiment}\ }\textbf {\bibinfo {volume} {2006}},\ \bibinfo {pages} {P09006} (\bibinfo {year} {2006})}\BibitemShut {NoStop}%
\bibitem [{\citenamefont {Henley}(2004)}]{CLHenley_2004}%
  \BibitemOpen
  \bibfield  {author} {\bibinfo {author} {\bibfnamefont {C.~L.}\ \bibnamefont {Henley}},\ }\bibfield  {title} {\bibinfo {title} {{From classical to quantum dynamics at Rokhsar–Kivelson points}},\ }\href {https://doi.org/10.1088/0953-8984/16/11/045} {\bibfield  {journal} {\bibinfo  {journal} {Journal of Physics: Condensed Matter}\ }\textbf {\bibinfo {volume} {16}},\ \bibinfo {pages} {S891} (\bibinfo {year} {2004})}\BibitemShut {NoStop}%
\bibitem [{\citenamefont {Chang}\ \emph {et~al.}(2025)\citenamefont {Chang}, \citenamefont {Dommes}, \citenamefont {Erramilli}, \citenamefont {Homrich}, \citenamefont {Kravchuk}, \citenamefont {Liu}, \citenamefont {Mitchell}, \citenamefont {Poland},\ and\ \citenamefont {Simmons-Duffin}}]{Chang2025}%
  \BibitemOpen
  \bibfield  {author} {\bibinfo {author} {\bibfnamefont {C.-H.}\ \bibnamefont {Chang}}, \bibinfo {author} {\bibfnamefont {V.}~\bibnamefont {Dommes}}, \bibinfo {author} {\bibfnamefont {R.~S.}\ \bibnamefont {Erramilli}}, \bibinfo {author} {\bibfnamefont {A.}~\bibnamefont {Homrich}}, \bibinfo {author} {\bibfnamefont {P.}~\bibnamefont {Kravchuk}}, \bibinfo {author} {\bibfnamefont {A.}~\bibnamefont {Liu}}, \bibinfo {author} {\bibfnamefont {M.~S.}\ \bibnamefont {Mitchell}}, \bibinfo {author} {\bibfnamefont {D.}~\bibnamefont {Poland}},\ and\ \bibinfo {author} {\bibfnamefont {D.}~\bibnamefont {Simmons-Duffin}},\ }\bibfield  {title} {\bibinfo {title} {{Bootstrapping the 3d Ising stress tensor}},\ }\href {https://doi.org/10.1007/JHEP03(2025)136} {\bibfield  {journal} {\bibinfo  {journal} {Journal of High Energy Physics}\ }\textbf {\bibinfo {volume} {2025}},\ \bibinfo {pages} {136} (\bibinfo {year} {2025})}\BibitemShut {NoStop}%
\bibitem [{\citenamefont {El-Showk}\ \emph {et~al.}(2012)\citenamefont {El-Showk}, \citenamefont {Paulos}, \citenamefont {Poland}, \citenamefont {Rychkov}, \citenamefont {Simmons-Duffin},\ and\ \citenamefont {Vichi}}]{PhysRevD.86.025022}%
  \BibitemOpen
  \bibfield  {author} {\bibinfo {author} {\bibfnamefont {S.}~\bibnamefont {El-Showk}}, \bibinfo {author} {\bibfnamefont {M.~F.}\ \bibnamefont {Paulos}}, \bibinfo {author} {\bibfnamefont {D.}~\bibnamefont {Poland}}, \bibinfo {author} {\bibfnamefont {S.}~\bibnamefont {Rychkov}}, \bibinfo {author} {\bibfnamefont {D.}~\bibnamefont {Simmons-Duffin}},\ and\ \bibinfo {author} {\bibfnamefont {A.}~\bibnamefont {Vichi}},\ }\bibfield  {title} {\bibinfo {title} {{Solving the 3D Ising model with the conformal bootstrap}},\ }\href {https://doi.org/10.1103/PhysRevD.86.025022} {\bibfield  {journal} {\bibinfo  {journal} {Phys. Rev. D}\ }\textbf {\bibinfo {volume} {86}},\ \bibinfo {pages} {025022} (\bibinfo {year} {2012})}\BibitemShut {NoStop}%
\bibitem [{Note2()}]{Note2}%
  \BibitemOpen
  \bibinfo {note} {Note that the logarithmic corrections to scaling {of} the spin-spin correlation function at the (unmeasured) {classical} Ising critical point in $D=4$ dimensions, will also appear in the EA correlation function {at} the {{\protect \it higher} Nishimori critical point} {in the $D=4$ dimensional} learning phase diagram.}\BibitemShut {Stop}%
\bibitem [{\citenamefont {P\"utz}\ \emph {et~al.}(2025{\natexlab{b}})\citenamefont {P\"utz}, \citenamefont {Vasseur}, \citenamefont {Ludwig}, \citenamefont {Trebst},\ and\ \citenamefont {Zhu}}]{putz2025flownishimoriuniversalityweakly}%
  \BibitemOpen
  \bibfield  {author} {\bibinfo {author} {\bibfnamefont {M.}~\bibnamefont {P\"utz}}, \bibinfo {author} {\bibfnamefont {R.}~\bibnamefont {Vasseur}}, \bibinfo {author} {\bibfnamefont {A.~W.}\ \bibnamefont {Ludwig}}, \bibinfo {author} {\bibfnamefont {S.}~\bibnamefont {Trebst}},\ and\ \bibinfo {author} {\bibfnamefont {G.-Y.}\ \bibnamefont {Zhu}},\ }\bibfield  {title} {\bibinfo {title} {{Flow to Nishimori Universality in Weakly Monitored Quantum Circuits with Qubit Loss}},\ }\href {https://doi.org/10.1103/ygfz-crvp} {\bibfield  {journal} {\bibinfo  {journal} {PRX Quantum}\ }\textbf {\bibinfo {volume} {6}},\ \bibinfo {pages} {040372} (\bibinfo {year} {2025}{\natexlab{b}})}\BibitemShut {NoStop}%
\bibitem [{Note3()}]{Note3}%
  \BibitemOpen
  \bibinfo {note} {{The (undeformed) toric code wavefunction can be written as an equal weight superposition of certain eigenstates of the $\{\protect \hat {\sigma }^{z}_{ij}\}$ operators~\cite {KITAEV20032}, where each eigenstate in the superposition corresponds to a loop configuration on the dual square lattice {where} the loops cut-through the links $\langle ij\rangle $ with $\protect \hat {\sigma }^z_{ij}=-1$ on the original lattice. The factor of $e^{\protect \frac {\beta }{2}\DOTSB \sum@ \slimits@ _{\langle ij\rangle }{\protect \hat {\sigma }^{z}_{ij}} }$ acting on the (undeformed) toric code ground state in Eq~\protect \eqref {EqDeformedToricCodeWF} assigns (up to an overall constant) a factor of $e^{-\beta \times (\protect \text {length of the loop})}$ to each loop in every loop configuration state. Each loop configuration state can be associated to a domain wall configuration of classical spins $\{\sigma _i\}$ defined on the sites (instead of the links) of the original square lattice, where {the}
  configuration of spins $\{\sigma _i\}$ is uniquely defined up to a global {$\protect \mathbb {Z}_2$ flip}, and the weight $e^{-{\beta } (\protect \text {length of the loop})}$ is precisely the Boltzmann weight of the domain wall {in} the $2D$ {classical} Ising {model} in Eq.~\protect \eqref {Eq2DIsingHamiltonian} at (inverse) temperature $\beta /2$. Therefore, the deformed toric code wavefunction in Eq.~\protect \eqref {EqDeformedToricCodeWF} and the RK wavefunction in Eq.~\protect \eqref {EqRKWF} for the $2D$ classical Ising model are exactly dual to each other. Finally, following the above mapping, it is also clear that the operator $\protect \hat {Z}_{ij}$ in the deformed toric code wavefunction in Eq.~\protect \eqref {EqDeformedToricCodeWF} corresponds to the bond-energy operator $\protect \hat {\sigma }^z_{i}\protect \hat {\sigma }^z_{j}$ for the RK wavefunction in Eq.~\protect \eqref {EqRKWF}.}}\BibitemShut {Stop}%
\bibitem [{Note4()}]{Note4}%
  \BibitemOpen
  \bibinfo {note} {{The operator $\protect \hat {\sigma }^z_{k}$ and {its} eigenstates $\vert \{\sigma _i\}\rangle $ are naturally defined as $\protect \hat {\sigma }^z_k\vert \{\sigma _i\}\rangle =\sigma _k\vert \{\sigma _i\}\rangle $.}}\BibitemShut {Stop}%
\bibitem [{Note5()}]{Note5}%
  \BibitemOpen
  \bibinfo {note} {{Although, we note that the boundary conditions decide the ground state of the toric code on the cylinder (surface code), and hence are important in evaluations of properties like coherent information (see the discussion in Ref.~\cite {PutzGarrattNishimoriTrebstZhu}). In this work, we will be only interested in calculating bulk correlation functions that distinguish different phases in the phase diagram in Fig.~\ref {fig:2DGaussianMeasurements}}}\BibitemShut {NoStop}%
\bibitem [{\citenamefont {Sourlas}(1994)}]{Sourlas_1994}%
  \BibitemOpen
  \bibfield  {author} {\bibinfo {author} {\bibfnamefont {N.}~\bibnamefont {Sourlas}},\ }\bibfield  {title} {\bibinfo {title} {{Spin Glasses, Error-Correcting Codes and Finite-Temperature Decoding}},\ }\href {https://doi.org/10.1209/0295-5075/25/3/001} {\bibfield  {journal} {\bibinfo  {journal} {Europhysics Letters}\ }\textbf {\bibinfo {volume} {25}},\ \bibinfo {pages} {159} (\bibinfo {year} {1994})}\BibitemShut {NoStop}%
\bibitem [{\citenamefont {Gullans}\ and\ \citenamefont {Huse}(2020)}]{gullans2020scalable}%
  \BibitemOpen
  \bibfield  {author} {\bibinfo {author} {\bibfnamefont {M.~J.}\ \bibnamefont {Gullans}}\ and\ \bibinfo {author} {\bibfnamefont {D.~A.}\ \bibnamefont {Huse}},\ }\bibfield  {title} {\bibinfo {title} {{Scalable Probes of Measurement-Induced Criticality}},\ }\href {https://doi.org/10.1103/PhysRevLett.125.070606} {\bibfield  {journal} {\bibinfo  {journal} {Phys. Rev. Lett.}\ }\textbf {\bibinfo {volume} {125}},\ \bibinfo {pages} {070606} (\bibinfo {year} {2020})}\BibitemShut {NoStop}%
\bibitem [{\citenamefont {Zhang}\ \emph {et~al.}(2025)\citenamefont {Zhang}, \citenamefont {Xu}, \citenamefont {Zhang}, \citenamefont {Xu}, \citenamefont {Bi},\ and\ \citenamefont {Luo}}]{zhang2025strongtoweakspontaneousbreaking1form}%
  \BibitemOpen
  \bibfield  {author} {\bibinfo {author} {\bibfnamefont {C.}~\bibnamefont {Zhang}}, \bibinfo {author} {\bibfnamefont {Y.}~\bibnamefont {Xu}}, \bibinfo {author} {\bibfnamefont {J.-H.}\ \bibnamefont {Zhang}}, \bibinfo {author} {\bibfnamefont {C.}~\bibnamefont {Xu}}, \bibinfo {author} {\bibfnamefont {Z.}~\bibnamefont {Bi}},\ and\ \bibinfo {author} {\bibfnamefont {Z.-X.}\ \bibnamefont {Luo}},\ }\bibfield  {title} {\bibinfo {title} {{Strong-to-weak spontaneous breaking of 1-form symmetry and intrinsically mixed topological order}},\ }\href {https://doi.org/10.1103/PhysRevB.111.115137} {\bibfield  {journal} {\bibinfo  {journal} {Phys. Rev. B}\ }\textbf {\bibinfo {volume} {111}},\ \bibinfo {pages} {115137} (\bibinfo {year} {2025})}\BibitemShut {NoStop}%
\bibitem [{Note6()}]{Note6}%
  \BibitemOpen
  \bibinfo {note} {{The line of transitions between the dephased `classical memory' (`spin glass') phase and the trivial `no memory' (ferromagnet) phase [gray line in Fig.~\ref {fig:2DGaussianMeasurements}] also meets the other two line of transitions at the tricritical point. This line of transitions flows under RG to the point of {perfect} (projective) measurements corresponding to $\Delta =\infty $ at the $2D$ classical Ising critical point~\cite {PutzGarrattNishimoriTrebstZhu,NahumJacobsen}.}}\BibitemShut {Stop}%
\bibitem [{Note7()}]{Note7}%
  \BibitemOpen
  \bibinfo {note} {Note that {\begin {align*} &\protect \overline {\langle \protect \hat {\protect \mathcal {O}}_1\rangle _{\protect \vec {m}}}=\DOTSB \sum@ \slimits@ _{\protect \vec {m}}\protect \tilde {P}(\protect \vec {m})\langle \Psi _{\protect \vec {m}}\vert \protect \hat {\protect \mathcal {O}}_1\vert \Psi _{\protect \vec {m}}\rangle \\&=\DOTSB \sum@ \slimits@ _{\protect \vec {m}}\langle RK\vert \protect \hat {K}^{\dagger }_{\protect \vec {m}}\protect \hat {K}_{\protect \vec {m}}\vert RK\rangle \protect \frac {\langle RK\vert \protect \hat {K}^{\dagger }_{\protect \vec {m}}\protect \hat {\protect \mathcal {O}}_1\protect \hat {K}_{\protect \vec {m}}\vert RK\rangle }{\langle RK\vert \protect \hat {K}^{\dagger }_{\protect \vec {m}}\protect \hat {K}_{\protect \vec {m}}\vert RK\rangle }\\ &=\DOTSB \sum@ \slimits@ _{\protect \vec {m}}{\langle RK\vert \protect \hat {K}^{\dagger }_{\protect \vec {m}}\protect \hat {\protect \mathcal {O}}_1\protect \hat {K}_{\protect \vec {m}}\vert RK\rangle }\\&={\langle RK\vert
  \protect \hat {\protect \mathcal {O}}_1\DOTSB \sum@ \slimits@ _{\protect \vec {m}}\protect \hat {K}^{\dagger }_{\protect \vec {m}}\protect \hat {K}_{\protect \vec {m}}\vert RK\rangle }={\langle RK\vert \protect \hat {\protect \mathcal {O}}_1\vert RK\rangle }=\langle \protect \hat {\protect \mathcal {O}}_1\rangle , \end {align*} where we have commuted $\protect \hat {\protect \mathcal {O}}_1$ through the Kraus operator $\protect \hat {K}_{\protect \vec {m}}$, since both are diagonal in $\vert \{\sigma _i\}\rangle $ basis, and {then} used the POVM condition~[Eq.~\protect \eqref {EqFirstMomentIsEqualToTheUnmeasured}].}}\BibitemShut {Stop}%
\bibitem [{\citenamefont {Kramers}\ and\ \citenamefont {Wannier}(1941)}]{KramersWannier1941}%
  \BibitemOpen
  \bibfield  {author} {\bibinfo {author} {\bibfnamefont {H.~A.}\ \bibnamefont {Kramers}}\ and\ \bibinfo {author} {\bibfnamefont {G.~H.}\ \bibnamefont {Wannier}},\ }\bibfield  {title} {\bibinfo {title} {{Statistics of the Two-Dimensional Ferromagnet}},\ }\href {https://doi.org/10.1103/PhysRev.60.252} {\bibfield  {journal} {\bibinfo  {journal} {Phys. Rev.}\ }\textbf {\bibinfo {volume} {60}},\ \bibinfo {pages} {252} (\bibinfo {year} {1941})},\ \bibinfo {note} {parts I and II; see also Phys. Rev. \textbf{60}, 263 (1941)}\BibitemShut {NoStop}%
\bibitem [{Note8()}]{Note8}%
  \BibitemOpen
  \bibinfo {note} {{The location $(\beta =0,\Delta \approx 1)$ of the ordinary Nishimori critical point can be obtained by exact mapping of the replica $R\rightarrow 1$ theory in Eq.~\protect \eqref {EqLatticeReplicaTheory} on the $\beta =0$ line to the Nishimori line in the replica $R\rightarrow 0$ RBIM with \protect \textit {Gaussian} bond disorder~\cite {McMillan,PiccoPujolHonecher2006}. This should be contrasted with the location of the ordinary Nishimori critical point in the phase diagram of Ref.~\cite {PutzGarrattNishimoriTrebstZhu} for binary measurements, where it occurs at $(\beta =0, \protect \tilde {\gamma }\approx 0.782)$. The latter difference in the locations of the critical point is of course non-universal.}}\BibitemShut {Stop}%
\bibitem [{\citenamefont {McMillan}(1984)}]{McMillan}%
  \BibitemOpen
  \bibfield  {author} {\bibinfo {author} {\bibfnamefont {W.~L.}\ \bibnamefont {McMillan}},\ }\bibfield  {title} {\bibinfo {title} {{Domain-wall renormalization-group study of the two-dimensional random Ising model}},\ }\href {https://doi.org/10.1103/PhysRevB.29.4026} {\bibfield  {journal} {\bibinfo  {journal} {Phys. Rev. B}\ }\textbf {\bibinfo {volume} {29}},\ \bibinfo {pages} {4026} (\bibinfo {year} {1984})}\BibitemShut {NoStop}%
\bibitem [{\citenamefont {de~Queiroz}\ and\ \citenamefont {Stinchcombe}(2003)}]{deQueirozStinchcombe2003}%
  \BibitemOpen
  \bibfield  {author} {\bibinfo {author} {\bibfnamefont {S.~L.~A.}\ \bibnamefont {de~Queiroz}}\ and\ \bibinfo {author} {\bibfnamefont {R.~B.}\ \bibnamefont {Stinchcombe}},\ }\bibfield  {title} {\bibinfo {title} {{Correlation-function distributions at the Nishimori point of two-dimensional Ising spin glasses}},\ }\href {https://doi.org/10.1103/PhysRevB.68.144414} {\bibfield  {journal} {\bibinfo  {journal} {Phys. Rev. B}\ }\textbf {\bibinfo {volume} {68}},\ \bibinfo {pages} {144414} (\bibinfo {year} {2003})}\BibitemShut {NoStop}%
\bibitem [{\citenamefont {Feller}(1966)}]{fellerintroduction}%
  \BibitemOpen
  \bibfield  {author} {\bibinfo {author} {\bibfnamefont {W.}~\bibnamefont {Feller}},\ }\href@noop {} {\emph {\bibinfo {title} {{An Introduction to Probability Theory and Its Applications, Volume 2}}}}\ (\bibinfo  {publisher} {J.Wiley and Sons, New York},\ \bibinfo {year} {1966})\BibitemShut {NoStop}%
\bibitem [{\citenamefont {Ludwig}\ and\ \citenamefont {Cardy}(1987)}]{LUDWIGCardy}%
  \BibitemOpen
  \bibfield  {author} {\bibinfo {author} {\bibfnamefont {A.~W.}\ \bibnamefont {Ludwig}}\ and\ \bibinfo {author} {\bibfnamefont {J.~L.}\ \bibnamefont {Cardy}},\ }\bibfield  {title} {\bibinfo {title} {{Perturbative evaluation of the conformal anomaly at new critical points with applications to random systems}},\ }\href {https://doi.org/https://doi.org/10.1016/0550-3213(87)90362-2} {\bibfield  {journal} {\bibinfo  {journal} {Nuclear Physics B}\ }\textbf {\bibinfo {volume} {285}},\ \bibinfo {pages} {687} (\bibinfo {year} {1987})}\BibitemShut {NoStop}%
\bibitem [{\citenamefont {Cardy}(2013)}]{CardyLog2013}%
  \BibitemOpen
  \bibfield  {author} {\bibinfo {author} {\bibfnamefont {J.}~\bibnamefont {Cardy}},\ }\bibfield  {title} {\bibinfo {title} {{Logarithmic conformal field theories as limits of ordinary CFTs and some physical applications}},\ }\href {https://doi.org/10.1088/1751-8113/46/49/494001} {\bibfield  {journal} {\bibinfo  {journal} {Journal of Physics A: Mathematical and Theoretical}\ }\textbf {\bibinfo {volume} {46}},\ \bibinfo {pages} {494001} (\bibinfo {year} {2013})}\BibitemShut {NoStop}%
\bibitem [{\citenamefont {Georges}\ \emph {et~al.}(1985)\citenamefont {Georges}, \citenamefont {Hansel}, \citenamefont {Le~Doussal},\ and\ \citenamefont {Bouchaud}}]{Georges1985b}%
  \BibitemOpen
  \bibfield  {author} {\bibinfo {author} {\bibfnamefont {A.}~\bibnamefont {Georges}}, \bibinfo {author} {\bibfnamefont {D.}~\bibnamefont {Hansel}}, \bibinfo {author} {\bibfnamefont {P.}~\bibnamefont {Le~Doussal}},\ and\ \bibinfo {author} {\bibfnamefont {J.-P.}\ \bibnamefont {Bouchaud}},\ }\bibfield  {title} {\bibinfo {title} {{Exact properties of spin glasses. II. Nishimori's line : new results and physical implications}},\ }\href {https://doi.org/10.1051/jphys:0198500460110182700} {\bibfield  {journal} {\bibinfo  {journal} {J. Phys. France}\ }\textbf {\bibinfo {volume} {46}},\ \bibinfo {pages} {1827} (\bibinfo {year} {1985})}\BibitemShut {NoStop}%
\bibitem [{\citenamefont {Patil}\ and\ \citenamefont {Ludwig}(2026)}]{PatilLudwig20262}%
  \BibitemOpen
  \bibfield  {author} {\bibinfo {author} {\bibfnamefont {R.~A.}\ \bibnamefont {Patil}}\ and\ \bibinfo {author} {\bibfnamefont {A.~W.~W.}\ \bibnamefont {Ludwig}},\ }\href@noop {} {} (\bibinfo {year} {2026}),\ \bibinfo {note} {in preparation}\BibitemShut {NoStop}%
\bibitem [{\citenamefont {Read}\ and\ \citenamefont {Ludwig}(2000)}]{ReadLudwig2001}%
  \BibitemOpen
  \bibfield  {author} {\bibinfo {author} {\bibfnamefont {N.}~\bibnamefont {Read}}\ and\ \bibinfo {author} {\bibfnamefont {A.~W.~W.}\ \bibnamefont {Ludwig}},\ }\bibfield  {title} {\bibinfo {title} {{Absence of a metallic phase in random-bond Ising models in two dimensions: Applications to disordered superconductors and paired quantum Hall states}},\ }\href {https://doi.org/10.1103/PhysRevB.63.024404} {\bibfield  {journal} {\bibinfo  {journal} {Phys. Rev. B}\ }\textbf {\bibinfo {volume} {63}},\ \bibinfo {pages} {024404} (\bibinfo {year} {2000})}\BibitemShut {NoStop}%
\bibitem [{Note9()}]{Note9}%
  \BibitemOpen
  \bibinfo {note} {{This is because the zeroth measurement-averaged moment {trivially} has zero scaling dimension.}}\BibitemShut {Stop}%
\bibitem [{\citenamefont {McCoy}\ and\ \citenamefont {Wu}(2014)}]{mccoy2014two}%
  \BibitemOpen
  \bibfield  {author} {\bibinfo {author} {\bibfnamefont {B.}~\bibnamefont {McCoy}}\ and\ \bibinfo {author} {\bibfnamefont {T.}~\bibnamefont {Wu}},\ }\href {https://books.google.de/books?id=nB_-AgAAQBAJ} {\emph {\bibinfo {title} {{The Two-Dimensional Ising Model: Second Edition}}}},\ Dover books on physics\ (\bibinfo  {publisher} {Dover Publications},\ \bibinfo {year} {2014})\BibitemShut {NoStop}%
\bibitem [{\citenamefont {Schultz}\ \emph {et~al.}(1964)\citenamefont {Schultz}, \citenamefont {Mattis},\ and\ \citenamefont {Lieb}}]{SML64isingfermion}%
  \BibitemOpen
  \bibfield  {author} {\bibinfo {author} {\bibfnamefont {T.}~\bibnamefont {Schultz}}, \bibinfo {author} {\bibfnamefont {D.}~\bibnamefont {Mattis}},\ and\ \bibinfo {author} {\bibfnamefont {E.}~\bibnamefont {Lieb}},\ }\bibfield  {title} {\bibinfo {title} {{Two-Dimensional Ising Model as a Soluble Problem of Many Fermions}},\ }\href {https://doi.org/10.1103/RevModPhys.36.856} {\bibfield  {journal} {\bibinfo  {journal} {Rev. Mod. Phys.}\ }\textbf {\bibinfo {volume} {36}},\ \bibinfo {pages} {856} (\bibinfo {year} {1964})}\BibitemShut {NoStop}%
\bibitem [{\citenamefont {Cardy}(1996)}]{Cardy_book}%
  \BibitemOpen
  \bibfield  {author} {\bibinfo {author} {\bibfnamefont {J.}~\bibnamefont {Cardy}},\ }\bibinfo {title} {Ch. 8, random systems},\ in\ \href@noop {} {\emph {\bibinfo {booktitle} {Scaling and Renormalization in Statistical Physics}}},\ \bibinfo {series and number} {Cambridge Lecture Notes in Physics}\ (\bibinfo  {publisher} {Cambridge University Press},\ \bibinfo {year} {1996})\ p.\ \bibinfo {pages} {145–168}\BibitemShut {NoStop}%
\bibitem [{\citenamefont {Komargodski}\ and\ \citenamefont {Simmons-Duffin}(2017)}]{KomargodskiSimmons-Duffin}%
  \BibitemOpen
  \bibfield  {author} {\bibinfo {author} {\bibfnamefont {Z.}~\bibnamefont {Komargodski}}\ and\ \bibinfo {author} {\bibfnamefont {D.}~\bibnamefont {Simmons-Duffin}},\ }\bibfield  {title} {\bibinfo {title} {{The random-bond Ising model in 2.01 and 3 dimensions}},\ }\href {https://doi.org/10.1088/1751-8121/aa6087} {\bibfield  {journal} {\bibinfo  {journal} {Journal of Physics A: Mathematical and Theoretical}\ }\textbf {\bibinfo {volume} {50}},\ \bibinfo {pages} {154001} (\bibinfo {year} {2017})}\BibitemShut {NoStop}%
\bibitem [{Note10()}]{Note10}%
  \BibitemOpen
  \bibinfo {note} {{One way to see this distinction is to note that at both of the perturbative fixed points the scaling dimension for the measurement-averaged second moment is given, in the respective epsilon expansions, by $2X^{D}_{\sigma }+O(\epsilon )$, where $X^{D}_{\sigma }$ is the scaling dimension of the spin operator at the unmeasured $D$-dimensional classical Ising critical point. While, on the other hand, we know \protect \textit {exactly} that the scaling dimension of the measurement-averaged second moment of the spin-spin correlation function is equal to just $X^{D}_{\sigma }$ at the {\protect \it higher} Nishimori critical point in $D$-dimensions.}}\BibitemShut {Stop}%
\bibitem [{\citenamefont {Patil}\ and\ \citenamefont {Ludwig}(2024)}]{PatilLudwig2024}%
  \BibitemOpen
  \bibfield  {author} {\bibinfo {author} {\bibfnamefont {R.~A.}\ \bibnamefont {Patil}}\ and\ \bibinfo {author} {\bibfnamefont {A.~W.~W.}\ \bibnamefont {Ludwig}},\ }\href@noop {} {\bibinfo {title} {{Highly complex novel critical behavior from the intrinsic randomness of quantum mechanical measurements on critical ground states -- a controlled renormalization group analysis}}} (\bibinfo {year} {2024}),\ \Eprint {https://arxiv.org/abs/2409.02107} {arXiv:2409.02107 [cond-mat.stat-mech]} \BibitemShut {NoStop}%
\bibitem [{\citenamefont {Li}\ \emph {et~al.}(2024)\citenamefont {Li}, \citenamefont {Vasseur}, \citenamefont {Fisher},\ and\ \citenamefont {Ludwig}}]{LiVasseurFisherLudwig}%
  \BibitemOpen
  \bibfield  {author} {\bibinfo {author} {\bibfnamefont {Y.}~\bibnamefont {Li}}, \bibinfo {author} {\bibfnamefont {R.}~\bibnamefont {Vasseur}}, \bibinfo {author} {\bibfnamefont {M.~P.~A.}\ \bibnamefont {Fisher}},\ and\ \bibinfo {author} {\bibfnamefont {A.~W.~W.}\ \bibnamefont {Ludwig}},\ }\bibfield  {title} {\bibinfo {title} {{Statistical mechanics model for Clifford random tensor networks and monitored quantum circuits}},\ }\href {https://doi.org/10.1103/PhysRevB.109.174307} {\bibfield  {journal} {\bibinfo  {journal} {Phys. Rev. B}\ }\textbf {\bibinfo {volume} {109}},\ \bibinfo {pages} {174307} (\bibinfo {year} {2024})}\BibitemShut {NoStop}%
\bibitem [{\citenamefont {Jian}\ \emph {et~al.}(2023)\citenamefont {Jian}, \citenamefont {Shapourian}, \citenamefont {Bauer},\ and\ \citenamefont {Ludwig}}]{JianShapourianBauerLudwig}%
  \BibitemOpen
  \bibfield  {author} {\bibinfo {author} {\bibfnamefont {C.-M.}\ \bibnamefont {Jian}}, \bibinfo {author} {\bibfnamefont {H.}~\bibnamefont {Shapourian}}, \bibinfo {author} {\bibfnamefont {B.}~\bibnamefont {Bauer}},\ and\ \bibinfo {author} {\bibfnamefont {A.~W.~W.}\ \bibnamefont {Ludwig}},\ }\href {https://arxiv.org/abs/2302.09094} {\bibinfo {title} {{Measurement-induced entanglement transitions in quantum circuits of non-interacting fermions: Born-rule versus forced measurements}}} (\bibinfo {year} {2023}),\ \Eprint {https://arxiv.org/abs/2302.09094} {arXiv:2302.09094 [cond-mat.stat-mech]} \BibitemShut {NoStop}%
\bibitem [{\citenamefont {Nishimori}\ and\ \citenamefont {Stephen}(1983)}]{NishimoriStephan}%
  \BibitemOpen
  \bibfield  {author} {\bibinfo {author} {\bibfnamefont {H.}~\bibnamefont {Nishimori}}\ and\ \bibinfo {author} {\bibfnamefont {M.~J.}\ \bibnamefont {Stephen}},\ }\bibfield  {title} {\bibinfo {title} {{Gauge-invariant frustrated Potts spin-glass}},\ }\href {https://doi.org/10.1103/PhysRevB.27.5644} {\bibfield  {journal} {\bibinfo  {journal} {Phys. Rev. B}\ }\textbf {\bibinfo {volume} {27}},\ \bibinfo {pages} {5644} (\bibinfo {year} {1983})}\BibitemShut {NoStop}%
\bibitem [{\citenamefont {Ozeki}\ and\ \citenamefont {Nishimori}(1993)}]{OzekiNishimori}%
  \BibitemOpen
  \bibfield  {author} {\bibinfo {author} {\bibfnamefont {Y.}~\bibnamefont {Ozeki}}\ and\ \bibinfo {author} {\bibfnamefont {H.}~\bibnamefont {Nishimori}},\ }\bibfield  {title} {\bibinfo {title} {{Phase diagram of gauge glasses}},\ }\href {https://doi.org/10.1088/0305-4470/26/14/009} {\bibfield  {journal} {\bibinfo  {journal} {Journal of Physics A: Mathematical and General}\ }\textbf {\bibinfo {volume} {26}},\ \bibinfo {pages} {3399} (\bibinfo {year} {1993})}\BibitemShut {NoStop}%
\bibitem [{\citenamefont {Jacobsen}\ and\ \citenamefont {Picco}(2002)}]{JacobsenPicco}%
  \BibitemOpen
  \bibfield  {author} {\bibinfo {author} {\bibfnamefont {J.~L.}\ \bibnamefont {Jacobsen}}\ and\ \bibinfo {author} {\bibfnamefont {M.}~\bibnamefont {Picco}},\ }\bibfield  {title} {\bibinfo {title} {{Phase diagram and critical exponents of a Potts gauge glass}},\ }\href {https://doi.org/10.1103/PhysRevE.65.026113} {\bibfield  {journal} {\bibinfo  {journal} {Phys. Rev. E}\ }\textbf {\bibinfo {volume} {65}},\ \bibinfo {pages} {026113} (\bibinfo {year} {2002})}\BibitemShut {NoStop}%
\bibitem [{\citenamefont {Honecker}\ \emph {et~al.}(2002)\citenamefont {Honecker}, \citenamefont {Jacobsen}, \citenamefont {Picco},\ and\ \citenamefont {Pujol}}]{HoneckerJacobsenPiccoPujol}%
  \BibitemOpen
  \bibfield  {author} {\bibinfo {author} {\bibfnamefont {A.}~\bibnamefont {Honecker}}, \bibinfo {author} {\bibfnamefont {J.~L.}\ \bibnamefont {Jacobsen}}, \bibinfo {author} {\bibfnamefont {M.}~\bibnamefont {Picco}},\ and\ \bibinfo {author} {\bibfnamefont {P.}~\bibnamefont {Pujol}},\ }\bibinfo {title} {{Nishimori Point in Random-Bond Ising and Potts Models in 2D}},\ in\ \href {https://doi.org/10.1007/978-94-010-0514-2_23} {\emph {\bibinfo {booktitle} {Statistical Field Theories}}},\ \bibinfo {editor} {edited by\ \bibinfo {editor} {\bibfnamefont {A.}~\bibnamefont {Cappelli}}\ and\ \bibinfo {editor} {\bibfnamefont {G.}~\bibnamefont {Mussardo}}}\ (\bibinfo  {publisher} {Springer Netherlands},\ \bibinfo {address} {Dordrecht},\ \bibinfo {year} {2002})\ pp.\ \bibinfo {pages} {251--261}\BibitemShut {NoStop}%
\bibitem [{\citenamefont {Patil}\ \emph {et~al.}(2026)\citenamefont {Patil}, \citenamefont {P\"utz}, \citenamefont {Trebst}, \citenamefont {Zhu},\ and\ \citenamefont {Ludwig}}]{zenodo_higher_nishimori}%
  \BibitemOpen
  \bibfield  {author} {\bibinfo {author} {\bibfnamefont {R.~A.}\ \bibnamefont {Patil}}, \bibinfo {author} {\bibfnamefont {M.}~\bibnamefont {P\"utz}}, \bibinfo {author} {\bibfnamefont {S.}~\bibnamefont {Trebst}}, \bibinfo {author} {\bibfnamefont {G.-Y.}\ \bibnamefont {Zhu}},\ and\ \bibinfo {author} {\bibfnamefont {A.~W.~W.}\ \bibnamefont {Ludwig}},\ }\bibfield  {title} {\bibinfo {title} {{Data for ``Higher Nishimori Criticality and Exact Results at the Learning Transition of Deformed Toric Codes''}},\ }\bibfield  {journal} {\bibinfo  {journal} {Zenodo}\ }\href {https://doi.org/10.5281/zenodo.19348702} {10.5281/zenodo.19348702} (\bibinfo {year} {2026})\BibitemShut {NoStop}%
\bibitem [{Note11()}]{Note11}%
  \BibitemOpen
  \bibinfo {note} {{Note that the expression in Eq.~\protect \eqref {EqEACorrelationFunctionGaussianMeasurementsAfterChangeOfVariables} \protect \textit {almost} resembles the definition of the EA correlation function for the RBIM with Gaussian bond disorder, where $\beta $ is the mean bond strength and $\Delta $ quantifies the variance in bond strengths. The only difference is that, unlike the EA correlation function for RBIM, we do not have a square of $\DOTSB \sum@ \slimits@ _{\{\sigma _i\}}e^{\DOTSB \sum@ \slimits@ _{\langle ij\rangle } \protect \mathfrak {n}_{ij} \sigma _i\sigma _j}$ in the denominator on the RHS of Eq.~\protect \eqref {EqEACorrelationFunctionGaussianMeasurementsAfterChangeOfVariables}. This {distinction} is, of course, crucial for the difference in corresponding replica limits $R\rightarrow 0$ and $R\rightarrow 1$ for the RBIM and {the} $2D$ Ising measurement/learning problem, respectively.}}\BibitemShut {Stop}%
\bibitem [{Note12()}]{Note12}%
  \BibitemOpen
  \bibinfo {note} {Note that by its definition in Eq.~\protect \eqref {EqDualCorrFunInAGivenMeasTrajectory}, $\langle \mu _{\protect \tilde {x}}\mu _{\protect \tilde {y}}\rangle _{\protect \vec {\protect \mathfrak {m}}}$ is non-negative in every {measurement} trajectory $\protect \vec {\protect \mathfrak {m}}$.}\BibitemShut {Stop}%
\bibitem [{\citenamefont {Cardy}(2008)}]{CardyLesHouches2008}%
  \BibitemOpen
  \bibfield  {author} {\bibinfo {author} {\bibfnamefont {J.}~\bibnamefont {Cardy}},\ }\href {https://arxiv.org/abs/0807.3472} {\bibinfo {title} {{Conformal Field Theory and Statistical Mechanics}}} (\bibinfo {year} {2008}),\ \Eprint {https://arxiv.org/abs/0807.3472} {arXiv:0807.3472 [cond-mat.stat-mech]} \BibitemShut {NoStop}%
\bibitem [{\citenamefont {Lukyanov}\ and\ \citenamefont {Terras}(2003)}]{LUKYANOVTERRAS}%
  \BibitemOpen
  \bibfield  {author} {\bibinfo {author} {\bibfnamefont {S.}~\bibnamefont {Lukyanov}}\ and\ \bibinfo {author} {\bibfnamefont {V.}~\bibnamefont {Terras}},\ }\bibfield  {title} {\bibinfo {title} {{Long-distance asymptotics of spin–spin correlation functions for the XXZ spin chain}},\ }\href {https://doi.org/https://doi.org/10.1016/S0550-3213(02)01141-0} {\bibfield  {journal} {\bibinfo  {journal} {Nuclear Physics B}\ }\textbf {\bibinfo {volume} {654}},\ \bibinfo {pages} {323} (\bibinfo {year} {2003})}\BibitemShut {NoStop}%
\bibitem [{\citenamefont {{Zamolodchikov}}(1986)}]{Zamolodchikovctheorem}%
  \BibitemOpen
  \bibfield  {author} {\bibinfo {author} {\bibfnamefont {A.~B.}\ \bibnamefont {{Zamolodchikov}}},\ }\bibfield  {title} {\bibinfo {title} {{``Irreversibility'' of the flux of the renormalization group in a 2D field theory}},\ }\href@noop {} {\bibfield  {journal} {\bibinfo  {journal} {Soviet Journal of Experimental and Theoretical Physics Letters}\ }\textbf {\bibinfo {volume} {43}},\ \bibinfo {pages} {730} (\bibinfo {year} {1986})}\BibitemShut {NoStop}%
\bibitem [{\citenamefont {Cardy}(1988)}]{Cardy1988}%
  \BibitemOpen
  \bibfield  {author} {\bibinfo {author} {\bibfnamefont {J.~L.}\ \bibnamefont {Cardy}},\ }\bibfield  {title} {\bibinfo {title} {{Central Charge and Universal Combinations of Amplitudes in Two-Dimensional Theories Away from Criticality}},\ }\href {https://doi.org/10.1103/PhysRevLett.60.2709} {\bibfield  {journal} {\bibinfo  {journal} {Phys. Rev. Lett.}\ }\textbf {\bibinfo {volume} {60}},\ \bibinfo {pages} {2709} (\bibinfo {year} {1988})}\BibitemShut {NoStop}%
\bibitem [{Note13()}]{Note13}%
  \BibitemOpen
  \bibinfo {note} {See also Ref.~\cite {NambiKhannaAllocaIadecolaHickeyVasseurWilson}, where an analogous argument was very recently made in the case of $(1+1)D$ measurement-induced phase transitions.}\BibitemShut {Stop}%
\bibitem [{Note14()}]{Note14}%
  \BibitemOpen
  \bibinfo {note} {Equivalently, note that there are simply \protect \textit {no} replicas at $R=0$.}\BibitemShut {Stop}%
\bibitem [{Note15()}]{Note15}%
  \BibitemOpen
  \bibinfo {note} {Since $f(R)$ is zero at $R=0$ and $R=1$, and \protect \textit {if} it does not have any other zeroes in $R\in (0,1)$, then $f(R)$ does not change sign in the interval $R\in (0,1)$. Since the slope of $f(R)$ defined $(df(R)/dR)|_{R=1}=\lim _{\delta \rightarrow 0^{+}}\protect \frac {(f(1)-f(1-\delta ))}{\delta }=-\lim _{\delta \rightarrow 0^{+}}\protect \frac {(f(1-\delta ))}{\delta }$ is negative at $R=1$ (See Eqs. \protect \eqref {EqCEffIsGreaterThanHalf} and \protect \eqref {EqDefF(R)}), the slope $(df(R)/dR)|_{R=0}=\lim _{\delta \rightarrow 0^{+}}\protect \frac {(f(0+\delta )-f(0))}{\delta }=\lim _{\delta \rightarrow 0^{+}}\protect \frac {(f(0+\delta ))}{\delta }$ at $R=0$ must be non-negative. {This is because $f(1-\delta )$ and $f(0+\delta )$ should be both positive for $\delta >0$} [We are assuming that $f(R)$ is differentiable in the closed interval $R\in [0,1]$.]}\BibitemShut {NoStop}%
\bibitem [{Note16()}]{Note16}%
  \BibitemOpen
  \bibinfo {note} {Factorization of a general matrix into a product of an orthogonal (or unitary) and an upper triangular matrix.}\BibitemShut {Stop}%
\bibitem [{\citenamefont {Nambi}\ \emph {et~al.}(2026)\citenamefont {Nambi}, \citenamefont {Khanna}, \citenamefont {Allocca}, \citenamefont {Iadecola}, \citenamefont {Hickey}, \citenamefont {Vasseur},\ and\ \citenamefont {Wilson}}]{NambiKhannaAllocaIadecolaHickeyVasseurWilson}%
  \BibitemOpen
  \bibfield  {author} {\bibinfo {author} {\bibfnamefont {D.}~\bibnamefont {Nambi}}, \bibinfo {author} {\bibfnamefont {K.}~\bibnamefont {Khanna}}, \bibinfo {author} {\bibfnamefont {A.}~\bibnamefont {Allocca}}, \bibinfo {author} {\bibfnamefont {T.}~\bibnamefont {Iadecola}}, \bibinfo {author} {\bibfnamefont {C.}~\bibnamefont {Hickey}}, \bibinfo {author} {\bibfnamefont {R.}~\bibnamefont {Vasseur}},\ and\ \bibinfo {author} {\bibfnamefont {J.~H.}\ \bibnamefont {Wilson}},\ }\href {https://arxiv.org/abs/2603.15744} {\bibinfo {title} {{Post-selected Criticality in Measurement-induced Phase Transitions}}} (\bibinfo {year} {2026}),\ \Eprint {https://arxiv.org/abs/2603.15744} {arXiv:2603.15744 [quant-ph]} \BibitemShut {NoStop}%
\end{thebibliography}%

\end{document}